\documentclass[twocolumn, tighten, trackchanges]{aastex62}
\usepackage{xspace}
\usepackage{graphicx}
\usepackage{amssymb}
\usepackage{amsmath}
\usepackage{subfigure}
\usepackage{hyperref}
\usepackage{footmisc}

\usepackage{natbib} 

\newcommand{\myemail}{zpace@astro.wisc.edu}
\newcommand{\Dn}{D$_n$4000\xspace}
\newcommand{\HdeltaA}{H${\delta}_A$\xspace}
\newcommand{\HgammaA}{H${\gamma}_A$\xspace}
\newcommand{\mplv}{MPL-8\xspace}
\newcommand{\mplvngal}{6779\xspace}
\newcommand{\mplvfull}{MaNGA Product Launch 8\xspace}
\newcommand{\ntestgalaxies}{473\xspace}
\newcommand{\ntestspaxels}{537333\xspace}
\newcommand{\nrungalaxies}{1773\xspace}
\newcommand{\nSFHs}{4000\xspace}
\newcommand{\nsubsample}{10\xspace}

\newcommand{\logml}[1]{\ensuremath{\log \Upsilon^*_{#1}}}
\newcommand{\qtydev}[1]{\ensuremath{\Delta #1}}
\newcommand{\qtywid}[1]{\ensuremath{\sigma_{#1}}}
\newcommand{\qtydevwid}[1]{\ensuremath{\frac{\qtydev{#1}}{\qtywid{#1}}}}

\defcitealias{pace_19b_pca}{Paper \textsc{ii}}
\defcitealias{chen_pca}{C12}

\revised{\today}
\submitjournal{ApJS}

\shorttitle{PCA Stellar Masses: Paper \textsc{i}}
\shortauthors{Pace et al.}

\begin{document}

\title{Resolved and Integrated Stellar Masses in the SDSS-\textsc{iv}/\textsc{MaNGA} Survey, Paper \textsc{i}: \\ PCA spectral fitting \& stellar mass-to-light ratio estimates}

\correspondingauthor{Zachary J. Pace}
\email{\myemail}

\author[0000-0003-4843-4185]{Zachary J. {Pace}}
\affil{Department of Astronomy, University of Wisconsin-Madison,
       475 N Charter St., Madison, WI 53706}

\author[0000-0003-3097-5178]{Christy {Tremonti}}
\affil{Department of Astronomy, University of Wisconsin-Madison,
       475 N Charter St., Madison, WI 53706}

\author{Yanmei {Chen}}
\affil{Department of Astronomy, Nanjing University,
       Nanjing 210093, China}

\author{Adam L. {Schaefer}}
\affil{Department of Astronomy, University of Wisconsin-Madison,
       475 N Charter St., Madison, WI 53706}

\author[0000-0002-3131-4374]{Matthew A. {Bershady}}
\affil{Department of Astronomy, University of Wisconsin-Madison,
       475 N Charter St., Madison, WI 53706}
\affil{South African Astronomical Observatory,
       P.O. Box 9, Observatory 7935, Cape Town South Africa}

\author[0000-0003-1809-6920]{Kyle B. {Westfall}}
\affil{UCO/Lick Observatory, University of California, Santa Cruz,
       1156 High St., Santa Cruz, CA 95064}

\author[0000-0003-0946-6176]{M\'{e}d\'{e}ric Boquien}
\affil{Centro de Astronom\'{i}a, Universidad de Antofagasta,
       Avenida Angamos 601, Antofagasta 1270300, Chile}

\author[0000-0001-7883-8434]{Kate {Rowlands}}
\affil{Center for Astrophysical Sciences, Department of Physics and Astronomy,
       Johns Hopkins University \\
       3400 North Charles Street, Baltimore, MD 21218, USA}

\author[0000-0001-8085-5890]{Brett {Andrews}}
\affil{TT PACC, Department of Physics and Astronomy, University of Pittsburgh,
       Pittsburgh, PA 15260, USA}

\author[0000-0002-8725-1069]{Joel R. {Brownstein}}
\affil{Department of Physics and Astronomy, University of Utah,
       115 S. 1400 E., Salt Lake City, UT 84112, USA}

\author[0000-0002-7339-3170]{Niv {Drory}}
\affil{McDonald Observatory, The University of Texas at Austin,
       1 University Station, Austin, TX 78712, USA}

\author[0000-0002-6047-1010]{David {Wake}}
\affil{Department of Physics, University of North Carolina Asheville,
       One University Heights, Asheville, NC 28804, USA}



\begin{abstract}
We present a method of fitting optical spectra of galaxies using a basis set of six vectors obtained from principal component analysis (PCA) of a library of synthetic spectra of 40000 star formation histories (SFHs). Using this library, we provide estimates of resolved effective stellar mass-to-light ratio ($\Upsilon^*$) for thousands of galaxies from the SDSS-IV/MaNGA integral-field spectroscopic survey. Using a testing framework built on additional synthetic SFHs, we show that the estimates of \logml{i} are reliable (as are their uncertainties) at a variety of signal-to-noise ratios, stellar metallicities, and dust attenuation conditions. Finally, we describe the future release of the resolved stellar mass-to-light ratios as a SDSS-IV/MaNGA Value-Added Catalog (VAC) and provide a link to the software used to conduct this analysis\footnote{\label{footnote:software_link}The software can be found at \url{https://github.com/zpace/pcay}.}
\end{abstract}



\section{Introduction}
\label{sec:intro}

A galaxy's stellar mass is one of its most important physical properties, reflecting its current evolutionary state and future pathway. On the whole, more massive systems tend to possess older stellar populations \citep{gallazzi_05_age_metallicity, gallazzi_06} with very little current star formation \citep{kauffmann_heckman_white_03, balogh_04_cmd, baldry_06_massquenching}, a small gas mass fraction \citep{mcgaugh_de-blok_97}, higher gas-phase metallicity \citep{tremonti_mz}, and stellar populations enhanced in $\alpha$-elements relative to iron \citep{thomas_maraston_korn_04, thomas_05}. Fundamentally, a galaxy's stellar mass indicates the total mass of the dark matter halo in which it is embedded \citep{yang_03_smhmr, behroozi_wechsler_conroy_13, somerville_behroozi_2018}: the higher the mass of the dark matter halo, the more evolved the galaxy tends to be, and the lesser the galaxy's capacity for future star formation.

Traditionally, two methods have been used to estimate galaxy stellar mass: kinematics and stellar population analysis. By measuring the average motions of stars, the dynamical mass (a distinct but related property which includes both baryonic and dark matter), can be determined. The DiskMass Survey \citep[DMS,][]{diskmass_i} used measurements of the vertical stellar and gas velocity field and stellar velocity dispersion $\sigma^*_z$, in concert with inferred values of disk scale height $h_z$ to estimate the azimuthally-averaged dynamical mass surface density $\Sigma^{\rm dyn}$ of 30 local, low-inclination disk galaxies within several radial bins \citep{diskmass_vii}. However, dynamical measurements are subject to systematics related to the vertical distribution and scale height of stars, how the vertical velocity is measured \citep{aniyan_freeman_16, aniyan_freeman_18}, and the typical assumption of a constant stellar mass-to-light ratio used in Jeans-based estimates \citep{bernardi_sheth_17}.

The second method of stellar mass estimation relies on comparing photometry or spectroscopy of galaxies to stellar population synthesis (SPS) models. SPS weds theoretical stellar isochrones to theoretical model atmospheres or observed libraries of stellar spectra, under the constraint of the stellar initial mass function (IMF), in order to obtain an estimate of the mass-to-light ratio, and therefore, the mass. \citet{tinsley_72,tinsley_73} defined the fundamentals of this method, combining the analytic expressions for the stellar IMF, star formation rates (SFR), and theory of chemical enrichment. \citet{bell_dejong_01} and \citet{bell_03} later took existing stellar models and described empirical relationships between optical colors and stellar mass-to-light ratios. Other approaches infer a star formation history (SFH) from broadband, multi-wavelength spectral energy distributions (SEDs): in such a case, the starlight itself can be observed in many bands \citep{shapley_05_seds}, or its indirect consequences can also be considered, such as infrared dust emission that arises after stars form \citep{dale_01_infrared_dust}. Software libraries such as \texttt{MagPhys} \citep{da-cunha_charlot_elbaz_08,da-cunha_charlot_11_magphys}, \texttt{Cigale} \citep{burgarella_05, giovannoli_11, serra_11}, and \texttt{Prospector} \citep{leja_johnson_conroy_van-dokkum_17_prospector} take this approach, often (but not always) after adopting a family of SFHs. In short, estimates of stellar mass-to-light are generally made by finding the combination of simple stellar populations (SSPs; i.e., stars of a single age and metallicity) that produces the best match to an observed galaxy spectrum or photometry.

Simple SFH scenarios, such as \citet{bell_03}, produce almost-linear relationships (often referred to as color-mass-to-light relations, or CMLRs) between optical color and the logarithm of stellar mass-to-light ratio. This can be a convenient first tool, but there are significant systematics associated with stellar IMF, metallicity, and attenuation by dust (see Section \ref{subsec:cmlrs}). Often, different CMLRs produce extremely contradictory mass-to-light estimates \citep{mcgaugh_schombert_schombert_14}. We demonstrate below that inferring stellar mass-to-light ratio from optical spectra offers some improvements over CMLRs.

Additionally, certain spectroscopic features---such as the strength of the 4000$\mbox{\AA}$ break \cite[\Dn:][]{bruzual_83, balogh_99, balogh_00}, equivalent width of the H$\delta$ absorption line \citep[\HdeltaA:][]{worthey_ottaviani_97}, and several other atomic and molecular indices \citep[e.g. CN, Mg$b$, NaD:][]{worthey_94}---have been used to estimate mean stellar age, metallicity, activity of recent starbursts, and stellar mass-to-light \citep{kauffmann_heckman_white_03,gallazzi_05_age_metallicity,silchenko_06,wild_kauffmann_07_pca}. Spectral indices are akin to optical colors in that they are a lower-dimensional view on a galaxy's spectrum, but a view designed to effectively capture an informative phase of stellar evolution.

The advent of large spectroscopic surveys with good spectrophotometric calibration has enabled more widespread use of full-spectral fitting: spectra spanning a large fraction of the visible wavelength range offer a much more detailed view on a galaxy's SED, albeit within a smaller overall wavelength range than techniques which simultaneously examine UV, optical, infrared, and radio domains. Many software libraries exist for performing such analysis, including such as \texttt{FIREFLY} \citep{firefly_wilkinson_pmanga}, \texttt{STECKMAP} \citep{steckmap}, \texttt{VESPA} \citep{vespa_tojeiro}, \texttt{pPXF} \citep{cappellari_ppxf}, \texttt{STARLIGHT} \citep{starlight}, and \texttt{Pipe3D} \citep{pipe3d_i, pipe3d_ii}. Very recent developments include techniques which simultaneously consider spectroscopic and photometric measurements \citep{chevallard_charlot_16_beagle, thomas_le-fevre_17, fossati_mendel_18}.

The reliability of the resulting spectral fits is hampered by four main factors. First, certain phases of stellar evolution, such as the thermally-pulsating asymptotic giant branch (TP-AGB) stage, are still poorly understood, and this causes troublesome systematics \citep{maraston_06, marigo_08}. Second, due to the degeneracy between stellar population age and metallicity, it is difficult to map spectral features uniquely to a combination of stellar populations. Modern spectroscopic surveys alleviate this somewhat with the inclusion of the NIR Ca\textsc{ii} triplet \citep{terlevich_caII_89, vazdekis_caII_03}, a feature that is sensitive to Calcium abundance (and secondarily, overall metal abundance) in stars older than 2 Gyr \citep{usher_beckwith_18_CaT}, as well as other spectral indices \citep{spiniello_trager_12, spiniello_trager_14}, but care is still required. Third, the process of stellar population-synthesis relies on fully-populating the parameter space of temperature, surface gravity, metal abundance, and ${\rm [\frac{\alpha}{Fe}] }$ (usually combining multiple stellar libraries, interpolating across un-sampled regions of parameter space, or patching with theoretical libraries). Fourth, it is unclear how to best recover the information contained in spectra: some approaches continue to model spectra as the sum of simple mono-age, mono-abundance SSPs; but it is possible that the resulting numerical freedom may produce un-physical results.

This work's spectral-fitting technique follows \citet[][hereafter \citetalias{chen_pca}]{chen_pca}: in \citetalias{chen_pca}, principal component analysis (PCA) was performed on a ``training library" of synthetic optical spectra (the synthetic spectra were themselves produced using a stochastically-generated family of SFHs), yielding a set of orthogonal ``principal component" (PC) vectors. PCA is a method of finding structure in a high-dimensional point process \citep{jolliffe_1986_pca}, which has been applied in the field of astronomy to such problems as spectral-fitting \citep{budavari_09}, photometric redshifts \citep{cabanac_02}, and classification of quasars \citep{yip_connolly_04, suzuki_06}. PCA transforms data in many dimensions into fewer dimensions in a way that minimizes information loss. If a spectrum containing measurements of flux at $l$ wavelengths is interpreted as a single data point in $l$-dimensional space, a group of $n$ spectra forms a cloud of $n$ points. PCA finds the $q$ vectors (also called ``eigenvectors", ``eigenspectra", or ``PCs") that are best able to mimic the full, $l$-dimensional data. Equivalently, PCA finds a vector space in which the covariance matrix is diagonalized for those $n$ spectra.

In the PCA-based spectral-fitting paradigm, a set of 4--10 eigenspectra were used as a reduced basis set for fitting observed spectra at low-to-moderate signal-to-noise. That is, the best representation was found for each observed spectrum in terms of the eigenspectra. The goodness-of-fit was evaluated in PC space for each model spectrum in the training library: this was used as a weight in constructing a posterior PDF for stellar mass-to-light ratio. In this paradigm, the ``samples" of the PDF are simply the full set of CSP models, which have known stellar mass-to-light ratio. In other words, the training library both defines the eigenspectra and provides samples for a quantity of interest. This process can be carried out on thousands of observed spectra (using tens of thousands of models) simultaneously, and is Bayesian-like because the training library acts as a prior on the ``allowed" values of, for example, stellar mass-to-light ratio. \citet{tinker_mstar} utilized the stellar mass-to-light ratio estimates from \citetalias{chen_pca} to measure the intrinsic scatter of stellar mass at a fixed halo mass for high-mass BOSS galaxies, finding that PCA-based estimates of stellar mass correlated with total halo mass \emph{better} than photometric mass estimates, a result suggesting that the PCA estimates are more accurate.

Though this work uses a method very similar to \citetalias{chen_pca}, we also integrate new model spectra generated with modern isochrones and stellar atmospheres. We apply this method to galaxy spectra from SDSS-IV/MaNGA \citep[Mapping Nearby Galaxies at Apache Point Observatory,][]{bundy15_manga}, an integral-field spectroscopic (IFS) survey of 10,000 nearby galaxies ($z \lesssim 0.15$). The MaNGA survey is designed to enhance the current understanding of galaxy growth and self-regulation by observing galaxies with a wide variety of stellar masses, specific star formation rates, and environments. We produce resolved maps of stellar mass-to-light ratio for a significant fraction of the \mplvngal galaxies included in \mplvfull (\mplv).

The structure of this paper is as follows: in Section \ref{sec:data}, we discuss the MaNGA IFS data; in Section \ref{sec:SFHs}, we detail the procedural generation of star-formation histories and their optical spectra (the ``training data"), a \Dn--\HdeltaA comparison between the model library \& actual observations, and why CMLRs do not recover sufficient detail about the underlying stellar mass-to-light ratio; in Section \ref{sec:method}, we review the underlying mathematics of PCA, present its application to parameter estimation---concentrating, in particular, on why we might expect improvement over traditional methods in the case of IFS data---and examine the reliability of the resulting estimates of stellar mass-to-light ratio in relation to degenerate parameters like metallicity and attenuation; and in Section \ref{sec:discussion}, we provide example maps of stellar mass-to-light ratio for a selection of four late-type galaxies and three early-type galaxies, discussing possible future improvements to the spectral library, and outlining a future release of resolved stellar mass-to-light ratio maps. To complement this work's investigation of random errors in PCA-based estimates of stellar mass-to-light ratio, in \citet[][hereafter \citetalias{pace_19b_pca}]{pace_19b_pca}, we compare resolved maps of stellar mass surface density (derived from stellar mass-to-light ratio estimates made in this work) to estimates of dynamical mass surface density from the DiskMass Survey, in a view on the systematics of PCA-derived stellar mass-to-light ratios; and construct aperture-corrected, total stellar masses for a large sample of MaNGA galaxies.

\section{Data}
\label{sec:data}

This work employs IFS data from the MaNGA survey, part of SDSS-IV \citep{blanton_17_sdss-iv}. MaNGA is the most extensive IFS survey undertaken to date, targeting 10,000 galaxies in the local universe ($0.01 < z < 0.15$), with observations set to complete in early 2020 \citep{bundy15_manga}. MaNGA's instrument is built around the SDSS 2.5-meter telescope at Apache Point Observatory \citep{gunn_sdss_telescope} and the SDSS-BOSS spectrograph \citep{smee_boss_instrument,sdss_boss_dawson_13}, which has a wavelength range of 3600 to 10300 $\mbox{\AA}$ and spectral resolution $R \sim 2000$. The BOSS spectrograph is coupled to close-packed fiber hexabundles (also called integral-field units, or IFUs) with between 19 and 127 fibers subtending 2" apiece on the sky \citep{manga_inst}. The IFUs are secured to the focal plane with a plugplate \citep{sdss_summary}, and are exposed simultaneously. Flux-calibration is accomplished with 12 seven-fiber ``mini-bundles" which observe standard stars simultaneous to science observations \citep{manga_spectrophot}, and sky-subtraction uses 92 single fibers spread across the three-degree focal plane.

MaNGA observations use only dark time, and require three sets of exposures, which are accumulated until a specified threshold signal-to-noise is achieved \citep{manga_progress_yan_16}. Additionally, all constituents of each set of exposures must be taken under similar conditions. A three-point dither pattern is used to increase the spatial sampling such that 99\% of the area within the IFU is exposed to within 1.2\% of the mean exposure time \citep{manga_obs}. This also accomplishes a more uniform sampling of the plane of the sky than non-dithered observations, and gives a closer match to the fiber point-spread function: a typical fiber-convolved point-spread function has FWHM of 2.5" \citep{manga_obs}.

The MaNGA survey primarily targets galaxies from the NASA-Sloan Atlas \texttt{v1\_0\_1} \citep[NSA, ][]{blanton_11_nsa}. The survey's science goals guide the specific target choices made: in particular, two-thirds of targets (the ``Primary+ sample") are covered to 1.5 effective radii ($R_e$), and one-third (the ``Secondary sample") are covered to 2.5 $R_e$. MaNGA targets are selected uniformly in SDSS $i$-band absolute magnitude \citep{fukugita_96_sdss_photo, doi_2010_sdssresponse}, which will result in an approximately-flat distribution in the log of stellar mass \citep{manga_sample_wake_17}. Further, within a prescribed redshift range corresponding to a given absolute magnitude, the MaNGA sample is selected to be volume-limited. Absolute magnitudes have been calculated using K-corrections computed with the \texttt{kcorrect v4\_2} software package \citep{blanton_roweis_07}, assuming a \citet{chabrier03} stellar initial mass function and \citet{BC03} SSPs, and are tabulated in the \texttt{DRPALL} catalog file.

The MaNGA Data Reduction Pipeline \citep[DRP:][]{manga_drp} reduces individual integrations into both row-stacked spectra (RSS), which behave like collections of single-fiber pointings; and rectified data-cubes (CUBE), which are constructed from the RSS with a modified Shepard's algorithm to produce a spatially-uniform grid on the plane of the sky with spaxels measuring 0.5 arcsecond on a side. This work uses CUBE products with logarithmically-uniform wavelength spacing ($d\log \lambda = 10^{-4}$, $d\ln \lambda \approx 2.3 \times 10^{-4}$)\footnote{In this work, the notation $\log$ denotes a base-10 logarithm, and $\ln$ denotes a base-$e$ logarithm.}, also called LOGCUBE products. LOGCUBE products are then analyzed further, using the MaNGA Data Analysis Pipeline \citep[DAP:][]{manga_dap}, which produces resolved measurements (referred to as ``MAPS") of stellar and gas line-of-sight velocity, the stellar continuum, gaseous emission fluxes, and spectral indices.

The \nrungalaxies galaxies analyzed in this study are drawn randomly from \mplvfull (\mplv), an internally-released set of both reduced and analyzed observations numbering \mplvngal galaxies observed between March 2014 and July 2018. \mplv's reduced products number nearly 2100 more than SDSS DR15 \citep{sdss_dr15}, which was released in December 2018. 

\begin{figure}
    \centering
    \includegraphics[width=\columnwidth]{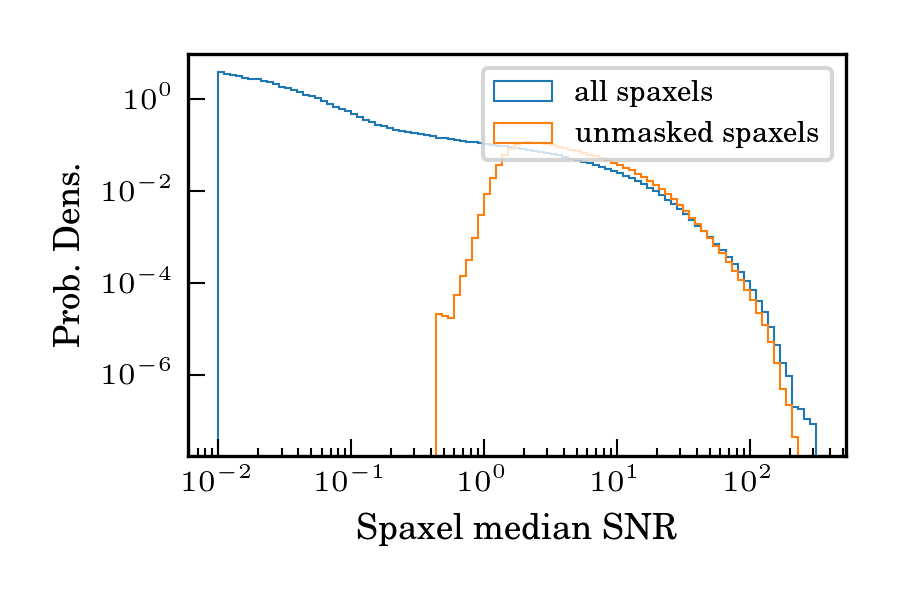}
    \caption{The distributions of median spectral signal-to-noise ratio for all MaNGA spaxels (blue) and those spaxels for which none of the MaNGA DRP data-quality flags indicate potential problems with the estimates obtained in this work (orange)---see Section \ref{subsec:data_quality} for descriptions of data-quality diagnostics, and how channel-specific quality flags inform reliability of mass-to-light ratio estimates.}
    \label{fig:spax_snr}
\end{figure}

This study uses both DRP-LOGCUBE and DAP-MAPS products. The DAP-MAPS products (at this time released only within the SDSS collaboration) are not spatially-binned for the stellar continuum fit \citep[see][]{manga_dap} (i.e., this work uses the ``SPX" products). We apply no sample cuts. The distribution of the median spectral signal-to-noise ratio of all MaNGA spaxels is shown in Figure \ref{fig:spax_snr}: spectral channels flagged at the MaNGA-DRP stage as either having low or no IFU coverage, or with known unreliable measurement, have had their inverse-variance weight set to zero (spaxels with such issues affecting their spectra form the low-SNR tail of the distribution).

\section{The Composite Stellar Population Library}
\label{sec:SFHs}

In order to generate the eigenvectors composing the principal component space, we first generate training data using theoretical models of SFHs. Training spectra are generated by passing a piecewise-continuous star formation history (according to a randomized prescription described below) through a stellar population synthesis library, after assuming a stellar initial mass function (IMF), a set of isochrones, and a stellar library. In this case, the \texttt{fortran} code \texttt{Flexible Stellar Population Synthesis} (\texttt{FSPS}) \citep{fsps_1, fsps_2, fsps_3} and its \texttt{python} bindings \citep{pyfsps_DFM} were used. Padova 2008 \citep{marigo_08} isochrones were adopted. 

The unpublished C3K library of theoretical stellar spectra \citetext{Conroy, in prep.} was used for the population synthesis: the library is based on the Kurucz frameworks (routines and line lists)edit2{, and} the native resolution is $R=10,000$ from 1500$\mbox{\AA}$ to 1.1${\rm \mu}$m. Though most similar studies employ empirical stellar libraries \citep[e.g. MILES or E-MILES:][]{vazdekis10, vazdekis_16_e-miles}, there is at this time no widely-used library which closely matches the MaNGA wavelength range and resolution while covering the full stellar age and metallicity range expected in MaNGA. The latest E-MILES stellar population models, for instance, match MaNGA resolution and wavelength range, but have few stars younger than 0.1 Gyr at solar metallicity and above \citep[see][Figures 5 and 6]{vazdekis_16_e-miles}.

\subsection{SFHs and stellar population properties}

In order for the PCA model to emulate observed galaxy spectra, it must first be ``trained" to recognize what they can look like (that is, PCA ``learns" which wavelengths tend to vary together, and how strongly). By generating a plausible library of SFHs and their associated spectra, we provide the initial guidelines informing how real, observed spectra are fit. Since any fit to an observed spectrum must fall within the circumscribed domain of the training data, it is important to make that training data permissive enough to encompass physical reality. With too restrictive a training set, fits would suffer from additional systematic bias. As such, our SFH template parameters are intended to have weakly informative priors \citep{simpson_rue_14_priors, gelman_hennig_15} which encompass the areas of parameter space that are physically allowable and in line with previous studies (to this point, see the below description of the mass weighted mean stellar age distribution, Figure \ref{fig:mwa_dist}), while allowing only a relatively small proportion of more complex models (e.g., those involving bursts or a transition in SFH behavior).

\begin{figure*}
    \centering
    \includegraphics[width=\textwidth]{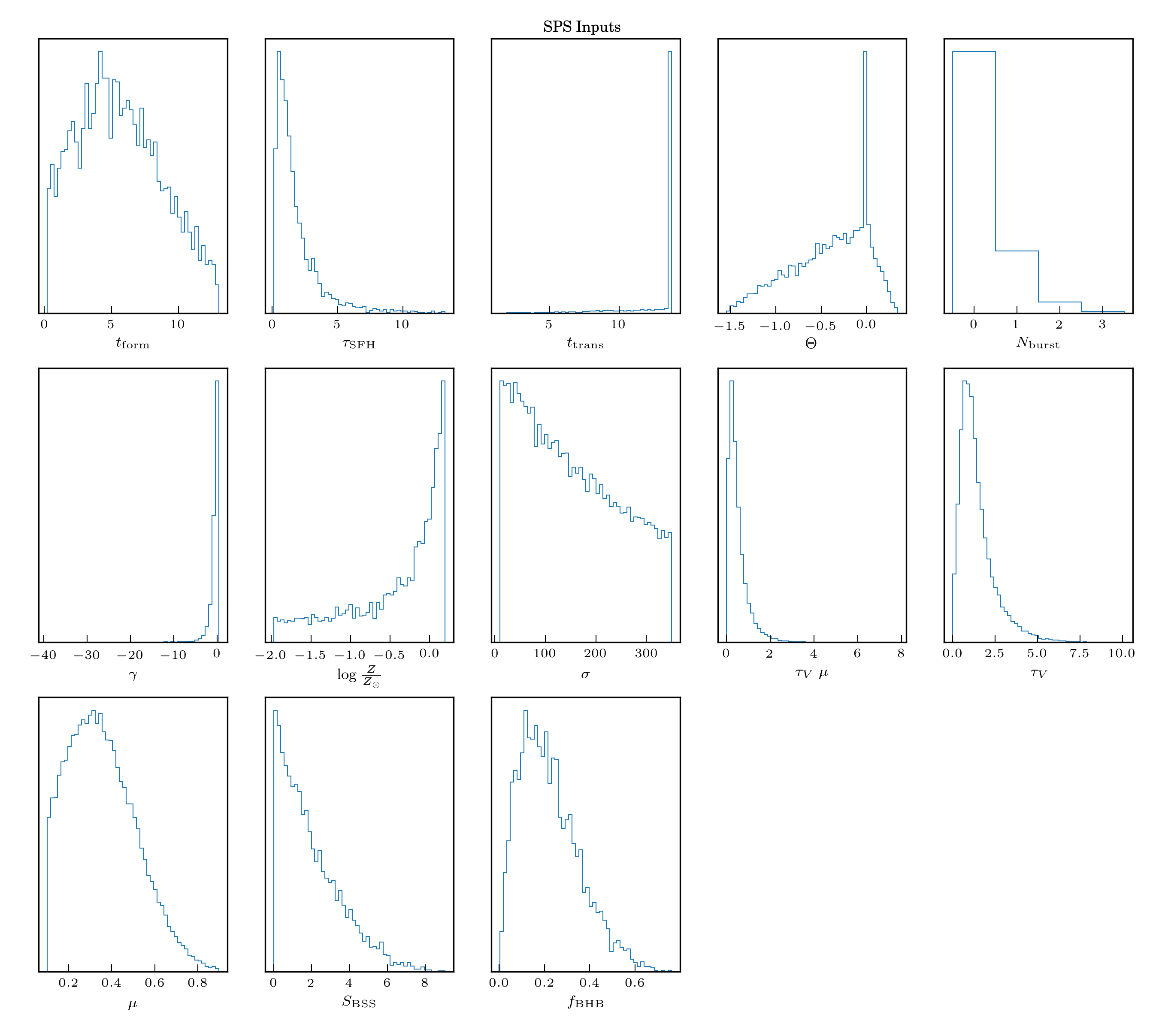}
    \caption{The distributions of the inputs provided to \texttt{FSPS} described in more detail in Section \ref{sec:SPS_params}, left to right and top to bottom: formation time, e-folding time, transition time, transition strength, transition slope, number of bursts, stellar metallicity, stellar velocity dispersion, attenuation, specific frequency of blue straggler stars, and fraction of blue horizontal branch stars. There is no covariance between these parameters.}
    \label{fig:prior_inputparams}
\end{figure*}

\begin{figure*}
    \centering
    \includegraphics[width=\textwidth]{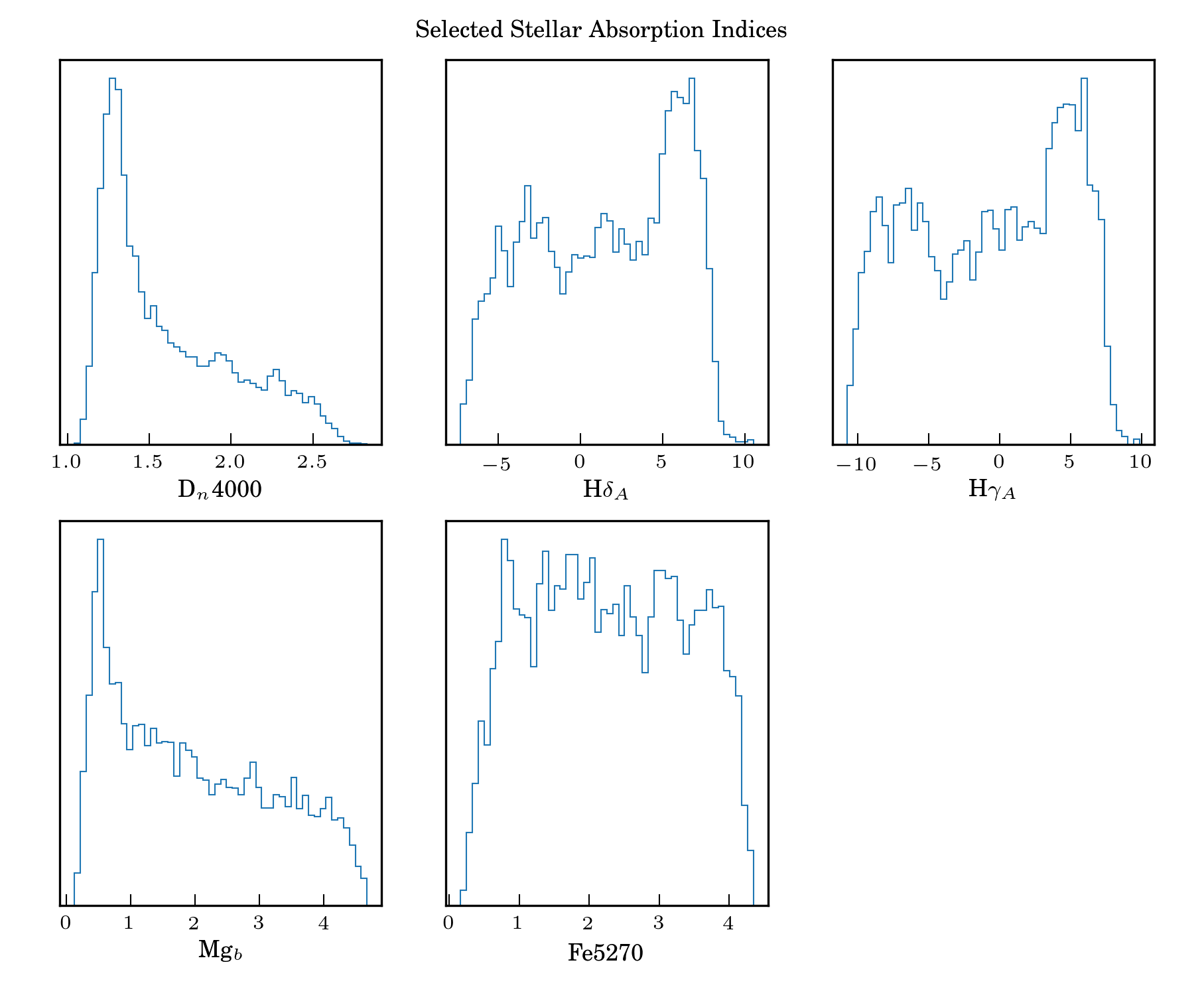}
    \caption{The distributions of five absorption indices in our synthetic training data: \Dn, \HdeltaA, \HgammaA, Mg$_b$, and Fe5270.}
    \label{fig:prior_specinds}
\end{figure*}

\begin{figure*}
    \centering
    \includegraphics[width=\textwidth]{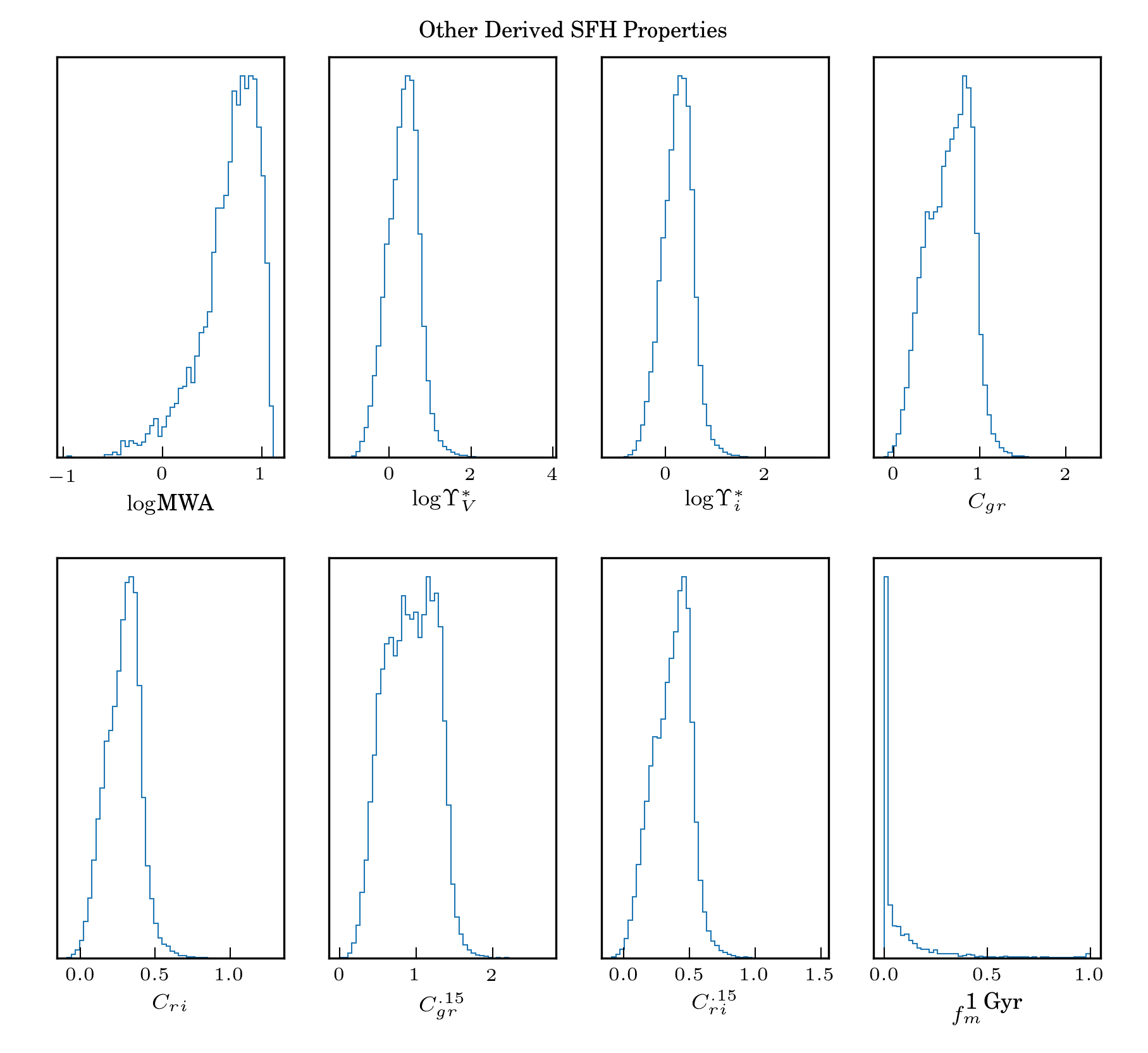}
    \caption{The distributions of eight derived parameters which are secondarily obtained from the SPS, using the distributions of inputs shown in Figure \ref{fig:prior_inputparams}: mass-weighted stellar age, stellar mass-to-light ratios in $V$ \& $i$ bands, rest-frame $g-r$ and $r-i$ color, the same colors at redshift of 0.15, and fraction of stellar mass formed in the past 1 Gyr.}
    \label{fig:prior_otherderived}
\end{figure*}

Distributions of most parameters provided to \texttt{FSPS}, selected resulting stellar absorption indices, and other derived parameters of interest are respectively shown in Figures \ref{fig:prior_inputparams}, \ref{fig:prior_specinds}, and \ref{fig:prior_otherderived}, in addition to the full-text descriptions provided below.

\subsubsection{SFH families: the delayed-$\tau$ model}
\label{sec:SPS_params}

\citetalias{chen_pca} based the adopted family of SFHs on a tau (declining exponential) model: additionally, one or more stochastic bursts were permitted, and a fraction of SFHs cut off rapidly, in order to emulate post-starburst galaxies at high redshifts \citep{kriek_06, kriek_09}. However, merely allowing a \emph{cutoff} in the SFH does a poor job at reproducing the more vigorously-star-forming outer regions of disk galaxies, which do have older stars, but whose SFHs are shown in both observations and cosmological simulations to rise through the present day \citep{pacifici_12_SFHs,pacifici_13_SFHs}. \citet{simha_14_SFHs} found that a more flexible delayed-$\tau$ model (also referred to as ``lin-exp") plus stochastic bursts and an optional subsequent ramp up (rejuvenation) or down (cutoff) effectively decouples late from early star-formation, and provides a better fit to photometric data. As such, we adopt this slightly more complicated framework.

The most basic delayed-$\tau$ model is parametrized by a starting time (before which the SFH is identically zero) and an $e$-folding timescale (which sets the shape of the SFH). Each are drawn from a smooth distribution:

\begin{itemize}
\itemsep0em
    \item \emph{Formation time} ($t_{form}$), drawn from a normal distribution with mean of 5 Gyr and width of 4 Gyr, and which truncates below 0.2 Gyr and above 13.0 Gyr. This broad distribution is similar to the uniform distribution adopted in \citetalias{chen_pca}.
    \item \emph{e-folding timescale of the continuous component} ($\textrm{EFTU}$), which has a log-normal distribution centered at $\log \frac{\mu}{\rm Gyr} = 0.4$ and with $\log \frac{\sigma}{\rm Gyr} = 0.4$. The distribution truncates below 0.1 Gyr and above 15 Gyr. Since the peak of the $t e^{-t/\tau}$ has its peak at an interval $\tau$ after formation, this broad distribution of e-folding times allows SFHs that form quickly, as well as those that continue to rise until the present day.
\end{itemize}

The prevalence, duration, and strength of merger-induced bursts have been investigated in Tree-SPH and N-body sticky-particles simulations \citep{dimatteo_08}. Though such simulations lack the resolution to model gas cooling to molecular-cloud temperatures, they are useful simply as order-of-magnitude guidance. \citet{dimatteo_08} also found that most merger-driven bursts last several $10^8$ years, with the vast majority lasting less than 1 Gyr, which informs the upper-limit for burst duration shown below. \citet{gallazzi_bell_09} conclude that recent bursts are necessary to reproduce the full space of stellar absorption features, simultaneously warning that \textit{overestimating} the number of bursts could result in systematically-low mass-to-light ratios for galaxies dominated by a continuous SFH.

We express the strength of a burst in terms relative to the peak of the underlying lin-exp model: that factor is simply added to the latent SFR at all times in the range $[t_{burst}, t_{burst} + dt_{burst}]$. Finally, we note that we do not model stochastic, short-timescale (several to tens of Myr) variations in SFR, primarily due to computational concerns. Conceivably, for sufficiently young ($<$ 1 Gyr) stellar populations, anomalously-steady models (i.e., SFHs that are \emph{too smooth}) could induce a negative systematic in stellar mass-to-light ratio (effecting an additional, intrinsic scatter in the SFH parameter space). The bursts are generated according to the following randomized prescription:

\begin{itemize}
    \item \emph{Number of bursts} ($n_{bursts}$, an integer) giving the number of starbursts. The value is generated by a Poisson distribution with a mean and variance $\frac{0.5 \times (t_0 - \min(\{t_{t}), t_{form}\}))}{t_0}$. That is, if a SFH were to initiate immediately at $t=0~{\rm Gyr}$ and not be cut off, the average number of bursts would be 0.5. Functionally, most SFHs experience zero stochastic bursts, and the mean number of bursts per SFH is 0.256.
    \item \emph{Burst amplitudes} ($A$, a list with $n_{bursts}$ elements). Individual values in $A$ are distributed log-normally between 0.1 and 10, and indicate the addition of $A$ times the maximum value of the pure delayed-$\tau$ model during the times when the burst is active.
    \item \emph{Burst times} ($t_{burst}$, a list), whose length is the same as $A$, and whose elements are uniformly distributed between $t_{form}$ and $t_0$ (or $t_{t}$, if there is a cutoff).
    \item \emph{Burst duration} ($dt_{burst}$, a list) which specifies the duration for each burst, uniformly distributed between 0.05 and 1 Gyr. 
\end{itemize}

The delayed-$\tau$ model with bursts is further modulated by conservative allowances for a cutoff/rejuvenation at late times:

\begin{itemize}
    \item \emph{Transition probability} ($p_t$), a 25\% chance of either a rejuvenation or a cutoff in the SFH at time $t_t$, under the assumption that most SFHs are smoothly-varying.
    \item \emph{Transition time} ($t_{t}$), after which star formation may occasionally (as dictated by $p_t$) cut off or revive. This may occur with equal probability after $t_{form} + EFTU / 4$ until the present day (at the smallest allowable value, such a SFH functions like a brief starburst which could last as little as 25 Myr). If $p_t$ dictates there is no burst, $t_t$ is set to the age of the universe, and thus never impacts the SPS.
    \item \emph{Transition strength} $\theta$, specifying an ``angle" in time-SFR space, such that $\theta = 0$ corresponds to a SFR held constant after $t_t$ and $\theta = \frac{\pi}{2}$ corresponds to an immediate cutoff in the SFH. $\theta$ is distributed according to a triangular distribution rising from zero in the domain $\frac{\pi}{2} < \theta \le 0$, and falling back to zero in the domain $0 < \theta \le \frac{\pi}{6}$. If $p_t$ dictates there is no burst, $\theta$ is set to zero, but does not impact the SPS because $t_t$ is set to the age of the universe.
    \item \emph{Cutoff slope} is an $\arctan$-re-parametrization of $\theta$, scaled in units of the maximum of a pure delayed-$\tau$ model $\Phi_{max}$ per Gyr. Therefore, $\gamma = 0$ corresponds to a perfectly constant SFR after $t_t$, and $\gamma = -1.0$ to a reduction in the SFR by $\Phi$ in 1 Gyr.
\end{itemize}

\begin{figure}
    \centering
    \includegraphics[width=\columnwidth]{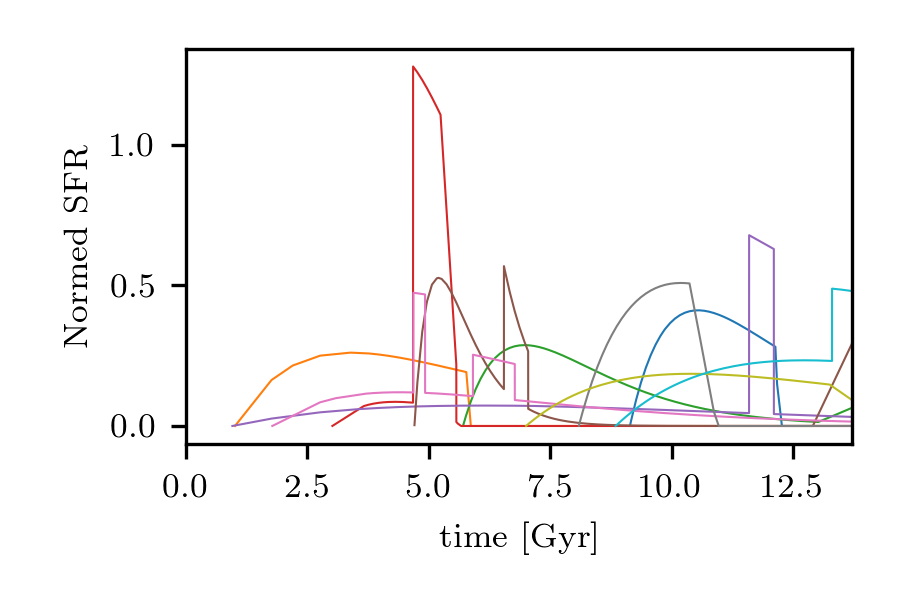}
    \caption{Ten sample SFHs generated using the random prescription given in Section \ref{sec:SFHs}.}
    \label{fig:sample_sfhs}
\end{figure}

These specific parameter distributions were chosen to produce the best match to the joint distribution of moderate- and high-signal-to-noise MaNGA spectra in \Dn-\HdeltaA space (see Section \ref{subsubsec:Dn4000-HdA_comp}). Compared with \citetalias{chen_pca}, $t_{form}$ is peaked more strongly at intermediate times, and ${\rm EFTU}$ is permitted to be less than 1 Gyr (and indeed, $\sim 40\%$ are). Additionally, starbursts occur on average less frequently in this work, but with stronger amplitude, since \citetalias{chen_pca} considered spatially-unresolved spectra (whose bursts have been ``spatially-averaged" to a greater probability, but lower mean strength). A sample of ten SFHs, drawn randomly from this prescription, is shown in Figure \ref{fig:sample_sfhs}.

\begin{figure}
    \centering
    \includegraphics[width=\columnwidth]{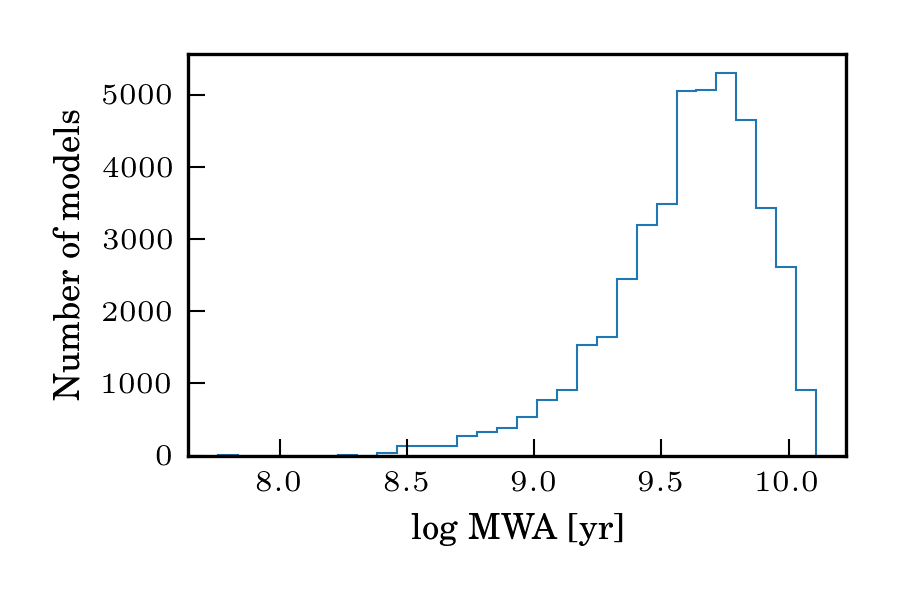}
    \caption{The distribution of the log of mass-weighted mean stellar age for all of the template SFHs. As in \citet{gallazzi_brinchmann_08}, the distribution has a broad peak around $\log {\rm MWA} \sim 9.8$. Unlike \citet{gallazzi_brinchmann_08}, though, the distribution extends with significant power below $\log {\rm MWA} \sim 9.0$, meaning that more recent star-formation is permitted.}
    \label{fig:mwa_dist}
\end{figure}

Previous analyses of SDSS central spectroscopy have derived distributions of mass-weighted mean stellar age (MWA) for galaxies in the nearby universe: for example, \citet{gallazzi_charlot_05}, following \citet{kauffmann_heckman_white_03}, reports a distribution of mass-weighted mean stellar age (MWA) derived from fits to high signal-to-noise spectra. The \citet{gallazzi_charlot_05} MWA distribution strongly resembles the distribution from this work's model library (Figure \ref{fig:mwa_dist}). This work's model library has a more probable low-MWA tail, and a significantly younger mode. As the galaxy disks sampled by MaNGA have a diversity of ages (both young and old) compared to the central regions sampled in \textsc{SDSS-i} spectroscopy, this is a positive characteristic. The model MWA distribution from this work also bears similarity to the MWA distribution derived from integrating the \citet{madau_dickinson_2014_csfrd} cosmic star formation rate density: \citet{madau_dickinson_2014_csfrd} report a young-age tail, which this work's prior easily encompasses, but has a mode at nearly 10Gyr (about twice as old as the mode of this work's model libraries). No value judgment is made here regarding a particular MWA distribution; that said, noting MWA distributions' changes in shape resulting from manipulating the CSP inputs has proven informative in constructing a flexible training library.

\definecolor{mplc0}{HTML}{1f77b4}
\definecolor{mplc1}{HTML}{ff7f0e}
\definecolor{mplc2}{HTML}{2ca02c}
\definecolor{mplc3}{HTML}{962728}
\definecolor{mplc4}{HTML}{9467bd}
\definecolor{mplc5}{HTML}{8c564b}
\definecolor{mplc6}{HTML}{e377c2}
\definecolor{mplc7}{HTML}{7f7f7f}
\definecolor{mplc8}{HTML}{bcbd22}
\definecolor{mplc9}{HTML}{17becf}

\begin{table*}
    \centering
    \begin{tabular}{||c|c|c|c|c||} \hline \hline
        Line color & $t_f$ & EFTU & $t_t$ & $\theta$ \\ \hline
        \textcolor{mplc0}{\texttt{C0}} (Tableau Blue) & 9.14 & 1.40 & 12.13 & -1.36 \\ \hline
        \textcolor{mplc1}{\texttt{C1}} (Tableau Orange) & 1.01 & 2.38 & 5.78 & -1.43 \\ \hline
        \textcolor{mplc2}{\texttt{C2}} (Tableau Green) & 5.71 & 1.27 & 13.01 & 0.13 \\ \hline
        \textcolor{mplc3}{\texttt{C3}} (Tableau Red) & 3.02 & 1.29 & 5.25 & -0.70 \\ \hline
        \textcolor{mplc4}{\texttt{C4}} (Tableau Purple) & 0.96 & 5.18 & 7.10 & -0.04 \\ \hline
        \textcolor{mplc5}{\texttt{C5}} (Tableau Brown) & 4.71 & 0.50 & 12.90 & 0.13 \\ \hline
        \textcolor{mplc6}{\texttt{C6}} (Tableau Pink) & 1.78 & 2.61 & - & - \\ \hline
        \textcolor{mplc7}{\texttt{C7}} (Tableau Gray) & 8.09 & 2.10 & 10.37 & -0.92 \\ \hline
        \textcolor{mplc8}{\texttt{C8}} (Tableau Olive) & 7.00 & 3.39 & 13.25 & -0.59 \\ \hline
        \textcolor{mplc9}{\texttt{C9}} (Tableau Cyan) & 8.85 & 3.88 & - & - \\ \hline
    \end{tabular}
    \caption{Selected CSP parameters for the SFHs shown in Figure \ref{fig:sample_sfhs}. For each CSP, we list the line color, the formation time $t_f$, the e-folding time of the continuous component (EFTU), the transition time $t_t$, and the transition strength $\theta$. Models with no transition behavior have the $t_t$ and $\theta$ columns left blank.}
    \label{tab:sfh_params}
\end{table*}

\subsubsection{Stellar composition \& velocities, attenuation, and uncertain stellar evolution}

Since the star-forming ISM is known to enrich with heavy elements as successive generations of stars form, it is most correct to consider both a metallicity history and a star formation history. Though gaseous emission captures the current enrichment state, it is subject to significant differences in interpretation, including concepts as fundamental as the zeropoint \citep{stasinska_07_review}. Certain stellar absorption indices, particularly those targeting magnesium \citep{barbuy_92}, reflect the average enrichment state of the stars. However, the Mg-based indices in particular are not reliable at low metallicity \citep{maraston_03b}. In addition, there is some evidence that when trying to fit a population with known-evolving metallicity using a single, non-evolving stellar metallicity, absorption index-based estimates of stellar mass-to-light ratio suffer from smaller biases than do color-based estimates. This is because stellar mass-to-light ratio varies in the \Dn--\HdeltaA plane in a very similar way, when fixed- and evolving-metallicity populations are compared \citep[see][Section 5]{gallazzi_bell_09}. Finally, in order to properly consider a SFH with an evolving metallicity, additional parameters must be introduced to capture inflows, outflows, and feedback \citep{matteucci_16_chemev}. Section \ref{subsubsec:Dn4000-HdA_comp} briefly outlines a comparison made between spectral indices such as \Dn and \HdeltaA measured in the models and in the observations, and supports the assertion that non-evolving metallicities suffice for our purposes.

\begin{figure}
    \centering
    \includegraphics[width=\columnwidth]{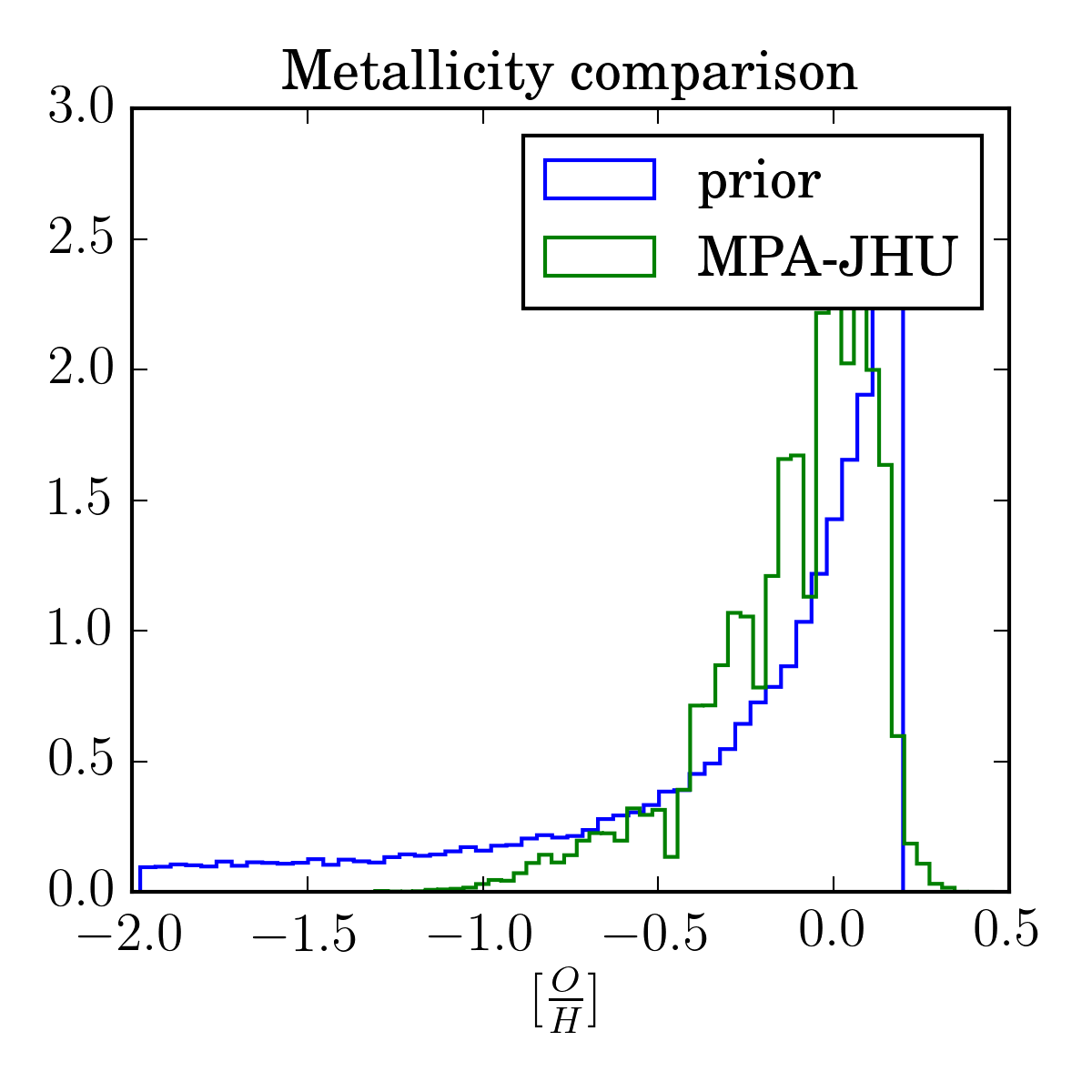}
    \caption{Comparison of the solar-normalized oxygen abundance ($[\frac{O}{H}]$) inferred from SDSS nebular emission (\textit{green}) with the adopted metallicity prior. An offset of approximately -.29 dex is applied to re-scale the  SFH library's metallicity range ($[Z]$, based on mass) to the number-abundance of oxygen from the SDSS data, since the two adopt different values of solar metallicity and oxygen abundance.}
    \label{fig:metallicity_prior}
\end{figure}

With these considerations in mind, we do not implement any chemical evolution prescription, instead adopting SFHs with non-evolving stellar metallicities. Each SFH model is assigned a single metallicity $[Z]$, constant through time, which has an 80\% chance to be drawn from a metallicity distribution that is linearly-uniform, and a 20\% chance to be drawn from a metallicity distribution that is logarithmically-uniform. This allows a small, but well-populated low-metallicity tail. Figure \ref{fig:metallicity_prior} compares the gas-phase oxygen abundance from the SDSS MPA-JHU catalog \citep{tremonti_mz} to the adopted metallicity prior, after a zeropoint normalization \citep{asplund_09}. These distributions should be (and are) roughly similar, since the chemical composition of the gas reflects how baryons are cycled through stars and enriched successively by several generations of star-formation.

Two uncertain phases of stellar evolution, blue horizontal branch (BHB) and blue straggler stars (BSS), are also modulated in prevalence: while they can affect stellar mass-to-light ratio estimates by $\sim$ 0.1 dex for the intermediate-age populations found in galaxy disks \citep{fsps_1}, there are precious few precise enough measurements of either of their abundances to inform this work's CSP library. Adopting smooth and permissive priors for these less-well-constrained parameters avoids unjustified restrictions on the resulting spectral fits. In reality, we are in most cases unable to further constrain these parameters based on our fits to spectra (see Section \ref{sec:method} and Appendix \ref{apdx:fakedata}), but we also lack the observational constraints from stellar-evolution to choose one value in particular.

While BHB stars are likely more common at low metallicity, it is inadvisable to neglect them for other cases \citet{fsps_1}. As such, we draw their fraction by number ($f_{BHB}$) from a beta-distribution with shape parameters $\alpha=2$ and $\beta=7$: this distribution is restricted to lie between zero and one, and represents a plausible range of BHB incidence rates. Specific BSS frequency ($S_{BSS}$, defined with respect to \emph{all} horizontal branch stars) is known to vary somewhat with environment, but is not constrained well in an absolute sense by observations \citep{santucci_2015_BSS}. The binary mass-transfer pathway for BSS formation \citep{gosnell_14_bluestraggler} implies that any factors (e.g., environment or metallicity) affecting star formation could also manifest in the BSS population. Furthermore, \citet{piotto_04_bluestraggler} noted that BSS frequency is lower in clusters than in the field, in a way not explained by the expected increased collision rates in clusters. As such, we adopt a broad distribution, 10 times the value of a draw from a beta distribution with shape parameters $\alpha=1$ and $\beta=4$---which allows the full range of 0.0--0.5 adopted by \citet{fsps_1}, is more permissive at the high end than the estimates of \citet{dorman_95}, and peaks at approximately 0.2.

Attenuation of the starlight is accomplished using a two-component dust model \citep{charlot_fall_00}. In this model, all stars are attenuated by diffuse dust with a V-band optical depth $\tau_V \mu$ and power-law slope of -0.7 \citep[as in][]{chevallard_charlot_16_beagle, da-cunha_charlot_elbaz_08}; while stars younger than 10 Myr are further attenuated by the dense ISM with optical depth $\tau_V (1 - \mu)$ and power-law slope of of -1.3. Therefore, a young stellar population will in total experience an optical depth of $\tau_V$, discounting effects resulting from different dust law slopes. In our schema, $\tau_V \mu$ and $\mu$ are randomized directly (rather than $\tau_V$ and $\mu$ individually), because all stellar populations experience $\tau_V \mu$, meaning that it is a more effective parameter most of the time:

\begin{itemize}
    \itemsep0em
    \item The \emph{product} $\tau_V \mu$ is drawn from a normal distribution with mean 0.4 and standard deviation 0.2, truncated at 0 and 1.2.
    \item \emph{Fractional optical depth of the diffuse ISM} ($\mu$), drawn from a normal distribution with mean of 0.3 and standard deviation of 0.2, truncated at 0.1 and 0.9.
    \item \emph{Optical depth of young birth clouds} ($\tau_V$), the quotient $\frac{\tau_V \mu}{\mu}$.
\end{itemize}

These distributions were chosen such that their means correspond roughly to the ``standard" values given in \citet{charlot_fall_00}, with significant latitude to allow for both unobscured and highly-obscured stellar populations. The overall distribution has a similar mode to, but is broader (i.e., more permissive at the high end) than the attenuation distribution for star-forming galaxies found in \citet{brinchmann_04_mpajhu} by fitting emission lines with photoionization models.

Stellar velocity dispersion ($\sigma$) is also varied, and is drawn from a truncated exponential distribution with lower-limit of 10 km s$^{-1}$, upper-limit of 350 km s$^{-1}$, and e-folding scale of 350 km s$^{-1}$. This is intended to populate both the low-$\sigma$ stellar disk and the high-$\sigma$ bulge. This does not include the wavelength- and redshift-varying contribution of the instrumental line-spread function (LSF), which is accounted for separately (see Appendix \ref{apdx:lsf}).

Each SFH is initialized with several different sets of dust properties ($\tau_V$ and $\mu$) and velocity dispersion ($\sigma$), for computational reasons. In addition to easing the processing load, this ensures a complete population of the parameter space at significantly lower computational cost (in total, \nSFHs CSPs were generated, each of which has \nsubsample combinations of attenuation and velocity dispersion.) Subsampling in velocity dispersion and attenuation becomes important to ensure that enough models fit the data well enough to perform good parameter inference: in Appendix \ref{apdx:fakedata}, we use additional synthetic data (i.e., ``held-out data" generated identically to the training spectra but not included in PCA training) to test the reliability of our stellar mass-to-light estimates against mock galaxies with known physical properties; and in Section \ref{subsubsec:Dn4000-HdA_comp}, we compare the distribution of all of the training data in \Dn and \HdeltaA to empirical measurements of the same indices from many thousands of spectra reported in the MaNGA DAP, demonstrating that other than one small, well-known systematic affecting \HdeltaA at moderate \Dn, the training models are distributed very similarly to empirical measurements of \Dn and \HdeltaA in observed spectra.

For each model, the SFH is stored (as it is used explicitly by \texttt{FSPS}), along with the strengths of several spectral absorption indices\footnote{All stellar absorption indices are computed on spectra with velocity dispersion of $\sigma$ = 65 km s$^{\rm -1}$, approximately equal to the difference in resolution between full-resolution model spectra and MaNGA data. This is preferable to employing correction factors which are not guaranteed to work for stellar populations younger than 3 Gyr \citep{kuntschner_04_lick_losvd_corr}.}, the mass-weighted age, and $V$- \& $i$-band mass-to-light ratios ($\Upsilon^*_V$ \& $\Upsilon^*_i$)\footnote{\emph{Effective} mass-to-light ratios are used, since they include only light that reaches the observer. In other words, these effective mass-to-light ratios are affected by dust. All subsequent references to mass-to-light ratio use this same abbreviation. For the purposes of estimating stellar mass, though, this convention suffices, because the bandpass flux which is multiplied by the mass-to-light ratio and the distance modulus \emph{also} is attenuated by dust, so the two dust contributions cancel.}.

\subsection{\Dn-\HdeltaA comparison of training library to MaNGA spaxels}
\label{subsubsec:Dn4000-HdA_comp}

\begin{figure*}
    \centering
    \includegraphics[width=\textwidth]{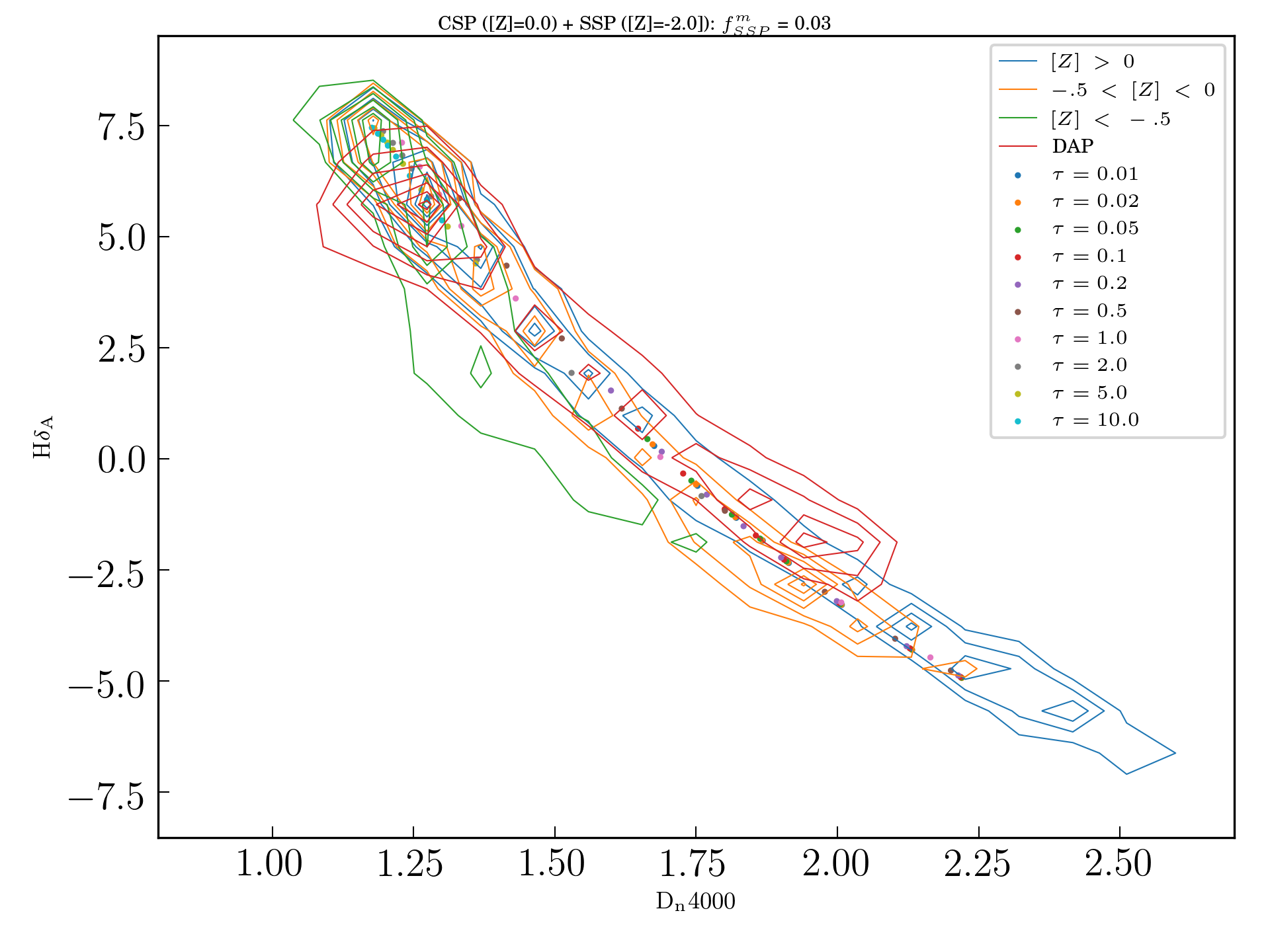}
    \caption{The distribution of the training models in \Dn-\HdeltaA space (separated by stellar metallicity: super-solar metallicity in blue contours, slightly sub-solar in orange contours, and very sub-solar in green contours); plus data points for individual models with composite metallicities. Each point signifies a delayed-tau model with time constant denoted by the color of point (but forming at a variety of times post-Big-Bang). The continuous portion of the model is fixed at solar metallicity. Added to this CSP is a SSP which forms instantaneously at the same time as the CSP begins to form stars, with a contribution to the current stellar mass of 3\%. The SSP has an extremely low metallicity (${\rm [Z]}$ = -2).}
    \label{fig:CSP+lowZ_SSP_mf03}
\end{figure*}

We evaluated the correspondence between the suite of synthetic models and the real MaNGA data by comparing the distribution of the \Dn and \HdeltaA absorption indices measured by the MaNGA DAP to those from the full suite of SFH models (the ``training data") used in this work (Figure \ref{fig:CSP+lowZ_SSP_mf03}). We observe an offset in \HdeltaA between synthetic models and observations at fixed \Dn greater than 1.5, consistent with previous work \citep[see][Figure 2]{kauffmann_heckman_white_03}. This effect has been attributed to stellar models, and the offset observed (which grows with \Dn, but remains less than 0.8 $\mbox{\AA}$) is well within the locus of previous measurements. Some degree of this offset may be attributable to $\alpha$-enhancement, which cannot be manipulated in the set of stellar atmospheres adopted for this work. It is likely that such a mismatch exists at all values of \Dn, but becomes apparent only at \Dn $>$ 1.5---that is, at ages of several Gyr (where CSPs with e-folding timescales shorter than 1Gyr begin to have similar spectra to SSPs).

Furthermore, though \citet{maraston_stromback_09} finds that superimposing approximately 3\% (by mass) of low-metallicity stars onto synthetic continuous stellar populations can resolve a color mismatch between synthetic CSP models and luminous red galaxies (LRGs), we find no evidence for a similar improvement in the case of \Dn and \HdeltaA (in Figure \ref{fig:CSP+lowZ_SSP_mf03}, we show the case where the mass fraction is 3\%). We observe, though we do not show, that as the mass fraction of the SSP increases, the value of \HdeltaA actually \emph{decreases} at fixed \Dn. That said, \citet{maraston_stromback_09} note that a potential astrophysical reason for the bluer-than-anticipated colors in metal-poor galaxies is an especially strong blue horizontal branch, which is manipulated separately in our population synthesis. Finally, since the fraction of MaNGA spaxels that lie in the centers of massive LRGs is low, any effect of mixed-metallicity populations may be subdominant to others which pertain to more star-forming systems.

An attempt to replace \HdeltaA with the sum of \HdeltaA and \HgammaA in lieu of just \HdeltaA, since a deficiency in \HgammaA has been noted to function opposite to a deficiency in \HdeltaA. In reality, the match is not greatly improved.

\subsection{Why not use CMLRs?}
\label{subsec:cmlrs}

\citet{bell_03} produced conversions between various optical colors and mass-to-light ratio, and we will re-evaluate this approach here. Table \ref{tab:bell_vs_here} compares the inputs to the stellar population synthesis modelling used to derive the \citet{bell_03} CMLRs, to the inputs used in this work. Salient differences include this work's modest allowances for starbursts, inclusion of attenuation---previously argued to be unimportant, due to the slope of the reddening vector being very similar to the CMLR \citep{bell_dejong_01, bell_03}---, and use of the \citet{kroupa_imf_01} IMF.

\begin{table*}
    \centering
    \begin{tabular}{||c||p{1.5in}|p{1.5in}|p{1.5in}||} \hline
        Input & \citet{bell_03} & \citet{taylor_gama_cmlrs} & This Work \\ \hline 
        Stellar models & P\'{E}GASE \citep{fioc_97_pegase} & \citet{BC03} & \citet{fsps_c3k} \\ \hline 
        Stellar IMF & ``Diet" \citet{salpeter_imf_55}---also see \citet{bell_dejong_01} & \citet{chabrier03} & \citet{kroupa_imf_01} \\ \hline 
        SFHs & delayed-$\tau$ & $\tau$-model, grid-sampled & Composite: delayed-$\tau$, burst(s), cutoff, rejuvenation \\ \hline 
        Attenuation & None & Uniform screen & Two-component \citep{charlot_fall_00} \\ \hline 
    \end{tabular}
    \caption{SPS inputs, compared between \citet{bell_03}, \citet{taylor_gama_cmlrs}, and this work.}
    \label{tab:bell_vs_here}
\end{table*}

\begin{figure*}
    \centering
    \includegraphics[width=\textwidth]{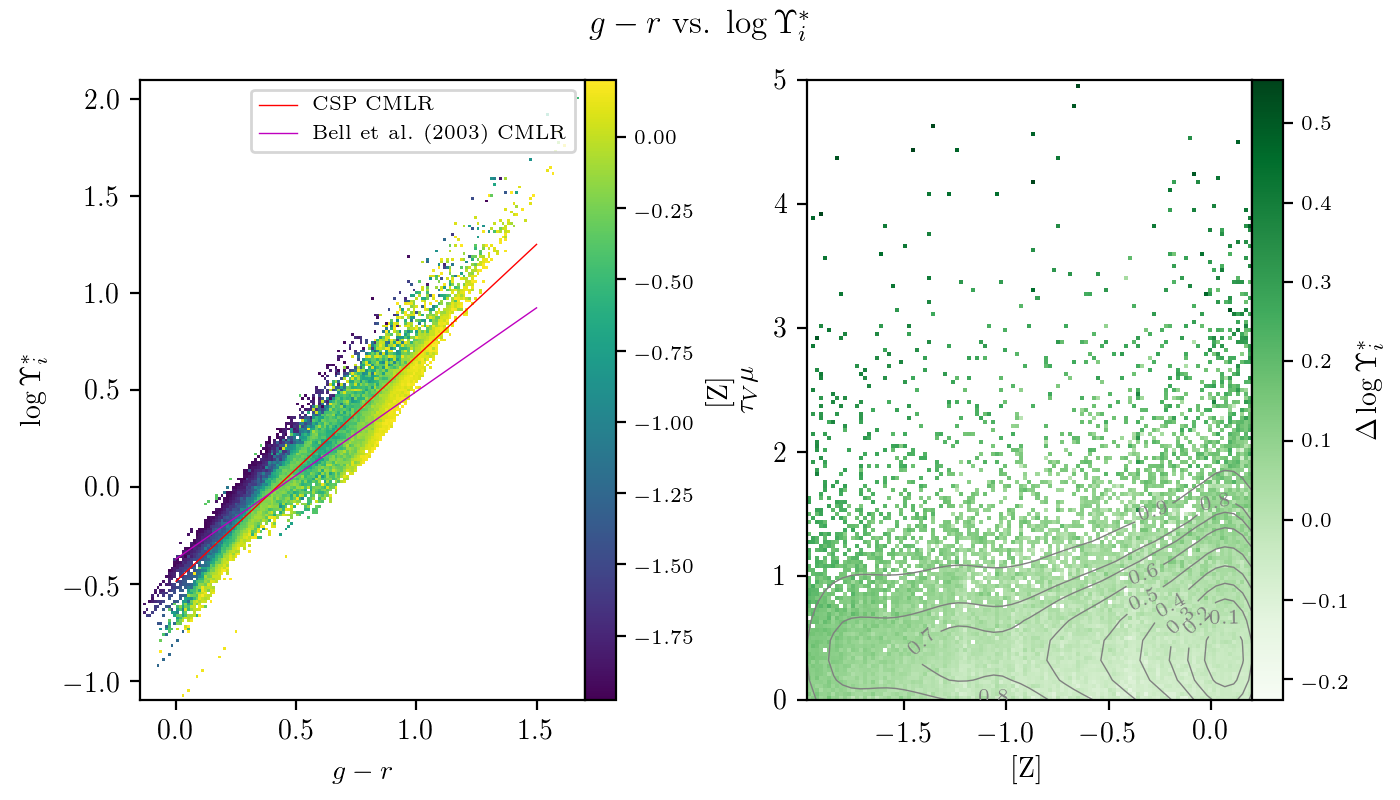}
    \caption{Left-hand panel: effective $i$-band stellar mass-to-light ratios (in solar units) for model spectra generated  above, plotted against rest-frame $g-r$ color, and colored by stellar metallicity; in \emph{red}, the CMLR obtained from a least-squares fit to the CSP library; in \emph{magenta}, the CMLR from \citet{bell_03}, after a -0.15 dex Salpeter-to-Kroupa IMF correction. Right-hand panel: a visualization of the typical difference between a SFH's true stellar mass-to-light ratio and the mean CMLR at that color: each image pixel is colored according to the median of the CMLR deviation for all CSPs in that small ${\rm [Z]}$--$\tau_V \mu$ bin (or white, if there are none). Overlaid in red contours is shown the approximate fraction of models within a given contour (derived from a two-dimensional kernel density estimation).}
    \label{fig:CMLR_Cgr-MLi-dev}
\end{figure*}

\begin{figure*}
    \centering
    \includegraphics[width=\textwidth]{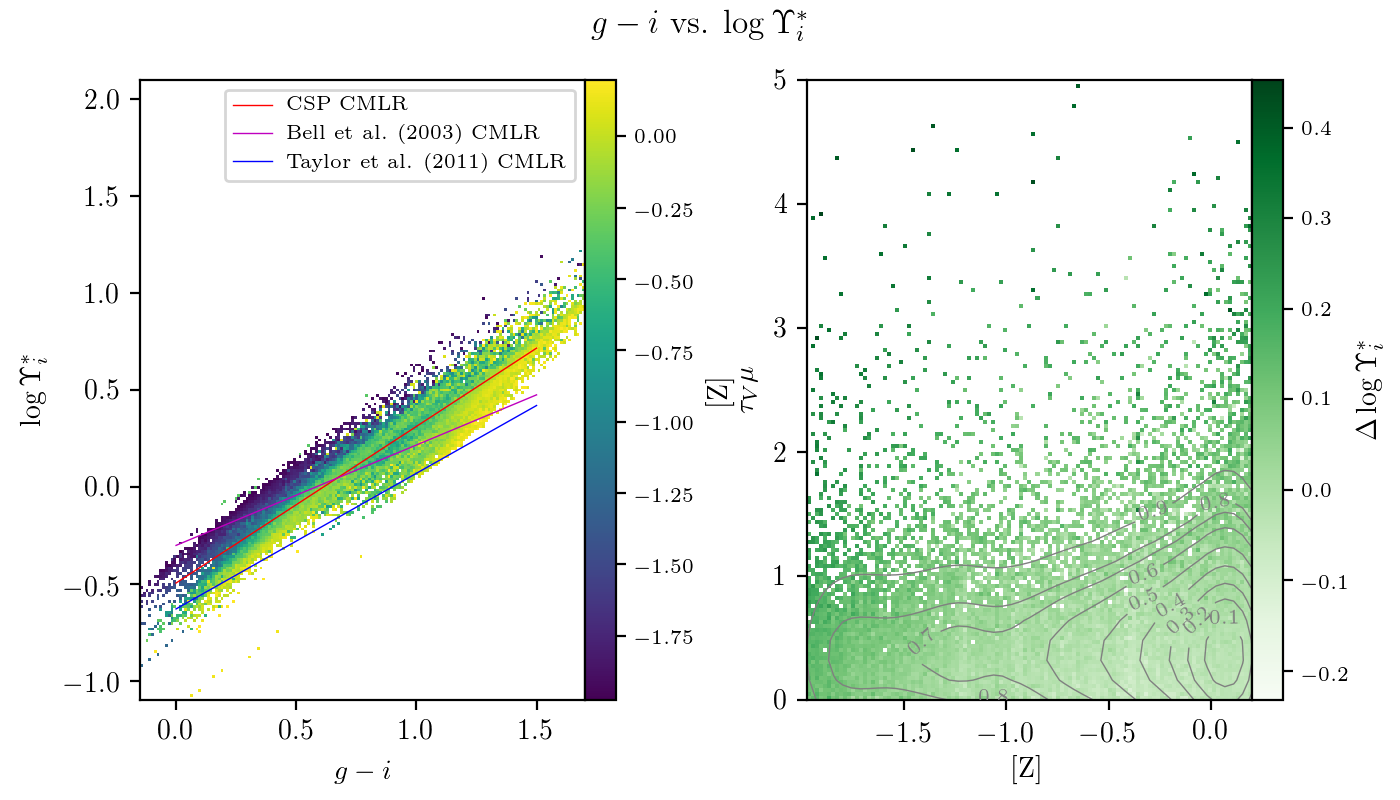}
    \caption{As Figure \ref{fig:CMLR_Cgr-MLi-dev}, except calibrating $\log \Upsilon^*_i$ against $g-i$ color, plus \citet{taylor_gama_cmlrs} CMLR in the left-hand panel in \emph{blue}. All mass normalizations are once again corrected to \citet{kroupa_imf_01} IMF.}
    \label{fig:CMLR_Cgi-MLi-dev}
\end{figure*}

Using the training data described above, we use a least-squares fit to find the optimal CMLR for $i$-band stellar mass-to-light ratio and both $g-r$ and $g-i$ colors---the latter being provided as a point of comparison to the GAMA survey \citep{taylor_gama_cmlrs}---and then examine the mean absolute deviation between the predicted and actual values of $\log \Upsilon^*_i$ (Figures \ref{fig:CMLR_Cgr-MLi-dev} and \ref{fig:CMLR_Cgi-MLi-dev}). As Figures \ref{fig:CMLR_Cgr-MLi-dev} and \ref{fig:CMLR_Cgi-MLi-dev} illustrate, our models follow a well-defined CMLR, but with a scatter of at 0.05--0.1 dex about the best-fit: scatter is lowest at modest values of stellar attenuation and sub-solar metallicities (the $g-i$ CMLR is slightly better in this respect). Furthermore, the CMLRs rely upon nearly perfect photometry, which in reality rarely improves to sub-0.02 levels at kiloparsec sampling scales for large surveys. That is, depending on the precise choice of CMLR, observational effects can very easily add further uncertainties of $\sim$0.05 dex. The differences between the \citet{bell_03}, \citet{taylor_gama_cmlrs}, and this work's CMLRs are not insubstantial: \citet{bell_03} CMLRs have uniformly smaller slope, meaning that they will produce mass estimates that are higher (lower) for blue (red) colors. In contrast, stellar mass-to-light ratios from the \citet{taylor_gama_cmlrs} CMLR will be uniformly lower than this work's, by 0.15--0.4 dex. This highlights the impact of the specific SFH family chosen, the stellar models, and even the choice of attenuation (see below).

Overall, one should note that the scatter about the CMLR is not entirely random. This means that even before systematics related to stellar model atmospheres and our fiducial SFH family, the functional lower-limit on stellar mass-to-light ratio uncertainty is about 0.1 dex. Especially in the case of vigorous recent star-formation, stellar metallicity is associated with pronounced departures from the ``mean" CMLR (the large number of blue, low-metallicity models at higher-than-predicted mass-to-light ratios is one noticeable example). Figures \ref{fig:CMLR_Cgr-MLi-dev} and \ref{fig:CMLR_Cgi-MLi-dev} show that while CSPs in the most common regions of parameter space have their stellar mass-to-light ratios described well by the best-fit CMLR, departures from the median case can cause troublesome systematics: for instance, at low metallicity, $\Delta \logml{i}$ can reach values of 0.2--0.3 dex, even at low attenuations; and higher optical depths ($\tau_V \mu \sim 3$) can boost this discrepancy as high as 0.4 dex. This is not simply a scatter about the CMLR, but is rather a true systematic. The effect is similar in ${\rm [Z]}-\tau_V (1 - \mu)$ space.

To illustrate the potential effects of attenuation in pulling a single SFH away from a CMLR, consider the following scenario: at fixed fractional optical depth $\mu = 0.4$, a SFH with $t_{form} = \tau = 2~{\rm Gyr}$ changes in $g-r$ color and $\log \Upsilon^*_i$ by +0.094 and +0.14 when $\tau_V$ is changed from 0.0 to 1.0, a considerably steeper slope than the fiducial CMLR. Furthermore, the slope of the attenuation vector in $(g-r)$--$\log \Upsilon^*_i$ space increases with $\mu$, matching the slope of the CMLR at $\mu \sim 0.14$ (recall that $\mu$ affects the balance between attenuation of young stars and old). In \citet{bell_dejong_01} and \citet{bell_03}, attenuation was explicitly ignored, because the attenuation vector lay almost parallel in color--mass-to-light space to the adopted CMLR. Depending on the exact value of $\mu$, this may not be true. So, for the SFH chosen above, the stellar mass-to-light ratio is under-estimated for most realistic values of $\mu$ and $\tau_V$.

Also a concern is the effect of the stellar IMF on the relative mass normalization. Using 1000 separately-randomized SFHs for three of the five stellar IMFs built into \texttt{FSPS}\footnote{It was computationally less costly to randomize each IMF's set of SFHs than it was to use the same SFHs using each of the three IMFs.}---\citet{salpeter_imf_55}, \citet{chabrier03}, and \citet{kroupa_imf_01}---, we have separately-determined the $\Upsilon^*_i$ normalization and its overall dependence on $g-r$ color (see Table \ref{tab:cmlr_fit}) for our training library. The effects of IMF on integrated colors are of course small (and likely attributable to differences in SFH randomization), but the overall mass normalizations differ quite strongly: \citet{kroupa_imf_01} and \citet{chabrier03} are offset respectively by $-0.209~{\rm dex}$ and $-0.252~{\rm dex}$, with respect to \citet{salpeter_imf_55}.

\begin{table}
    \centering
    \begin{tabular}{||c|c|c|c||} \hline
        IMF & $m$ & $b$ & $\sigma~{\rm [dex]}$ \\ \hline 
        \citet{salpeter_imf_55} & 1.145 & -0.286 & 9.38 $\times 10^{-2}$ \\ \hline 
        \citet{kroupa_imf_01} & 1.147 & -0.496 & 9.62 $\times 10^{-2}$ \\ \hline 
        \citet{chabrier03} & 1.155 & -0.538 & 9.44 $\times 10^{-2}$ \\ \hline
    \end{tabular}
    \caption{Linear fit relating $g-r$ color to $\log \Upsilon^*_i$, and the magnitude of the scatter about the best-fit line (all a little less than 0.1 dex).}
    \label{tab:cmlr_fit}
\end{table}

In summary, CMLRs do not capture the full range of information contained in a galaxy's SED; indeed, they can also be susceptible to degeneracies between age, metallicity, and attenuation. Specifically, even at infinite signal-to-noise, both CMLRs tested here suffer from intrinsic scatter above the 0.05 dex level in the very common stellar metallicity range of -0.5--0.2 and at diffuse ISM optical depths greater than 1.0. Deviations from a fiducial dust law can also induce changes in the effective stellar mass-to-light ratio which are \emph{not} parallel to the CMLR, as had been previously suggested \citep{bell_dejong_01, bell_03}. There is much more information to glean from galaxy SEDs than can be encoded in optical colors.

\section{Parameter Estimation in the PCA Framework}
\label{sec:method}

The goal of this analysis is to obtain estimates of physical quantities (especially resolved stellar mass) by reducing the dimensionality of observed spectra from a vector of length $\sim 4000$ to one of length $\sim 6$, and the overall approach to the analysis is very close to \citetalias{chen_pca}: for some observed spectrum, we then find its best representation in terms of linear combinations of the principal component vectors, taking into account covariate noise arising from imperfect spectrophotometry. Finally, we evaluate how well each training spectrum matches the observed spectrum \emph{in principal component space}, and assign weights to the training spectra accordingly. The weights are used to approximate probability density functions (PDFs) of interesting quantities such as stellar mass-to-light ratio ($\Upsilon^*$). Table \ref{tab:vars} provides a complete digest of the notation used in this section to describe the use of principal component analysis.

\begin{table*}
    \centering
    \begin{tabular}{||c|p{5in}|c||} \hline
        Symbol & Description & Dimension\footnote{if applicable}\\ \hline
        $n$ & Number of CSPs in training library & - \\ \hline
        $n'$ & Number of spaxels analyzed in a single MaNGA datacube & - \\ \hline
        $l$ & Number of spectral channels in each CSP and observed spectrum & - \\ \hline
        $p$ & Number of quantities (such as stellar mass-to-light ratio) stored for each CSP & - \\ \hline
        $q$ & Number of principal components retained for final dimensionality reduction & - \\ \hline
        $D$ & Training data (already normalized) & $(n,l)$ \\ \hline
        $D'_q$ & Training data, comprising only the first $q$ PCs (subscript often omitted for clarity) & $(n,l)$ \\ \hline
        $E$ & Eigenspectra obtained from the model library & $(q,l)$ \\ \hline
        $A$ & Principal Component amplitudes obtained by projecting spectra onto the eigenspectra & $(n,q)$ \\ \hline
        $R$ & Residual obtained by subtracting $D'_q$ from $D$ & $(n,l)$ \\ \hline
        $K_{th}$ & Theoretical covariance matrix, obtained from $R$ & $(q,q)$ \\ \hline
        $\{Y_i\}$ & Set of physical parameters that produced the set of model spectra (also notated simply $Y$, when referring to a matrix with rows representing model spectra) & $(n,p)$ \\ \hline
        $C$ ($Z$) & Linear regression coefficients (zeropoints) that link the values in $\{P\}$ to PC amplitudes $A$ & $(q,p)$ ($(p)$)\\ \hline
        $O$ & Observed spectra, in flux-density units & $(n',l)$ \\ \hline
        $a$ & Median value of observed spectra $O$ or training data $T$, used to normalize data & $(n')$ \\ \hline
        $M$ & Median spectrum, obtained by averaging all model spectra's values in a given spectral element & $(l)$ \\ \hline
        $S$ & Unity-normalized and median-spectrum-subtracted spectra, $O/a - M$ & $(n',l)$ \\ \hline
        $O'_q$ & Observed spectra, comprising only the first $q$ PCs (subscript often omitted for clarity) & $(n',l)$ \\ \hline
        $K^{obs}$ & Observational covariance matrix, obtained from multiply-observed MaNGA objects and unique to a given spectrum & $(n',l,l)$ \\ \hline
        $V$ & The variance of one spectrum, obtained directly from the reduced data products & $(l,l)$ \\ \hline
        $N^{lhs}_{obs}$, $N^{rhs}_{obs}$ & Assumed noise propagated from exact de-redshifting of observed spectra into the fixed, rest-frame eigenspectra wavelength grid & \\ \hline
        $K^{PC}$ & PC covariance matrix for a given spectrum & $(n',q,q)$ \\ \hline
        $P^{PC}$ & Inverse of $K^{PC}$, computed for each observed spectrum (sometimes referenced elsewhere as ``concentration" or ``precision") & $(n',q,q)$ \\ \hline
        $\chi^2$ & Chi-squared deviation between each observed spectrum's PC representation and each model's & $(n',n)$ \\ \hline
        $W$ & Weight of each model spectrum used to construct joint parameter PDF, computed according to Equation \ref{eqn:logl} & $(n',n)$ \\ \hline
        $\{F_{i,j}\}$ & Set of marginalized PDFs for each spectrum (indexed by $i$) and each parameter (indexed by $j$) & $(n',p,-)$ \\ \hline
    \end{tabular}
    \caption{Symbols used in this section for the mathematical description of the PCA method.}
    \label{tab:vars}
\end{table*}

\subsection{The PCA system}
\label{subsec:run_pca}

We first construct the PCA vector basis:
\begin{enumerate}
\itemsep0em
    \item Pre-process all model spectra: 
    \begin{enumerate}
        \item Convolve with Gaussian kernel of width $\sigma \sim$ 65 km s$^{\rm -1}$, to account for difference between C3K native resolution and MaNGA instrumental resolution\footnote{The line-spread function specifies the (wavelength-dependent) manner in which a spectrum is blurred by a spectrograph. In the case of the MaNGA data, this amounts to between 1 and 3 pixels on the spectrograph, depending on the wavelength. Details for how to compute this can be found in \citet{cappellari_17}, as well as in Appendix \ref{apdx:lsf}.}.
        \item Interpolate to a logarithmic wavelength grid from 3600--8800 $\mbox{\AA}$, with $d \log \lambda = 1.0 \times 10^{-4}$, yielding final model spectra ($D$).
        \item Normalize spectra, dividing by their median values \footnote{Normalizing by the median (rather than the mean) makes very little difference for the training data, but is less sensitive to the occasional un-flagged emission line or small discontinuity in the observed data.}.
    \end{enumerate}
    \item Compute and subtract from all model spectra the median spectrum of all models ($M$), yielding median-subtracted training spectra ($T$)
    \item Compute the eigen-decomposition of $T$ using ``covariance method", retaining the first $q$ vectors as ``principal components" ($E$).
    \item Project $T$ onto $E$, compute the residuals $R$, and compute the resulting covariance $K_{th}$.
\end{enumerate}

\begin{figure*}
    \centering
    \includegraphics[width=\textwidth]{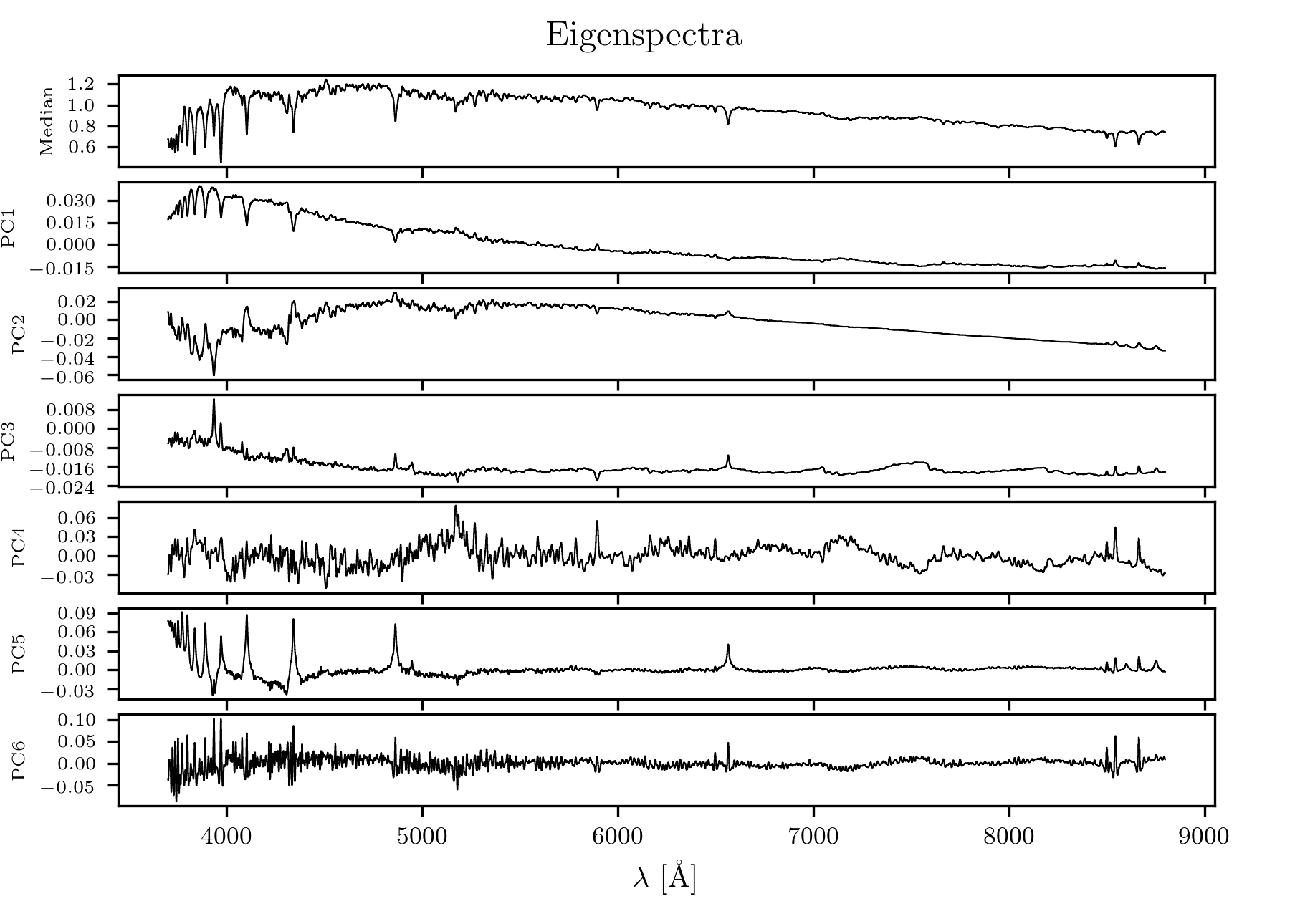}
    \caption{Top panel: the normalized mean spectrum of the training data. Panels 2--7: principal component vectors 1--6 of the training data.}
    \label{fig:eigenspectra}
\end{figure*}

Figure \ref{fig:eigenspectra} shows the normalized mean spectrum and each of the first six eigenspectra. Comparison with a ``broken-stick" model of marginal variance suggests that six is a suitable number (see Section \ref{subsec:picking_q} for more discussion). Conveniently, at this value of $q$, the remaining variance in the training data is well below typical random and spectrophotometric uncertainties, which means that the PC space should represent the complete view on the data within MaNGA's observational constraints. While the physical interpretation of the eigenspectra is not straightforward (and ``adding up" multiples of PCs is \emph{not} equivalent to ``adding up" stars to form a SSPs or SSPs to form a more complicated stellar population), we explore their correlations with physical properties in Section \ref{subsec:pc_meaning}.

If a single spectrum (with $l$ wavelength values) is a single point in $l$-dimensional space, then $n$ spectra form a cloud in $l$-dimensional space. In the case of the generated model spectra, we can claim to have constructed a space where $n \sim 20,000$ and $l \sim 5,000$. PCA will then find the orthogonal basis set that maximizes the amount of information retained, utilizing $q < l$ dimensions. PCA can be reduced to a singular value decomposition (SVD), but in our case, where $n > l$, it is equivalent and most efficient to compute as an eigenvalue problem on the covariance matrix. In particular, the training data $D$ has a dimension of $(n, l)$\footnote{We adopt the convention of a matrix with dimension $(a, b)$ having $a$ rows and $b$ columns. For such a matrix $A$, we would select the value in row $i$ and column $j$ as $A_{i,j}$, all of row $i$ as $A_{i,.}$, and all of column $j$ as $A_{.,j}$. For cases where subscripts could be mistaken for indices, we substitute superscripts.}, and we wish to reduce it to a set of eigenvectors $E$ (i.e., a subspace) of dimension $(q, l)$. The eigenvector that contains the most ``information" (corresponding to the vector in $l$-dimensional space that captures the most variation in the data) is the eigenvector of $C = \textrm{Cov}(D)$ with the largest eigenvalue.

To project all of the points in $D$ onto $E$, take the dot product of $D$ with the transpose of $E$, yielding a matrix of dimension $(n, q)$, whose $i^{\textrm{th}}$ row is the weights of each eigenvector used to construct $D_{i,.}$. Thus,
\begin{equation}
    A = D \cdot E^T
    \label{eqn:pc_downprojection_simple}
\end{equation}

Therefore, in order to reconstruct all of the training data $D$ in terms of their first $q$ PCs ($D'$), we take the dot product of $A$ and $E$

\begin{equation}
    T' = A \cdot E
\end{equation}
and define the residual
\begin{equation}
    R = D - D' = D - (A \cdot E)
\end{equation}
which is used to construct a theoretical covariance matrix $K_{th} = \textrm{Cov}(R)$, meant to account for all remaining variation in the models not captured by the first $q$ eigenspectra, and is used in addition to observational and spectrophotometric uncertainties in Section \ref{subsec:PC_unc} to compute weights on each model.

\subsection{Validating number of PCs retained: eigenvalues and the scree plot}
\label{subsec:picking_q}

The $i^{\rm th}$ eigenvalue $\lambda_i$ of a principal component system describes the fraction of the total variance in the system captured by PC $i$:

\begin{equation}
    V^f_i = \frac{\lambda_i}{\sum_j \lambda_j}
\end{equation}

\begin{figure}
    \centering
    \includegraphics[width=\columnwidth]{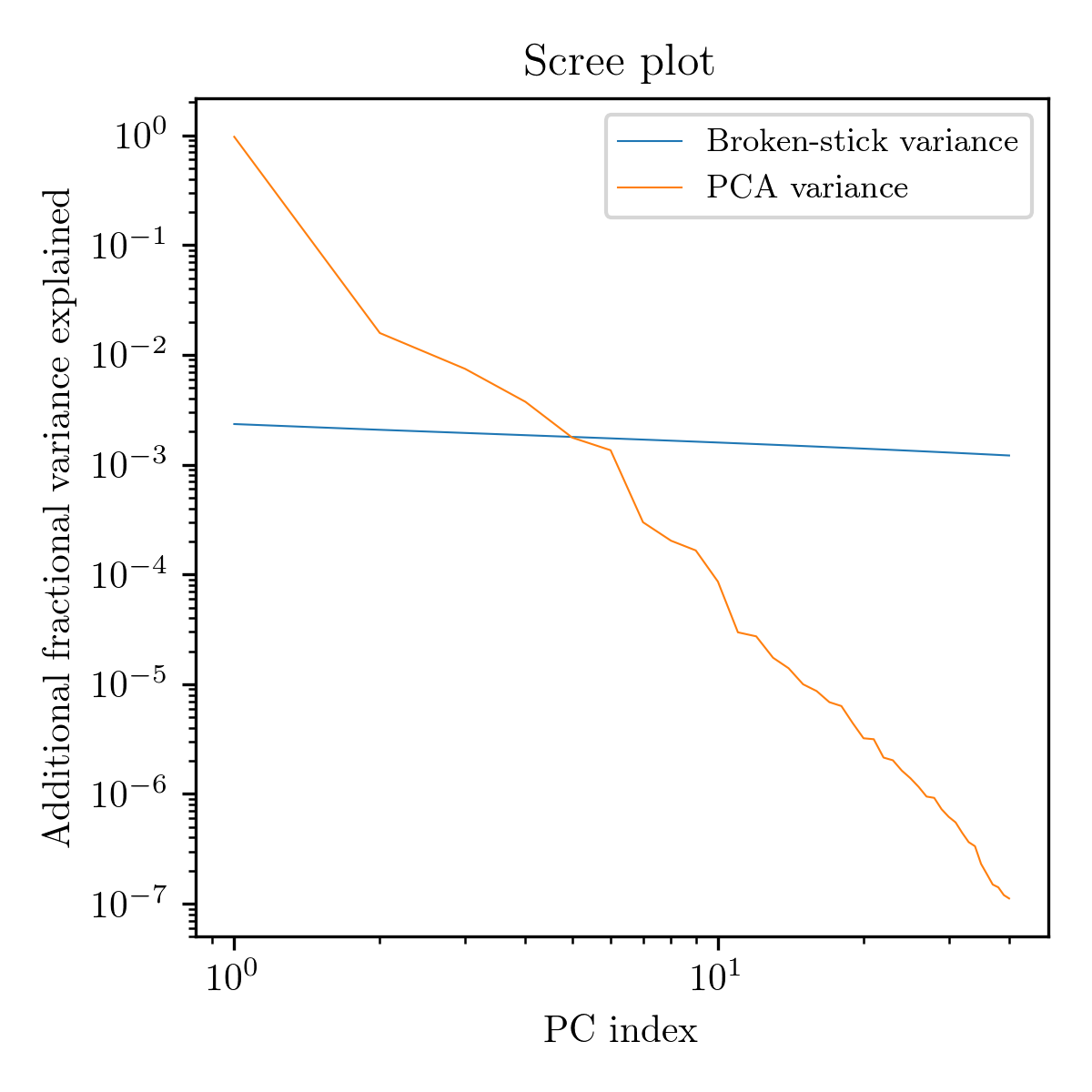}
    \caption{In blue: training data variance described by each successive principal component; in black: the fractional variance expected from the broken-stick method (randomly-apportioned variance).}
    \label{fig:screeplot}
\end{figure}

This is often visualized as a ``scree plot" (Fig. \ref{fig:screeplot}), in which a flattening of $V^f$ is used to indicate lessened marginal gains in fit quality per additional PC retained. \citet{jackson_pca_dim} recommends a heuristic based on the ``broken-stick" method, which assumes that the variance is split randomly into $N$ parts (that is, all spectral channels have uniform variance). In such a case, the $i^{\rm th}$-largest fractional variance will be 
\begin{equation}
    U^f_i = \frac{1}{N} \sum_{j=i}^N \frac{1}{j}
\end{equation}
The PC representation can be considered complete when $U^f_i$ exceeds $V^f_i$ (that is, when any improved fit quality can be ascribed entirely to adding a parameter to the fit). Fig. \ref{fig:screeplot} shows that $q = 6$ is safely in this regime.

\begin{figure}
    \centering
    \includegraphics[width=\columnwidth]{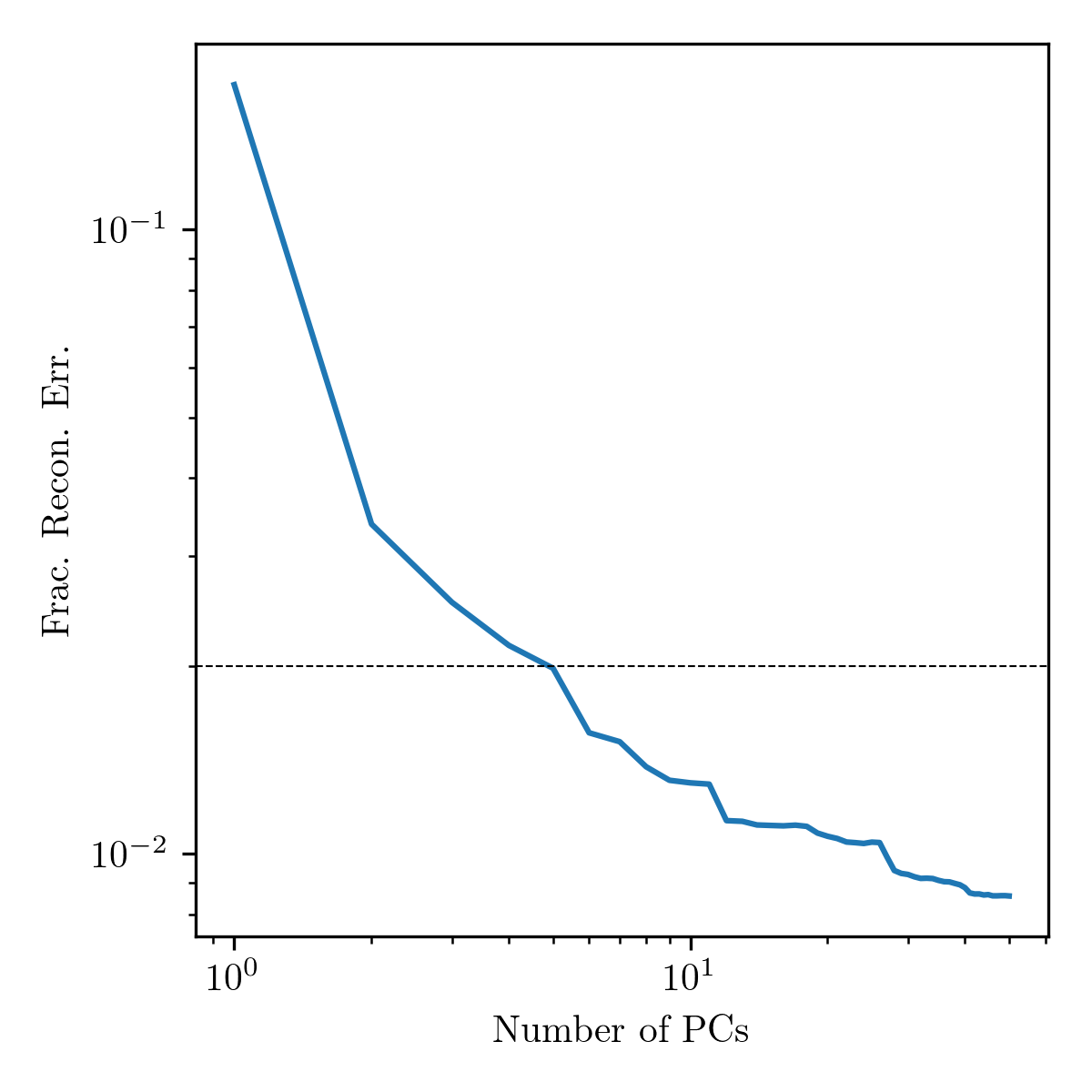}
    \caption{In light blue, the dependence of RMS reconstruction residual on the number of PCs retained. Reconstruction is carried out on a sample of 4000 held-out (``validation") spectra. The black dashed line denotes an RMS reconstruction error of 2\%.}
    \label{fig:xval_q}
\end{figure}

Furthermore, it is desirable to enforce a PCA solution that is general (i.e., the PCs should not lose substantial reliability on data not used to train the model). This can be thought of as the model simply memorizing the training data, and can be evaluated by examining fit quality on held-out (``validation") data generated identically to the training data. Over-fitting could arise from the training SFHs themselves, or from the three sub-sampled parameters ($\sigma$, $\tau_V$, and $\mu$). Fig. \ref{fig:xval_q} illustrates the root-mean-square (RMS) residual between the validation data and their PC representations.

The inclusion of noise in the observed spectra (but not in the CSPs used to construct the PCA model) means that lowering the ``down-projection error" for observed spectra (described by theoretical spectral covariance matrix $K^{th}$) will not substantially improve the fidelity of their reconstructions. In other words, setting the number of principal components retained to 6 means that noisy data will limit the quality of the down-projections (see Section \ref{sec:nmodels_vs_accuracy}) for spectra with median signal-to-noise ratio above approximately 20.

\subsubsection{Computational concerns}

The dimensionality $q$ of the chosen ``reduced" space (i.e., the number of eigenspectra with which we seek to reproduce some general observed spectra of dimension $l$) has a few additional important consequences from the computational perspective:
\begin{itemize}
    \itemsep0em
    \item Matrix multiplication of $A_{n \times m}$ and $B_{m \times p}$ generally is a $\mathcal{O}(m~n~p)$ operation, so minimizing the number of principal components retained will allow faster down-projection.
    \item Since the volume of a cube of $d$ dimensions and side length $2 r$ rises as $(2r)^d$; and the volume of a sphere of $d$ dimensions and radius $r$ rises as $\frac{2 r^d \pi^{d/2}}{d ~ \Gamma(d/2)}$, a sphere occupies a smaller fraction of the cube's volume as $d$ increases. A consequence of this ``curse of dimensionality" is observed when one arbitrarily increases the number of principal components retained, $q$: the distance between two points increases faster than the likelihood-weight can account for the increase, so model weights become extremely low. The likelihood scores used to compare each model to an observed spectrum only provide a \emph{point estimate} of the model likelihood, so seeing many models with nonzero likelihood scores will give confidence that a particular spectrum is well-characterized in PC space.
\end{itemize}

\subsection{Developing a physical intuition for principal components}
\label{subsec:pc_meaning}

\begin{figure*}
    \centering
    \includegraphics[height=0.925\textheight, width=0.925\textwidth, keepaspectratio]{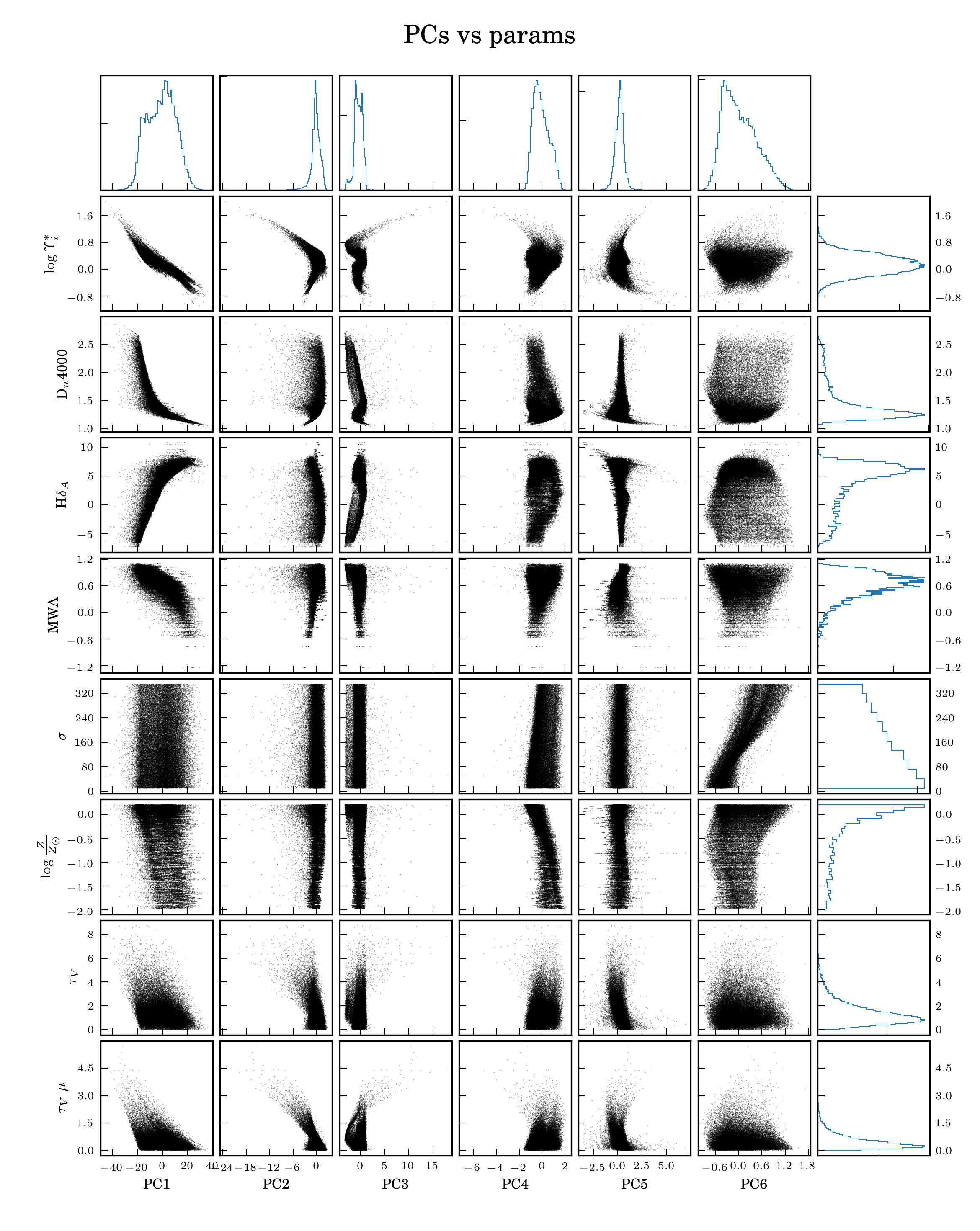}
    \caption{Selected directly-modelled parameters ($\sigma$, $\log \frac{Z}{Z_{\odot}}$, $\tau_V$, and $\tau_V \mu$) and derived parameters ($\log \Upsilon^*_i$, \Dn, \HdeltaA, and $\log$ mass-weighted stellar age), versus principal component amplitudes. Each scatter-subplot plots the amplitude of the PC corresponding to its column (on the x-axis) against the parameter corresponding to its row (on the y-axis). The right-most column and top-most row hold histograms of PC amplitudes and parameter values, respectively.}
    \label{fig:PCs_vs_params}
\end{figure*}

As in \citetalias{chen_pca}, we wish to develop an intuition for the physics encoded in each PC. Though easily understandable relationships between physical quantities and principal component amplitudes are not guaranteed, they do tend to emerge. These relationships can be visualized by plotting each model's PC amplitude against the set of parameters $\{P_i\}$ (see Fig. \ref{fig:PCs_vs_params}). For example, mass-to-light ratio in $r$, $i$, and $z$ bands are most correlated with the first PC. This is of course compatible with the overall shape of that eigenspectrum (see pane 2 of Figure \ref{fig:eigenspectra}). \citet{kong_01_pca} similarly noted (by performing PCA on SSPs) that a young stellar population is correlated with large coefficients on principal component 1. However, some of the information about stellar mass-to-light ratio is contained in higher PCs (which have smaller coefficients, on average), meaning that using just PC1 (as that study did) will never give better results than using all PCs. Another striking example is the correlation of velocity dispersion $\sigma$ with principal components 3 and 6.

However, caution must be used when interpreting the eigenspectra directly: these intuitive interpretations are made under the assumption that the training spectra represent reality both in individuals stars (not guaranteed, in the case of the fully-theoretical spectra used here); and in the adopted distributions of SFHs. That is, the training data and the PCA dimensionality-reduction must work in tandem.

\subsection{The observational spectral covariance matrix}
\label{subsec:cov}

There is an additional source of uncertainty in MaNGA spectra, beyond that provided in the LOGCUBE data products. Specifically, the spectrophotometric flux-calibration of individual exposures, followed by the compositing of those exposures into a regularly-gridded datacube, induces small \citep[$\sim 4\%$, according to][]{manga_drp}, wavelength-dependent irregularities in individual spectra. In part, this is because the exposures are taken under a wide variety of airmasses \& seeing conditions. The overall effect is that of a small covariance between spectral channels: \citetalias{chen_pca} found that accounting for this covariance is necessary for obtaining reliable estimates of stellar mass-to-light ratios and other quantities. The covariance is described by a matrix $K_{obs}$, which can be calculated by comparing multiple independent sets of observations of a single object (\citetalias{chen_pca}, Equation 9):

\begin{eqnarray}
    K_{obs}(\lambda_1,~\lambda_2) = \frac{1}{2 N_{pair}} \sum_{j=1}^{N_{pair}} \nonumber \\ \left[ (S_j^0(\lambda_1) - S_j^1(\lambda_1)) \times (S_j^0(\lambda_2) - S_j^1(\lambda_2)) \right]
\end{eqnarray}
where each element $K_{obs}(\lambda_1,~\lambda_2)$ denotes the covariance between observed-frame spectral elements $\lambda_1$ and $\lambda_2$, and is calculated using the difference between two spectra ($S_j^0$ and $S_j^1$) of a single object $j$.

\begin{figure}
    \centering
    \includegraphics[width=\columnwidth]{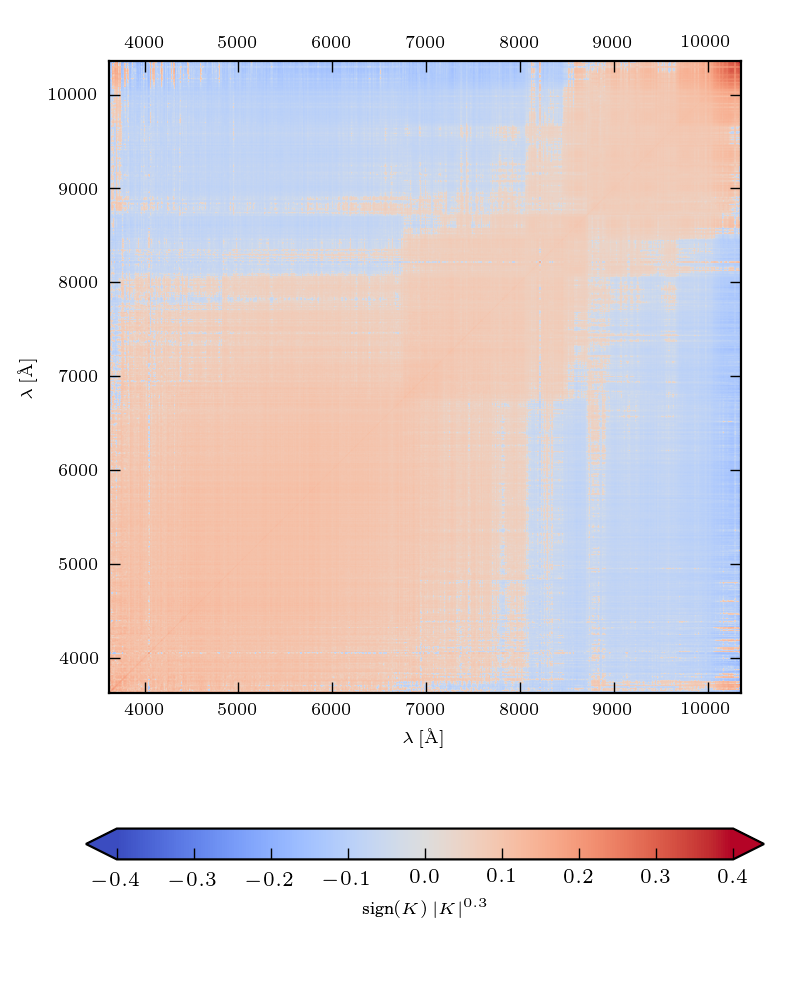}
    \caption{MaNGA's observational covariance matrix $K_{obs}$, which arises due to imperfect spectrophotometric flux-calibration of MaNGA spectra. See Figure 5 of \citetalias{chen_pca} for comparison.}
    \label{fig:cov_obs_manga}
\end{figure}

\begin{figure}
    \centering
    \includegraphics[width=\columnwidth]{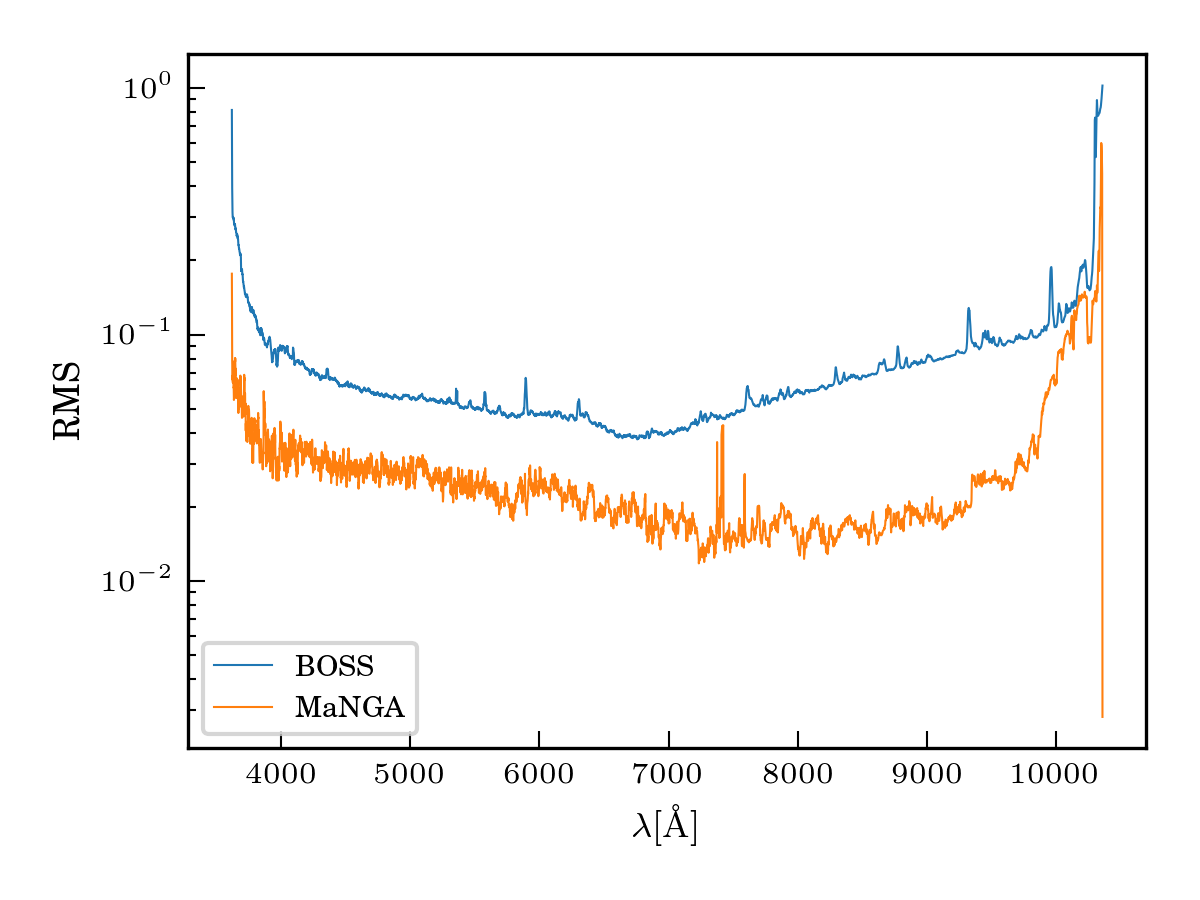}
    \caption{In blue, the RMS value of 10,000 noise vectors drawn from the BOSS covariance matrix; and in orange, the RMS value of 10,000 noise vectors drawn from the MaNGA covariance matrix.}
    \label{fig:boss_manga_cov}
\end{figure}

In \citetalias{chen_pca}, the spectral covariance matrix was found using reobserved objects from the SDSS(-III)/BOSS project. Since BOSS and MaNGA use the same spectrograph, the spectral covariances will be similar; however, the hexabundle construction of the MaNGA IFUs results in more precise compensation for atmospheric dispersion, which commensurately improves spectrophotometric calibration \citep{manga_spectrophot}. Therefore, we will recalculate $K_{obs}$ using multiply-observed MaNGA galaxies. Though the number of re-observed MaNGA galaxies is much lower than the number of re-observed BOSS sources, each MaNGA galaxy has hundreds or thousands of spectra that can be compared with their ``sister" locations. The result is shown in Figure \ref{fig:cov_obs_manga}, and as expected contains less off-diagonal power than the BOSS covariance. While the covariance should be smooth (since its main contributor is the multiplicative flux-calibration vector), there are some sharper features which manifest in the RMS of ten-thousand random draws from $K_{obs}$ (Figure \ref{fig:boss_manga_cov}): for instance, in the $\sim 7000-8000 \mbox{\AA}$ range. While such features could perhaps be attributed to poorly-compensated sky emission or telluric absorption, this appears not to be the case: we have examined both $K_{obs}$ itself and random draws from it, but found no consistent correspondence with typical telluric absorption or sky emission spectra.

$K_{obs}$ can be equivalently thought of as a multivariate-normal probability distribution (with each spectral channel being represented by one row and column in the covariance matrix) centered around zero, describing the noise profile for an ensemble of MaNGA spectra. This view offers a pathway towards comparing the covariance of MaNGA spectra with that of BOSS spectra. We draw 10,000 samples each from the BOSS covariance matrix (which was computed in \citetalias{chen_pca}) and the MaNGA covariance matrix. At each wavelength, the RMS value (which can be taken as the average RMS value of the noise in that spectral channel) is computed. The results of that computation are shown in Figure \ref{fig:boss_manga_cov}. As a general rule, the BOSS covariance matrix (computed and used in \citetalias{chen_pca}) has greater spectrophotometric uncertainty (generally by a factor of $\sim 5$ in the wavelength ranges employed in this work) than the MaNGA covariance matrix computed above.

$K_{obs}$ will be used in Section \ref{subsec:PC_unc} to obtain a PC amplitude covariance matrix and confidence bounds for parameters of interest.

\subsection{Fitting the observations with eigenspectra}
\label{sec:obs2pc}

Each observed spectrum can now be fit as a linear combination of ``eigenspectra" $E$, subject to a scaling ($a$) and an unknown (but constrained) noise vector ($N$), which comprises the incompleteness of the PCA decomposition and the imperfect spectrophotometry.

First, an observed spectrum is pre-processed:
\begin{enumerate}
\itemsep0em
    \item Galactic extinction is removed, assuming an \citet{odonnell_94_mw_extinction} extinction law and $R_V = 3.1$, and using the $E(B-V)$ color excess provided in the header of the reduced data products \citep{schlegel_finkbeiner_davis}. Both flux-density and its inverse-variance are corrected.
    \item The spectrum is brought into the rest frame, combining the systemic velocity obtained from the NSA with spatially-resolved stellar velocity field obtained from the MaNGA DAP results. Both the flux-density and its inverse-variance are de-redshifted and drizzled into the rest frame using an adaptation of \citet{carnall_17}\footnote{It is preferable to obtain a rest-frame spectrum with the exact same wavelength pixelization as the eigenspectra. Two wavelength solutions $f_l$ and $f_r$ are extracted, corresponding to the two closest integer-pixel mappings between the eigenspectra's wavelength grid and the spectral cube's ``exact" wavelength solution. $f_l$ and $f_r$ are combined with weights equal to the relative fraction that they subtend on the exact solution. The uncertainties in these two exact solutions are also propagated into a final, re-gridded solution. This approach was found to produce better fits to the observations than the integer-pixel solution, which tended to prefer a fit with broader absorption features.}.
    \item The spectrum is normalized by its median value ($a$), and the median spectrum ($M$) of the PCA system is subtracted.
    \item Spectral channels with likely contamination by emission lines are flagged for later replacement: any spectral channel within 1.5 times the line-width (velocity dispersion, as found in the MaNGA DAP) from the rest-frame line center is flagged. Which wavelength locations are masked is based on the equivalent width of the H$\alpha$ emission-line measured by the MaNGA DAP:
    \begin{itemize}
    \itemsep0em
        \item Always: H$\alpha$ through H$\epsilon$; [O\textsc{ii}]3726,28; [Ne\textsc{iii}]3869; [O\textsc{iii}]4959,5007; [O\textsc{I}]6300; [N\textsc{ii}]6548,84; [S\textsc{ii}]6716,30; [S\textsc{III}]9069; [S\textsc{III}]9531
        \item Where EW$(H\alpha) > 2~\mbox{\AA}$: Balmer lines through H$30$
        \item Where EW$(H\alpha) > 10~\mbox{\AA}$: Paschen series P$8$ through P$18$; He\textsc{i}3819; He\textsc{i}3889; He\textsc{i}4026; He\textsc{i}4388; He\textsc{i}4471; He\textsc{ii}4686; He\textsc{i}4922; He\textsc{i}5015; He\textsc{i}5047; He\textsc{i}5876; He\textsc{i}6678; He\textsc{i}7065; He\textsc{i}7281 [Ne\textsc{III}]3967; [S\textsc{II}]4069; [S\textsc{II}]4076; [O\textsc{III}]4363; [Fe\textsc{III}]4658; [Fe\textsc{III}]4702; [Fe\textsc{IV}]4704 [Ar\textsc{IV}]4711; [Ar\textsc{IV}]4740; [Fe\textsc{III}]4989; [N\textsc{I}]5197; [Fe\textsc{III}]5270; [Cl\textsc{III}]5518; [Cl\textsc{III}]5538; [N\textsc{II}]5755; [S\textsc{III}]6312; [O\textsc{I}]6363; [Ar\textsc{III}]7135; [O\textsc{II}]7319; [O\textsc{II}]7330; [Ar\textsc{III}]7751; [Ar\textsc{III}]8036; [O\textsc{I}]8446; [Cl\textsc{II}]8585; [N\textsc{I}]8683; [S\textsc{III}]8829
    \end{itemize}
    Flagged spectral channels are replaced (i.e., item-imputed) as the inverse-variance-weighted rolling-mean of the unmasked subset of the nearest 101 pixels. This approach is almost identical to that employed in \citetalias{chen_pca}. Since this step is performed on the normalized, median-subtracted spectrum, the replacement does not universally decrease, for instance, the depth of absorption-lines in the spectrum. As we discuss below, the more rigorous way of performing this calculation would involve \emph{re-computing} the geometric transformation that produces the PC amplitudes from the spectrum, an unacceptable loss in speed. Previous work has demonstrated that modestly-gappy data ($\sim 10\%$ of spectral channels masked) produces $\sim 2\%$ deviations (RMS) in principal-component amplitudes \citep[Figure 5 of][]{connolly_robust_pca}. Other possible frameworks for emission-line masking are discussed in Section \ref{subsec:balmer}.
    \item If the spectrum is more than 30\% masked by either data-quality flags or emission-line masks, the entire spectrum is presumed bad. Tests on further synthetic spectra (see Section \ref{subsec:mocks_tests} and Appendix \ref{apdx:fakedata} for more details about how such mock observations were prepared) suggest that in spectra \emph{un-contaminated} by bad data, fits with and without flags do not substantially change either PC amplitudes (i.e., a group of similar noise realizations of the same synthetic spectra, at a single SNR, does not, in a statistically-significant way, experience a change in its PC amplitudes) or the estimates of stellar mass-to-light ratio that emerge.
    \item The spectrum $S = \frac{O}{a} - M$ is now ready to be decomposed using the eigenspectra obtained in Section \ref{subsec:run_pca}.
\end{enumerate}

Transforming the discretely-sampled spectrum by a fraction of a pixel also induces a small, off-diagonal covariance $K_{obs}^{od}$. The exact magnitude of the covariance depends on the position within a rest-frame spectral bin of the boundary between the two nearest integer-pixel solutions. The position of this boundary, $f$, lies in the range 0--1 (in units of the width of a $\log \lambda$ bin), and the off-diagonal terms are the variances of the left-hand-side and right-hand-size, weighted by $f_{lhs}$ and $f_{rhs} = 1 - f_{lhs}$.
\begin{equation}
    K_{obs}^{od} = f_{lhs} N_{obs}^{lhs} + (1 - f_{lhs}) N_{obs}^{rhs}
\end{equation}
which depends only weakly on the precise rest-frame pixel boundary, so we fix $f_{lhs} = f_{rhs} = 0.5$, where the result is maximized for the case of constant noise.

\subsection{Towards optimal flagging and masking of Balmer emission-lines}
\label{subsec:balmer}

The Balmer absorption features in stellar population spectra are among the most important age diagnostics; however, in all but the most quiescent, gas-free environments, these features will be contaminated by gaseous emission. As stated above, in this work, we elect to flag all spectral elements within 1.5 times the velocity-dispersion of ${\rm H\alpha}$. Those flagged spectral elements of the median-subtracted spectrum $S$ are then replaced by the weighted mean of the nearest 101 spectral elements (hereafter notated as the ``WM101" or fidicual method). On one hand, this relatively narrow flagging region might induce a bias in the PC amplitudes for spectra with bright, high-velocity-dispersion gaseous emission; on the other hand, it is not desirable to sacrifice the information contained in these important spectral features. We address here two alternatives: work with emission line-subtracted spectra (as \citealt{gallazzi_charlot_05} does---the ``GC05" method), or explicitly exclude all flagged-and-replaced spectral channels (notated as the ``M" method, because it is equivalent to replacing flagged spectral channels with the median of the PC system).

It is perhaps most tempting to work with spectra where emission lines have already been subtracted, since the cores of the Balmer absorption lines are now uncontaminated (``GC05"). However, this requires having first executed a round of full-spectral fitting (which necessarily adopts a stellar library). Indeed, concurrent work with MaNGA IFS data has shown that emission line measurements can be sensitive to the particular SSP library used for fitting the stellar continuum: \citet[][Figure 9]{belfiore_19_dap-elines} indicates that as S/N rises beyond 10, changing spectral library from the hierarchically-clustered MILES library (MILES-HC, which is the DR15 fiducial) to MIUSCAT, M11-MILES, or BC03 induces a systematic uncertainty in emission-line flux comparable to the random uncertainties. In other words, the choice of stellar library is important.

One could also argue for the more conservative masking option, explicitly excluding all spectral channels suspected to be contaminated by emission-lines (``M" method). The case against that tactic is more subtle: first, the PC system used in this work is centered at zero, as a result of subtracting the median spectrum $M$ of the CSP library from each of the CSP spectra. When one ``eliminates" spectral channels thought to be unreliable, one implies that the values in those channels are identical to the corresponding value in $M$ (i.e, there is no further information beyond what the median spectrum of the SFH training library provides); in reality, the values in the spectrum in such channels are likely better approximated by an average of their near neighbors.

\subsubsection{Tension between flagged-and-replaced spectra and their fits?}

\begin{figure*}
    \centering
    \includegraphics[width=0.95\textwidth]{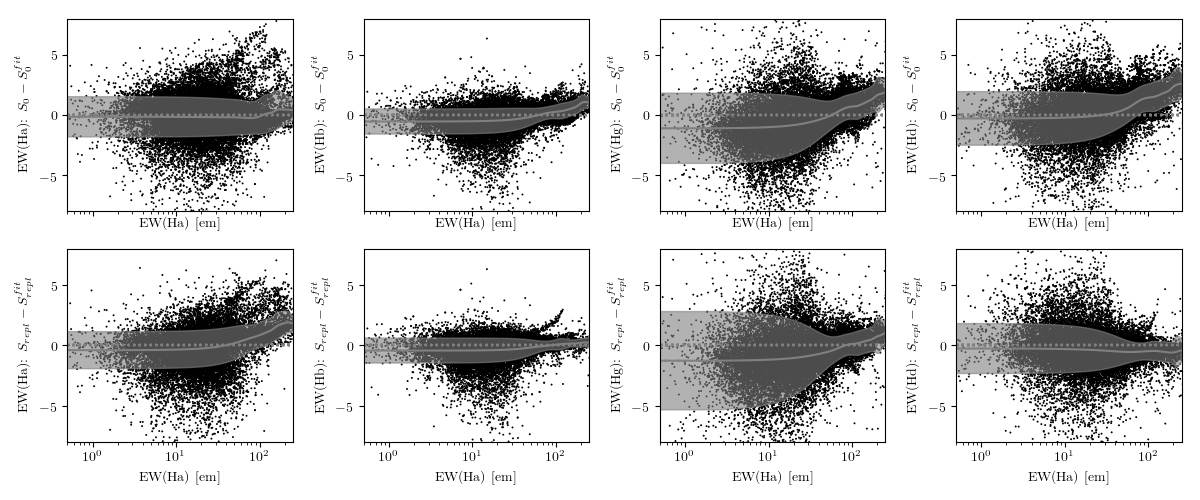}
    \caption{Each panel shows the difference in EW of Balmer absorption (left to right: ${\rm H\alpha}$, ${\rm H\beta}$, ${\rm H\gamma}$, ${\rm H\delta}$) between the corrected and corrected-then-fit spectra: in the top row, ``corrected" refers to flagged elements being replaced with the corresponding values in $M$ ($S_0$, equivalent to neglecting those spectral channels entirely); in the bottom row, ``corrected" refers to flagged elements replaced with the rolling mean of their neighbors ($S_{repl}$). The solid, gray line denotes the rolling median at fixed ${\rm EW_{em}(H\alpha)}$, and the gray band the dispersion at fixed ${\rm EW_{em}(H\alpha)}$.}
    \label{fig:balmermasking_test}
\end{figure*}

We show here a further test, which uses the 25 most extremely star-forming (but non-AGN) galaxies, based on total integrated ${\rm H\alpha}$ luminosity (from the MaNGA DAP). If the ``WM101" method neglects effects from emission wings, then we should see deficiencies in the stellar continuum fits around the Balmer lines as the equivalent width of ${\rm H\alpha}$ in emission increases. In other words, we want to know if unmasked emission wings cause a problem in our fiducial correction more than in the alternative ``M" method. We correct the 25 high-SFR galaxies using both methods, and fit them using the PCA basis set. Finally, for both correction methods, we measure \& compare equivalent width of four Balmer absorption lines (${\rm H\alpha}$, ${\rm H\beta}$, ${\rm H\gamma}$, and ${\rm H\delta}$) in both the corrected-observations and the fits to them (Figure \ref{fig:balmermasking_test}). If, as ${\rm EW_{em}(H\alpha)}$ increases, the ``fit" and ``corrected-then-fit" spectra produce significantly different ${\rm EW_{abs}(H\alpha)}$ values, then one would conclude that a strong Balmer emission line ``biases" the eventual spectral fit.

The result of these comparisons is shown in Figure \ref{fig:balmermasking_test}: each panel shows the difference in the equivalent widths of Balmer lines in absorption between the initial ``corrected" spectra and the fits to those spectra (in the top panels, correction is performed with the ``M" method; in the bottom panels, correction is performed with the ``WM101" method; and left to right, columns refer to ${\rm H\alpha}$, ${\rm H\beta}$, ${\rm H\gamma}$, and ${\rm H\delta}$). The differences between these cases are very slight, but at the most basic level, regardless of correction paradigm, stronger Balmer absorption in the corrected spectra than in their fits tends to correlate with increased ${\rm H\alpha}$ emission. However, replacement with $M$ tends to produce a stronger Balmer absorption deficit in the fits, regardless of which line is considered; the ``WM101" method behaves in a manner less dependent on ${\rm EW_{em}(H\alpha)}$ in the case of ${\rm H\beta}$ \& ${\rm H\delta}$ (little to no improvement is seen in the ${\rm H\alpha}$ and ${\rm H\gamma}$ cases). While it's clear that ``WM101" produces some tension between individual spectra and their fits, this basic test indicates that the performance in the vicinity of some Balmer absorption lines is more consistent than the ``M" method.

\subsubsection{Evaluating Balmer-masking with synthetic observations of PCA best-fits}

Here we produce and discuss an additional test of the two candidate replacement schemes: the fiducial (``WM101") and the alternative (``M"). For each of 200 randomly-selected galaxies, we perform a normal PCA fit of each spaxel (projecting each individual, observed spectrum onto the principal components obtained from the training data---see Section \ref{subsec:PC_unc}). The obtained principal component amplitudes $A$ are then used to reconstruct the best approximation of the observation, $O_{\rm true}$, which we treat as the ``known" spectrum. We also measure the equivalent width of the H$\beta$ absorption feature \citep{worthey_ottaviani_97} for $O_{\rm true}$. After applying instrumental noise to $O_{\rm true}$ (see Section \ref{subsec:cov} and Section \ref{subsec:mocks_tests} for more information about constructing synthetic observations), we fit $O_{\rm true}$ using each of the two correction methods, transform (as before) the resultant fit from PC space to spectral space ($O_{\rm fit}$), and once again measure H$\beta$ for each case.

\begin{figure*}
    \centering
    \includegraphics[width=0.95\textwidth]{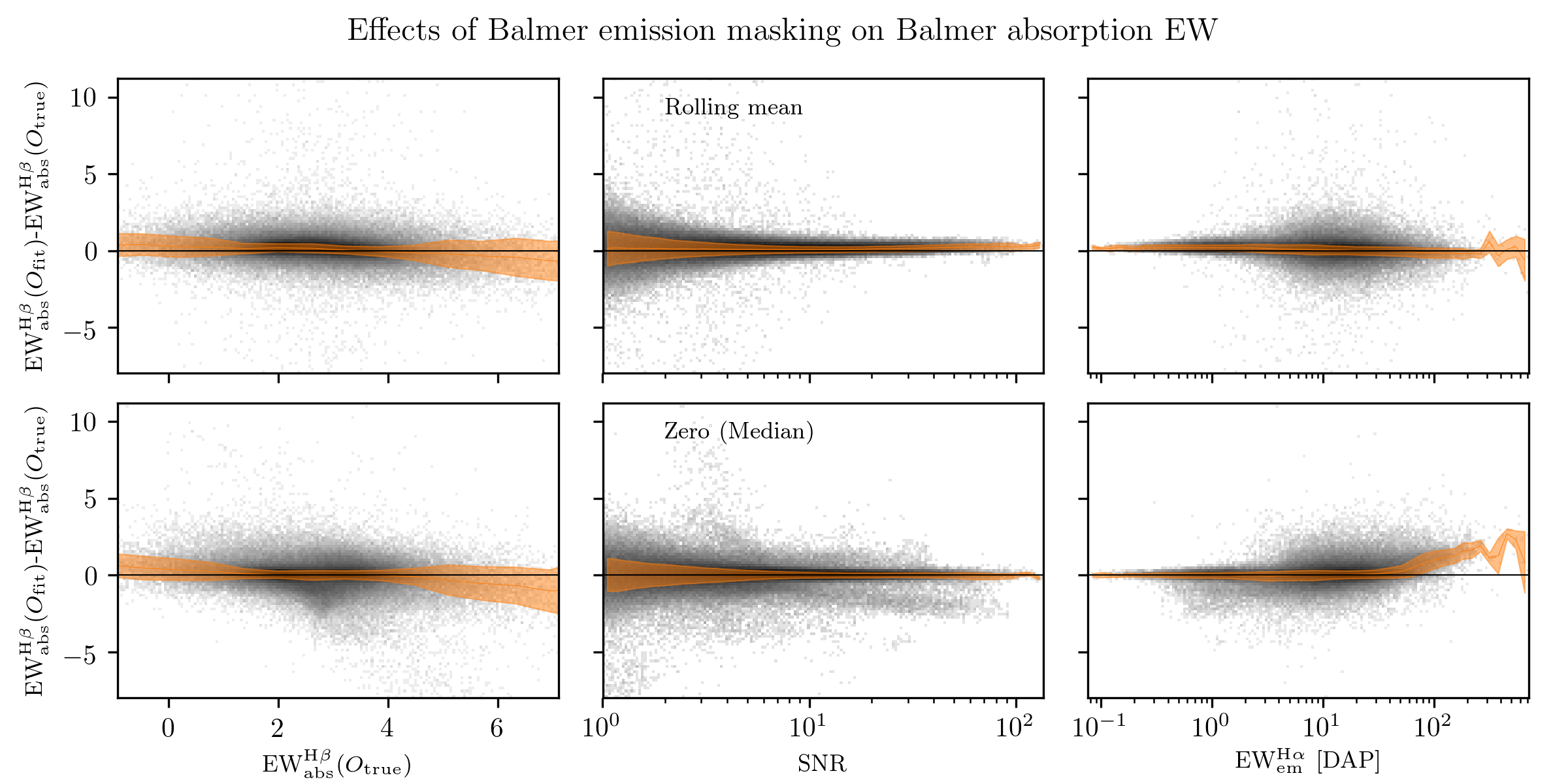}
    \caption{$d{\rm EW} = {\rm EW_{abs}^{H\beta}}(O_{\rm fit}) - {\rm EW_{abs}^{H\beta}}(O_{\rm true})$ versus (left to right) ${\rm EW_{abs}^{H\beta}}(O_{\rm true})$, median signal-to-noise ratio (SNR), and DAP equivalent width of the H$\alpha$, using the fiducial (top row) and the alternative (bottom row) strategies. Pixels are colored according to the logarithm of the number of spectra within. On each panel is overlaid the rolling median of $d{\rm EW}$ (red line) and the dispersion about the median calculated using the median absolute deviation (red band).}
    \label{fig:mocks_of_fits_ewHb}
\end{figure*}

Figure \ref{fig:mocks_of_fits_ewHb} shows the difference $d{\rm EW} = {\rm EW_{abs}^{H\beta}}(O_{\rm fit}) - {\rm EW_{abs}^{H\beta}}(O_{\rm true})$ for the rolling-mean replacement (top row) and the zero-replacement (bottom row). $d{\rm EW}$ is plotted against (from left to right) ${\rm EW_{abs}^{H\beta}}(O_{\rm true})$, median signal-to-noise ratio (SNR), and the equivalent width of the H$\alpha$ emission line as reported in the MaNGA DAP (spaxels with fractional uncertainty in H$\alpha$ emission equivalent width greater than $\frac{1}{3}$ are excluded). Broadly speaking, the two cases are very similar; but slight differences emerge in limiting cases. For instance, for the ``M" method seems to produce more outliers with $d{\rm EW} < 0 \mbox{\AA}$ at high ${\rm EW_{abs}^{H\beta}}(O_{\rm true})$, and vice-versa at low ${\rm EW_{abs}^{H\beta}}(O_{\rm true})$ (but behaves on average the same). Though this effect is small, it suggests that Balmer depths can be somewhat moderated by replacement with the median spectrum $M$ (which conceptually represents a medium-age stellar population).

Second, the ``WM101" method exhibits some unbalance of outliers having $d{\rm EW} \sim -1 \mbox{\AA}$ at moderate-to-high signal-to-noise. That said, it has a locus at $d{\rm EW} \sim 0.2 \mbox{\AA}$ at similar signal-to-noise. Finally, though the ``WM101" method is stable with respect to ${\rm EW^{H\alpha}_{em}} [DAP]$, the ``M" method becomes somewhat overzealous in its production of $d{\rm EW} > 0 \mbox{\AA}$ fits. The apparent ``bulging" of the two distributions at moderate ${\rm EW^{H\alpha}_{em}} [DAP]$ reflects that more spaxels reside in that neighborhood, rather than an intrinsic deficiency of the replacement schemas there. The effects we note here are subtle, and this test suggests that the two proposed replacement methods do not substantially differ except in the most extreme cases. Ultimately, the widths of $d{\rm EW}$ are small. That said, because the ``WM101" method behaves more uniformly with respect to ${\rm EW_{abs}^{H\beta}}(O_{\rm true})$ and ${\rm EW^{H\alpha}_{em}} [DAP]$, we believe it is the slightly preferable choice.

\subsubsection{Flag-and-replace stellar models}

We briefly explore here the effects of the ``WM101" method on the CSP model spectra themselves, and what influence that exerts on the eigenspectra. Since the Balmer absorption lines provide indications of stellar population age, smoothing over those features in the models should also suppress them in a resulting principal component basis set. Beginning with the set of CSPs described in Section \ref{sec:SFHs}, the ``WM101" method outlined in Section \ref{sec:obs2pc} is used to ``eliminate`` the influence of all spectral channels within 120 ${\rm km s^{-1}}$ of all Balmer lines\footnote{This velocity window is used as an illustration for the case of a reasonably wide emission line.}. After those adulterations, the model spectra are once again used to build a PC basis set.

\begin{figure*}
    \centering
    \includegraphics[angle=90,height=0.95\textheight]{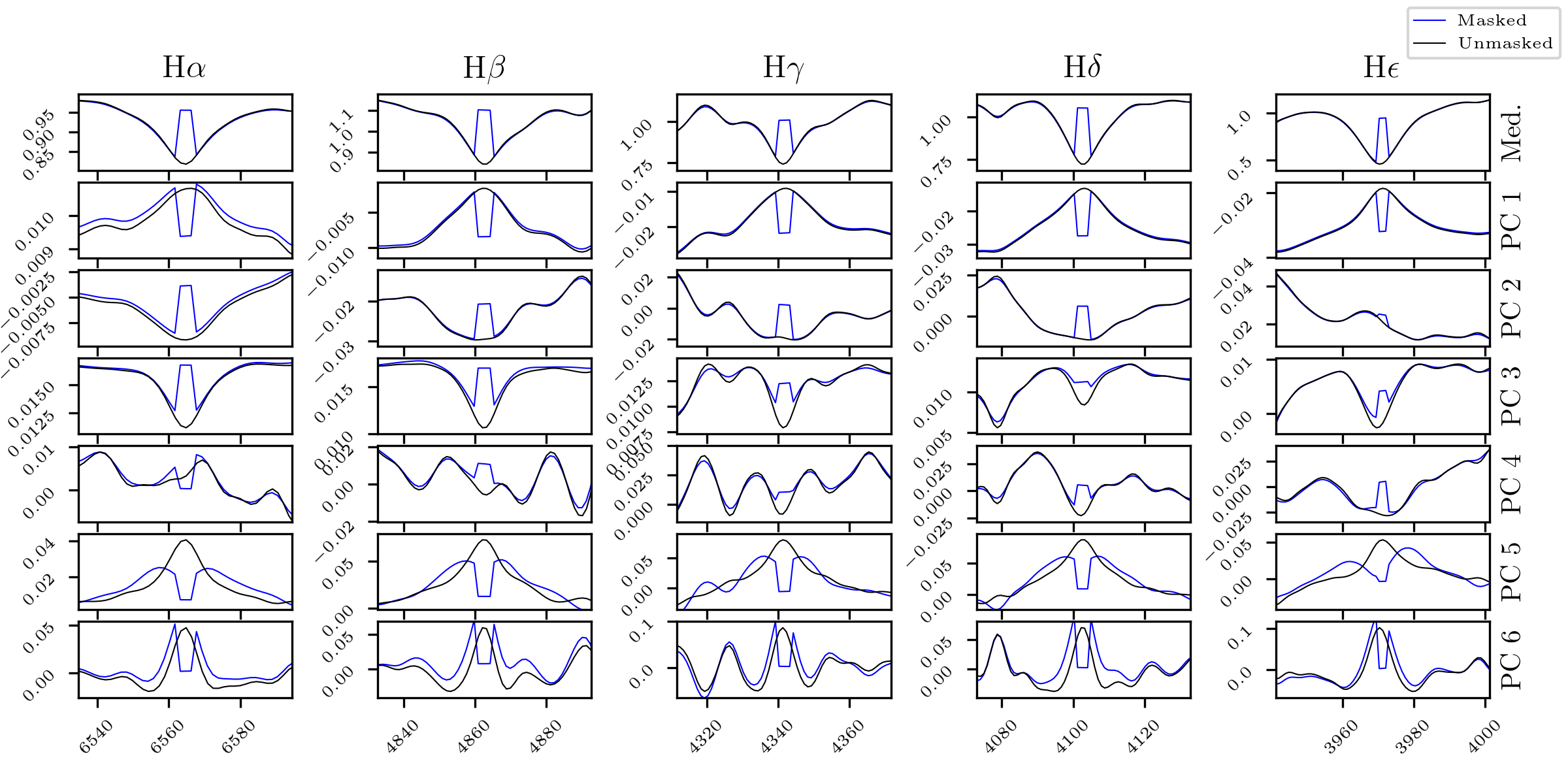}
    \caption{A comparison in the vicinity of the first five Balmer absorption lines (H$\alpha$, H$\beta$, H$\gamma$, H$\delta$, and H$\epsilon$) of the principal component basis set resulting from flag-and-replacement (\textit{blue}) and no flag-and-replacement (\textit{black}). The overall spectral shape is largely preserved, especially in the lower principal components; and the masked regions exert an influence opposite to the absorption features. Each row of panels signifies a principal component vector (or, for the top row, the median spectrum), and each column corresponds to one absorption feature).}
    \label{fig:pcs_withmasking_compare}
\end{figure*}

Figure \ref{fig:pcs_withmasking_compare} shows a comparison between the eigenspectra of our ``normal" PC basis set with that resulting from ``WM101" replacement of the \emph{model spectra themselves} in the vicinity of Balmer line centers. The shapes of the eigenspectra (i.e., neglecting the core of the absorption line affected by the mask) are conserved best in PC1--PC4 (as mass-to-light ratio follows PC1 most closely, this is a very desirable behavior). Furthermore, examining the shapes of the absorption features, if the fiducial PC basis set ``dips" in the core of the line, the flag-and-replaced version tends to ``rise" in the handful of spectral channels within the replacement area; and if the fiducial ``rises", then the replaced area ``dips". Note that the eigenspectra's masked spectral channels are \textit{not} necessarily drawn towards zero, simply in the opposite direction as the manifestation of the Balmer absorption feature.

\subsection{Estimating PC coefficients and uncertainties for observed spectra}
\label{subsec:PC_unc}

We now discuss finding the values and the uncertainty of the principal component coefficients $A$. Observed galaxy spectra $O$ previously had their missing data (where emission-line or data-quality masks are set) are then imputed by a rolling filter with a width of 101 pixels. If data cannot be imputed in this way, it is still possible to perform the calculation below by imputing missing values as zeros (introduces some bias), or by explicitly eliminating entries of columns of eigenspectra $E$ and both rows and columns of spectral covariance $K$ where data are flagged and replaced (much slower, as the projection matrix must be explicitly recalculated for each spectrum). Spectra $O$ are then normalized by dividing by their median value $a$ and subtracting the PCA median spectrum $M$, yielding a spectrum $S$.

The PC amplitudes $A$ are the solution to the linear system $E ~ A = S$, subject to covariate noise (assumed to be drawn from a multivariate-normal distribution with mean zero and covariance $K$). In particular, an individual observation $S$ includes the ``true" spectrum $S_0$; plus contributions from the ``theoretical" noise, $N_{th}$ (which accounts for the imperfect PCA decomposition), the error due to imperfect spectrophotometry $N_{obs}$ (discussed in Section \ref{subsec:cov}), the small off-diagonal covariance $K_{obs}^{od}$ resulting from the fractional-pixel rest-frame wavelength solution, and the photon-counting noise $N_{cube}$ reported in the datacube itself:
\begin{equation}
    \widetilde{S} = S + a ~ N_{th} + N_{obs} + N_{cube}
\end{equation}

The noise vectors $N_{th}$ and $N_{obs}$ are assumed to be drawn from their respective covariance matrices $K_{th}$ and $K_{obs}$, and $N_{cube}$ is the noise profile associated with the measured and reported inverse-variance of the data. $K_{th}$ was computed above as the covariance of the residual obtained in reconstructing the training data from the first $q$ PCs. $K_{obs}$ indicates the uncertainty manifested in the flux-calibration step of data reduction (see Section \ref{subsec:cov}).

$K_{obs}$ should be evaluated over the \emph{observed} wavelength range appropriate to particular observations, rather than over the corresponding rest-frame wavelength range. This produces a slightly different covariance matrix from spaxel to spaxel even within the same spectral cube, and potentially a very different covariance matrix from object to object. This is due to the different observed-frame positions of the same rest wavelength, as recessional velocity changes; as well as the varying surface brightness within a galaxy's physical extent. As $K_{obs}$ is assumed to be smooth on small wavelength scales, we use the nearest-pixel solution for each spectrum. We add a small ($\alpha \sim 10^{-3}$) regularization term to the main-diagonal of $K_{obs}$: this functions as a ``softening parameter", which maintains a minimum dispersion of $K_{PC}$ (only becoming important at high signal-to-noise). This small term allows for some marginal data-model mismatch (see Section \ref{sec:discussion})--but still allows for data-quality masks to be set in the case of PDFs which are an especially bad match for the prior (see Section \ref{subsec:data_quality} for more discussion of data-quality masks).

In order to solve the system $E ~ A = S$, we define the orthogonal projection matrix
\begin{equation}
    H = (E^T E)^{-1} E^T
\end{equation}
where $(E^T E)^{-1}$ is found only once through Cholesky decomposition \citep[one method of decomposing a Hermitian, positive-definite matrix into the product of a lower-triangular matrix and its conjugate transpose, ][]{numerical-recipes_1986} after regularizing on the main diagonal\footnote{The effect of the regularization's strength on the residual between original and ``reduced" spectrum is strongly subdominant to the effect of the dimensionality reduction itself, and the fit quality does not change noticeably over a wide range in regularization parameter $\alpha \sim [1 \times 10^{-6},~1]$}. Since $H$ depends only on the eigen-decomposition of the training data, it is not affected by the specific noise realization of an observation, and need not be calculated repeatedly, unless masked data wish to be explicitly discounted, rather than replaced with a local median as outlined above. The maximum-likelihood PC weights are then given by
\begin{equation}
    A = H ~ S
\end{equation}
and the principal component covariance matrix by
\begin{equation}
    K_{PC} = H^T K H
\end{equation}

\begin{figure*}
    \centering
    \includegraphics[width=\textwidth,height=0.9\textheight,keepaspectratio]{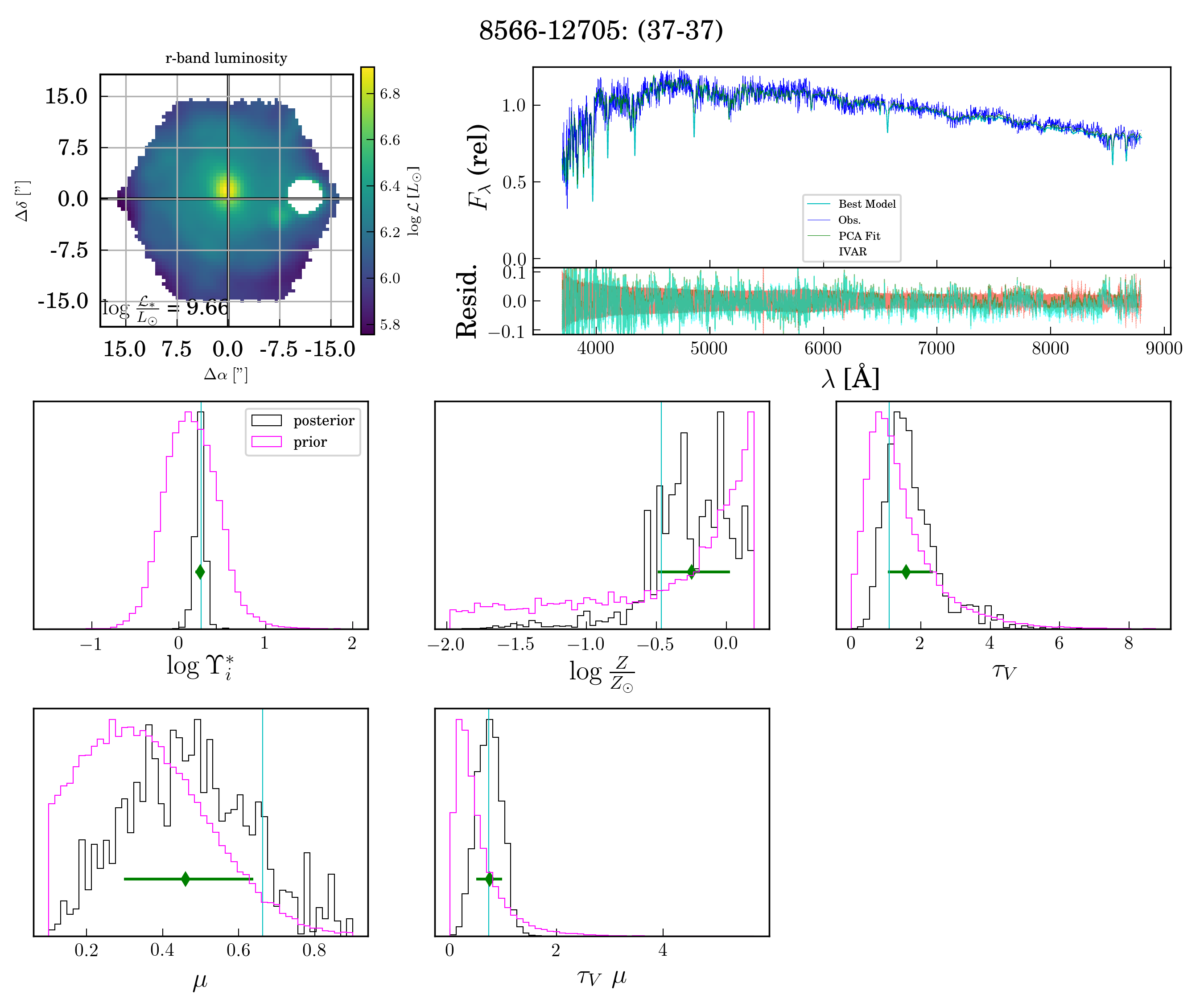}
    \caption{The standard diagnostic figure produced for the center spaxel (coordinates \texttt{37, 37}) of MaNGA galaxy 8566-12705. Top-left frame: map of the galaxy's $i$-band luminosity (the ``hole" in the map signifies where data have been masked due to either a foreground star or data-quality issues identified in the data reduction process). Top-right frame, top section: in \textit{blue}, the observed, median-normalized spectrum $\frac{O}{a}$; in \textit{green}, the spectrum reconstructed from the first 6 principal components of the model library; in \textit{cyan}, the highest-weighted model spectrum (flagged spectral channels are not displayed). Top-right frame, bottom section: in \textit{green}, the residual of the PC fit (with respect to the original spectrum); in \textit{cyan}, the residual of the best-fitting model (with respect to the original spectrum); flagged spectral channels are not displayed; in \textit{salmon}, the average fractional residual of the PCA fit, approximately 5\%, in this case (comparable to the typical spectrophotometric error budget of MaNGA spectra). Other frames: histograms of individual SPS input and derived parameters, with the full training model set (``prior") in magenta, the distribution after weighting by model likelihoods (in \emph{black}, see Equation \ref{eqn:logl}), and the highest-likelihood model as the vertical, cyan line. The 50$^{\rm th}$ percentile of the posterior is shown as a green diamond, and the 16$^{\rm th}$ to 84$^{\rm th}$ percentile range as a green, horizontal bar.}
    \label{fig:sample_fit}
\end{figure*}

The spectrum corresponding to the maximum-likelihood solution $A$ is therefore the inner product $E \cdot A$, and an example comparison between an observed spectrum and its maximum-likelihood PC representation is shown in Figure \ref{fig:sample_fit}, in the two top-right panels.

\subsubsection{Effects of sky residuals}
\label{subsubsec:sky_resid}

Since the spectral range considered here extends into the infrared wavelengths, it is important to consider the possible effects of badly-subtracted sky emission (properly-subtracted emission will have no effect apart from an increase in uncertainty). As of \mplv, 28 ``science" IFU frames have viewed just sky. These data provide a baseline for the types of sky residuals which might be present in typical science exposures.

\begin{figure*}
    \centering
    \includegraphics[width=\textwidth]{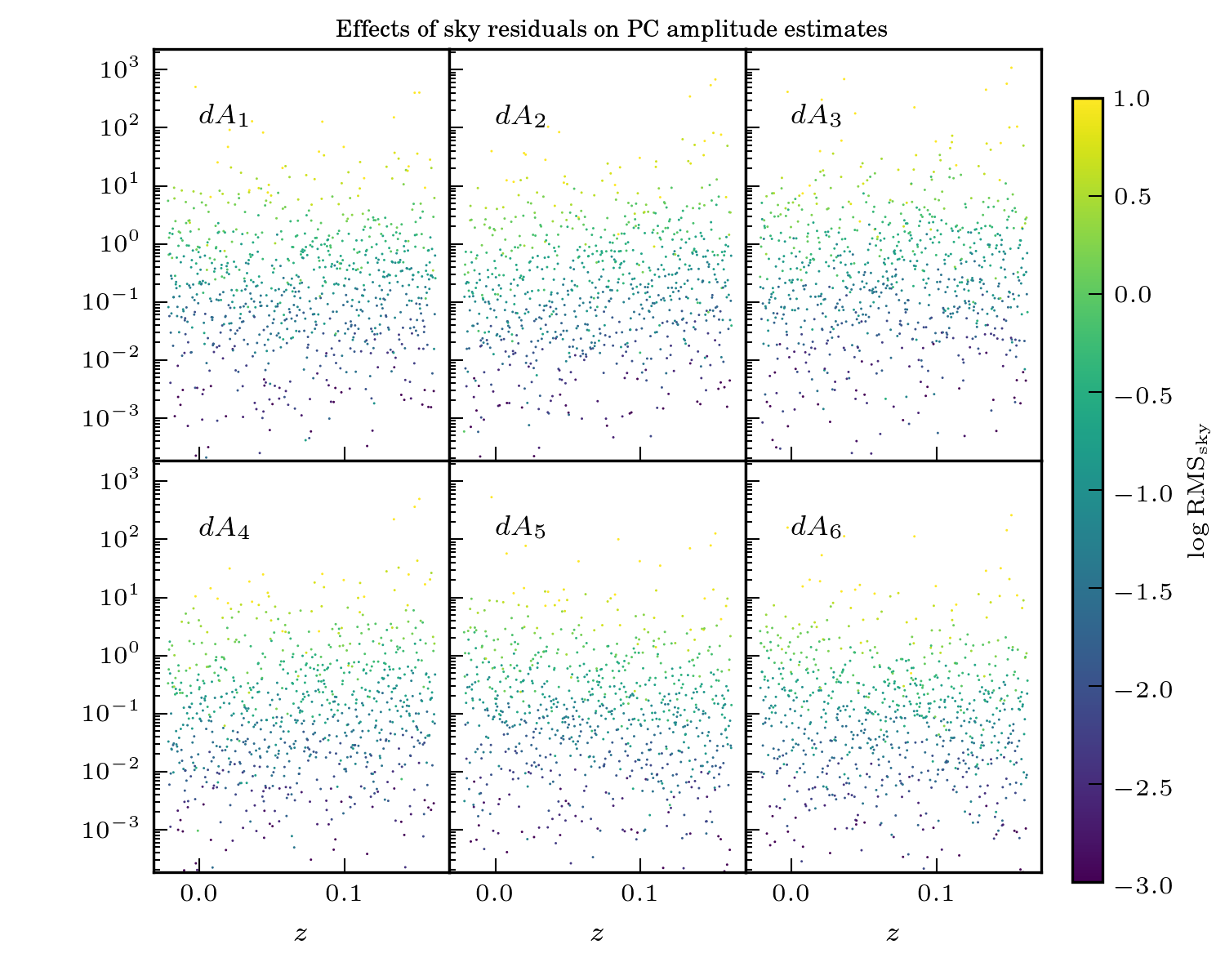}
    \caption{In each subplot, the change in principle component amplitude $dA_i$ (ordinate axis) induced by sky residuals at a level ${\rm RMS_{sky}}$ (color) relative to a normalized, observed spectrum, at some redshift (abscissa). At residual RMS below 10\%, the effects on PC amplitudes are generally small (also 10\%) or less. The PC amplitude perturbations very slightly increase with redshift.}
    \label{fig:sky_resid}
\end{figure*}

First, we test how incomplete sky-subtraction affects the estimates of principal component amplitudes themselves. In Figure \ref{fig:sky_resid}, we show the dependence of each of the first six principal components on sky residual RMS (relative to a given spectrum's normalization---so, a smaller sky residual in absolute terms will have a more severe effect in a low-surface-brightness spaxel). The weight vector associated with each spectrum is neglected, to emulate the worst-case of entirely un-subtracted sky. Redshift is varied along the abscissas: one observed-frame sky spectrum can probe a variety of rest-frame wavelengths, depending on the source redshift. Generally, at low residual RMS and low redshift, the effects on principal component amplitudes are small (less than .1). However, the impact of sky residuals rises with source redshift, since the observed-frame spectrum samples a redder wavelength range where there are more bands of sky emission. Estimates of mass-to-light ratio rely mainly on the first PC, whose amplitude is generally around 10, which makes deviations of $\sim .1$ relatively unimportant, when compared to the overall uncertainty budget.

Second, we create additional synthetic data by randomly sampling the sky-only IFU frames (as in Appendix \ref{apdx:fakedata}), and adding them to the ``mock" observations. This does not result in any noticeable change to the spectral fits, and the stellar mass-to-light ratios are consistent at .02 level, RMS.

\subsection{Quantity estimates}
\label{subsec:param_estimates}

In order to estimate a latent (i.e., unobserved) parameter or quantity of interest $P_i$ corresponding to some observed data $S$ in the lower-dimensional space defined by $E$, we find the likelihood $W_a$ of each model (where $a$ denotes an individual model) given $S$. We begin by finding the weighted-magnitude of the difference between a given model's PC coefficients $A_a$ and the PC down-projection of observations $A_o$, using the PC projection of the total spectral covariance matrix obtained above. That is, we calculate the Mahalanobis distance \citep{mahalanobis_36} between model and observations, subject to a distance metric defined by the covariance matrix:
\begin{equation}
    D^2_a = (A_a - A_o) \cdot P_{PC} \cdot (A_a - A_o)^T
    \label{eqn:model_chi2}
\end{equation}
where $D^2_a$ is the squared Mahalanobis distance between a model defined by PC coefficients $A_a$ and the PC down-projection of observations $A_o$, subject to the PC covariance matrix $K_{PC}$ and its inverse $P_{PC}$. The distance is immediately convertible to a model likelihood $W_a$ \citep{GIRI197749}:
\begin{equation}
    \log{W_a} = -\frac{1}{2} (\log{|K_{PC}|} - D^2_a - q \log{2 \pi})
    \label{eqn:logl}
\end{equation}

The likelihood is used as a weight, and accounts for theoretical degeneracies associated with any spectral fitting process (e.g., age-metallicity); as well as the effects of observational noise and spectrophotometric error. In reality, most desktop computers are capable of computing Equation \ref{eqn:model_chi2} simultaneously for all spaxels in a cube.

The lower panels of Figure \ref{fig:sample_fit} show example SPS-input and derived parameter distributions for the central spaxel of MaNGA galaxy 8566-12705. The magenta histograms show the distribution of training data used to build the PCA system and construct the parameter estimates, and the black histograms show the result of weighting by the individual model likelihoods yielded by Equation \ref{eqn:logl}. The best-characterized quantities are stellar mass-to-light ratio and dust optical depth affecting old stars $\tau_V \mu$ (in general, $\tau_V$ alone is best estimated when young stars are present; otherwise, $\tau_V \mu$ can be estimated more robustly). In contrast, stellar metallicity is only weakly constrained, and (though not displayed here) parameters scaling the BHB and BSS are not at all well-constrained.

An estimate for some parameter $Y_i$ can be obtained by computing $W_a$ (by evaluating Equation \ref{eqn:logl}) for all model spectra and a given observed spectrum, and then constructing a probability distribution based on those weights. Here, we quote the 16$^{\textrm{th}}$, 50$^{\textrm{th}}$, and 84$^{\textrm{th}}$ percentile values, with one-half the 16$^{\textrm{th}}$ to 84$^{\textrm{th}}$ percentile range as the ``distribution width". Below, we extensively evaluate the effectiveness of the PCA parameter estimation method in inferring stellar mass-to-light ratio, reddening, and stellar metallicity by using held-out ``test" data generated in the same way as the training data and (Appendix \ref{apdx:fakedata} describes in detail how the mock observations are created from synthetic spectra).

\subsubsection{Validating number of models against reliability of quantity estimates}
\label{subsec:pdf_population}

\begin{figure*}
    \centering
    \includegraphics{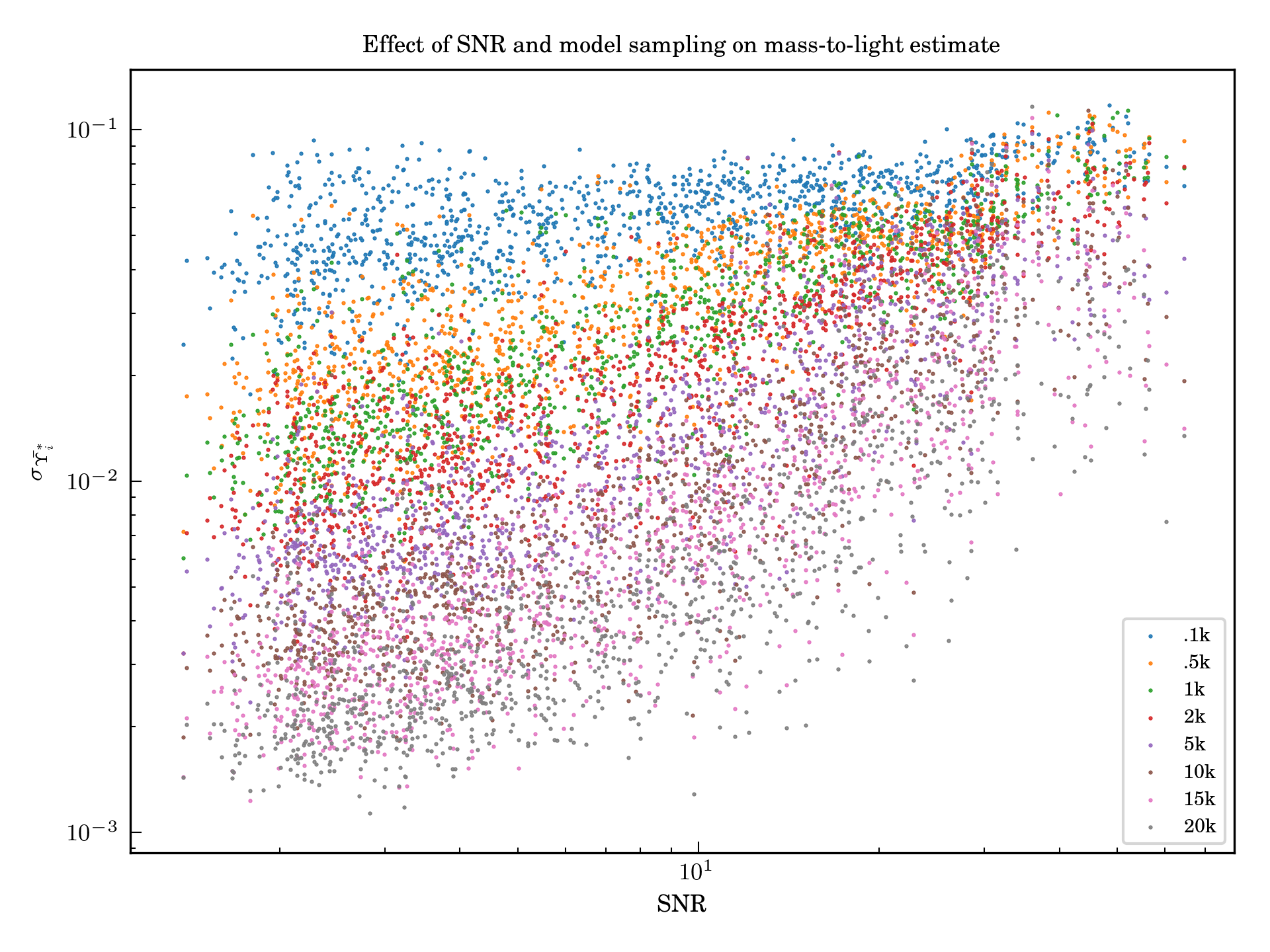}
    \caption{Variability in mass-to-light estimate (50th percentile of marginalized posterior PDF) associated with changing the number of models used to populate the distribution. Each color point represents a single spectrum with the specified number of models. At low signal-to-noise and high model count, this effect is negligible; however, more models help mitigate PDF under-population at S/N $>$ 10.}
    \label{fig:modelnumber_PDF}
\end{figure*}

Separate from the issue of PC decomposition, the number of training spectra may also impact the quality of the parameter estimates, since the PC-coefficient space must be well-sampled in the vicinity of the best-fit spectrum in order to build reliable parameter PDFs. To illustrate this, we generate estimates of $\Upsilon_i$ for all spaxels in a single galaxy after randomly selecting a fraction of the training data to use in building the PDF. We then calculate the standard deviation of that distribution. Fig. \ref{fig:modelnumber_PDF} illustrates the interplay of median spectral signal-to-noise ratio and number of models, which together affect parameter estimate accuracy. In particular, there is very little improvement that results from increasing the number of models beyond 15,000. Fig. \ref{fig:modelnumber_PDF} also shows that even using large numbers of models, at high signal-to-noise, estimates of $\log{\Upsilon^*_i}$ begin to be affected by under-population of the PDF, at the .01 (absolute) level.

\subsubsection{What limits our ability to infer quantities of interest?}
\label{sec:nmodels_vs_accuracy}

The question of number of training models can be further elucidated by the following example: suppose that a quantity of interest, $p$, has some unknown, linear dependence $B$ on principal component amplitudes $A$:

\begin{equation}
    P = A \cdot B + \epsilon
\end{equation}
where $\epsilon$ denotes a vector of white noise.

To illustrate this, we generate a vector $B$ from a $q$-dimensional unit Gaussian, and simulate the effects of sampling this ``placeholder quantity"'s PDF with a varying number of ``placeholder models", subject to covariate uncertainty in PC amplitude estimates. After fixing $B$, we randomize $N$ models ($N$ is allowed to vary from $10^1$ to $10^{6}$) distributed according to a $q$-dimensional unit Gaussian modulated by the eigenvalues of the PCA system derived from the CSP training library. A separate, ``correct" model PC amplitude vector and true quantity value $p_0$ are generated according to the same prescription. A PC amplitude covariance matrix $K_{PC}$ drawn at random from actual fits to MaNGA spectra (see Sections \ref{sec:obs2pc} and \ref{subsec:PC_unc}) is used to sample the posterior probability density function (PDF) of $Y$ (see Section \ref{subsec:param_estimates}), given an estimate of $A$ which is exactly correct. The median of this PDF, $\tilde{p}$, is taken as the fiducial estimate of $p$.

We proceed to evaluate how close $\tilde{p}$ is to the true value, $p_0$, normalizing the deviation $dp = \tilde{p} - p_0$ by the intrinsic width in the distribution of the quantity of interest in the set of placeholder models, $\sigma_p$. Under these assumptions, and setting $q = 6$, the critical number of models to achieve $\frac{dp}{\sigma_p} \lesssim .01$ is $N = 10^4$. Furthermore, as $N$ increases, this quantity of merit decreases further, though the most poorly-behaved cases ($\frac{dp}{\sigma_p} \sim 1$) arise with vanishingly-low frequency at $N \gtrsim 10^3$.

However, this does not tell the whole story, since we cannot exactly estimate $A$ for our observed spectra; an estimate of $A$ is more realistically drawn from a distribution centered at $A_0$ with covariance $K_{PC}$ (see Section \ref{subsec:cov} and \citealt{manga_spectrophot}). This erases many of the precision gains achieved at $N > 10^4$. In other words, imperfect spectrophotometry of the MaNGA data places a more stringent limit on the accuracy of quantity estimates.

\begin{figure*}
    \centering
    \includegraphics{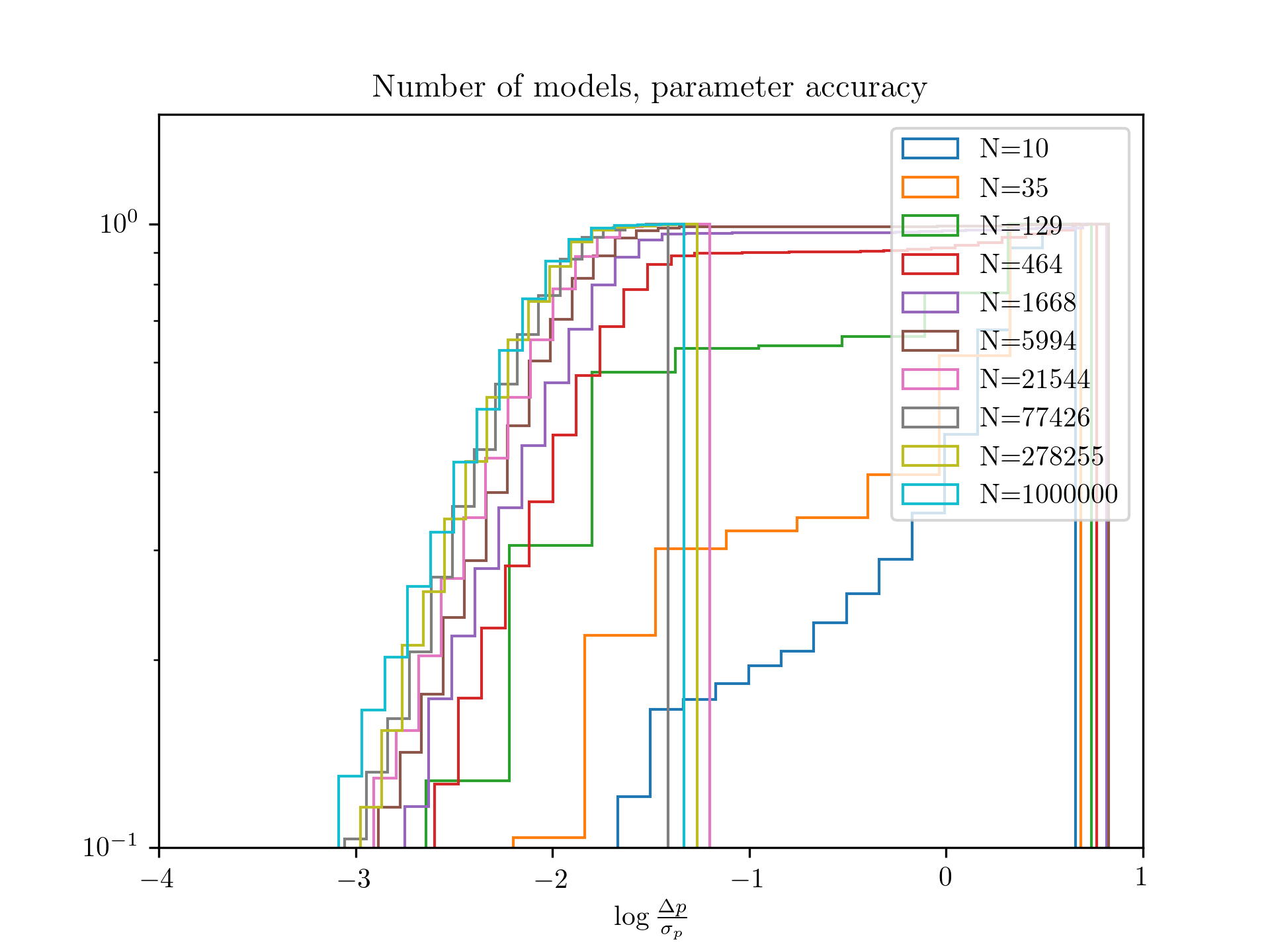}
    \caption{The cumulative distribution of $\log \frac{\Delta p}{\sigma_p}$ for values of $N$ between $10^1$ and $10^6$, under the assumption of imperfect estimation of $A$. The model library used in this work has $N = 40000$, reliably within the locus of trials with low $\frac{dp}{\sigma_p}$.}
    \label{fig:nmodels_paramerr_imperfectA}
\end{figure*}

Figure \ref{fig:nmodels_paramerr_imperfectA} shows the effect of varying $N$ from $10^1$ to $10^6$ on the cumulative distribution of $\log \frac{\Delta p}{\sigma_p}$. While at $N < 10^3$, these trials also exhibit some unreliability ($\log \frac{dp}{\sigma_p} \gtrsim 0$), there is almost no marginal benefit to adopting $N \gtrsim 10^4$.

This test indicates that while increasing the number of models brings some improvement in estimate quality for a generic quantity of interest, the benefit is diminished when the imperfect estimation of PC amplitudes $A$ (mediated by the spectrophotometric covariance of the data, via the PC covariance matrix $K_{PC}$) is accounted for. In order to realize meaningful benefits from increasing the number of CSP models, the spectrophotometry of the survey itself would have to improve by a significant margin.

\subsection{Data-quality and masking}
\label{subsec:data_quality}

We implement very basic data-quality cuts, intended to locate and mark spectra which might produce misleading measurements. Though no whole galaxies are neglected for data-quality concerns, spaxels with any of the following characteristics are presumed to have unreliable fits and parameter estimates:
\begin{itemize}
    \item More than 30\% of spectral pixels masked for any reason (combining the MaNGA DRP and emission-line flags)---see Section \ref{sec:obs2pc}
    \item Median signal-to-noise ratio below 0.1.
    \item Uncertainty in stellar line-of-sight velocity greater than 500 km/s
    \item Poorly-sampled posterior PDF: where the highest-likelihood model fit to a given spectrum is $W^*$, if less than a fraction $b$ of all models have likelihoods at least $d~W^*$, it is concluded that not enough models sample the important region of PC space to robustly estimate stellar mass-to-light ratio and other stellar population characteristics. Subsequent analysis in this work uses $b = 10^{-4}$ and $d = 10^{-1}$ (see Appendix \ref{subsec:pdf_population} for related discussion), but the associated data-product maps of $\Upsilon^*_i$ also have maps of the value of $b$ for $d = 0.01, 0.05, 0.1, 0.25, 0.5, 0.9$. In general, this mask is generally applied at high signal-to-noise ratio, and in cases where an observed spectrum differs from a typical galaxy spectrum (e.g., a broad-line AGN)
\end{itemize}

The effects of imperfect sky-subtraction on synthetic spectra are discussed in Appendix \ref{subsubsec:sky_resid}: stellar mass-to-light ratio is mostly informed by the first principal component (which typically has an amplitude of $\sim 10$), and the deviations in that principal component induced by un-subtracted sky at the 10\% (RMS)\footnote{In reality, this is a \emph{very significant} degree of sky-contamination, since the flux-density of the sky contamination is strongly bimodal.} level is $\le 0.1$, a 1\% perturbation. Taking into account all principal components, the logarithmic change induced for a stellar mass-to-light ratio estimate is $\le .02$.

\subsection{Tests on held-out, synthetic data} 
\label{subsec:mocks_tests}

We now address the reliability of the stellar mass-to-light ratio estimates obtained through PCA, by testing against synthetic data (referred to also as ```mock observations") intended to simulate real MaNGA observations. It is expected that reliability of \logml{i} estimates will increase with signal-to-noise ratio, before plateauing at in the range $10 \le \rm{SNR} \le 30$. At higher SNR, systematics related to the choice of model stellar atmospheres, SFHs, and other secondary factors will begin to adversely-influence the fit quality. Since a relatively small amount of recent star-formation can yield a blue spectrum, but most of the mass is contained in low-mass stars, blue spectra may have less accurate mass-to-light estimates. \logml{i} systematics with respect to stellar metallicity are possible for the same reason they are for pure CMLRs: in brief, stellar metallicity affects the evolution of single stars---changing, for example, main-sequence lifetimes at fixed initial mass, which significantly changes the integrated photometric properties (color and luminosity being the most salient) of the stellar population \citep[see][and related MESA/MIST works for a more thorough review]{choi_16_mist}. Finally, extreme attenuation could result in an under-estimate of \logml{i}, as in the CMLRs.

To evaluate the reliability of the \logml{i} fits with respect to color, known stellar metallicity, and known attenuation, we use test data that were generated identically to the rest of the CSP training library (see Section \ref{sec:SFHs}), but were not used to find the PCA system or for the parameter inference described in Section \ref{subsec:param_estimates}. Appendix \ref{apdx:fakedata} contains a complete description of the transformation from the test data to ``mock observations," which are intended to emulate an actual observation of such a spectrum. The mock observations are then pre-processed identically to real observations, and analyzed using the PCA framework previously described. The overall philosophy of the following tests is to bin simultaneously by median signal-to-noise ratio and either $g-r$ color, ${\rm [Z]}$, or $\tau_V$, to discover how those factors impact the reliability of inferred stellar mass-to-light ratios. We report in tabular format statistics of the stellar mass-to-light ratio estimates for both mock observations and real MaNGA galaxies. Table \ref{tab:diag_stats_summ} shows which tables and figures give diagnostic information for which quantities of merit, and binning by which spaxel properties.

\begin{table*}
    \centering
    \begin{tabular}{||c|c|c|c|c|c||} \hline
        Data Type & Bin Type 1 (subplot) & Bin Type 2 (line color) & Quantity of Merit & Table & Figure \\ \hline \hline
        Mock & SNR & $g - r$ & \qtydev{\logml{i}} & \ref{tab:mocks_snr_color_devMLi} & \ref{fig:mocks_snr_color_hist_devMLi} \\ \hline
        Mock & SNR & $g - r$ & \qtywid{\logml{i}} & \ref{tab:mocks_snr_color_widMLi} & \ref{fig:mocks_snr_color_hist_widMLi} \\ \hline
        Mock & SNR & $g - r$ & \qtydevwid{\logml{i}} & \ref{tab:mocks_snr_color_devwidMLi} & \ref{fig:mocks_snr_color_hist_devwidMLi} \\ \hline
        Mock & SNR & $\tau_V$ & \qtydev{\logml{i}} & \ref{tab:mocks_snr_tauV_devMLi} & \ref{fig:mocks_snr_tauV_hist_devMLi} \\ \hline
        Mock & SNR & ${\rm [Z]}$ & \qtydev{\logml{i}} & \ref{tab:mocks_snr_Z_devMLi} & \ref{fig:mocks_snr_Z_hist_devMLi} \\ \hline
        Obs. & SNR & $g - r$ & \qtywid{\logml{i}} & \ref{tab:obs_snr_color_widMLi} & \ref{fig:obs_snr_color_hist_widMLi} \\ \hline
    \end{tabular}
    \caption{Locations of figures \& summary statistics for mock observations \& real MaNGA data.}
    \label{tab:diag_stats_summ}
\end{table*}

In the general case, for some parameter $Y$, the known value intrinsic to one SFH is denoted $Y_0$, and the estimate as $\tilde{Y}$. We then define the ``deviation" between the two, 
\begin{equation}
    \qtydev{Y} = \tilde{Y} - Y_0
\end{equation}
and consider the dependence of deviation in stellar mass-to-light ratio on color, known stellar metallicity, and known attenuation.

We similarly define the ``uncertainty" of the distribution (\qtywid{Y}) as half the difference between the distribution's 16$^{\rm th}$ and 84$^{\rm th}$ percentiles. Finally, we define a parameter's ``normalized deviation" to be the deviation divided by the distribution half-width, \qtydevwid{Y}. Note the distinction between deviation (\qtydev{Y}), which relies on knowledge of the true parameter value $Y_0$ in comparison to the estimated value $\tilde{Y}$; and uncertainty (\qtywid{Y}), which is purely a description of the width of the posterior PDF of $Y$ given some observed spectrum.

\begin{figure}
    \centering
    \includegraphics{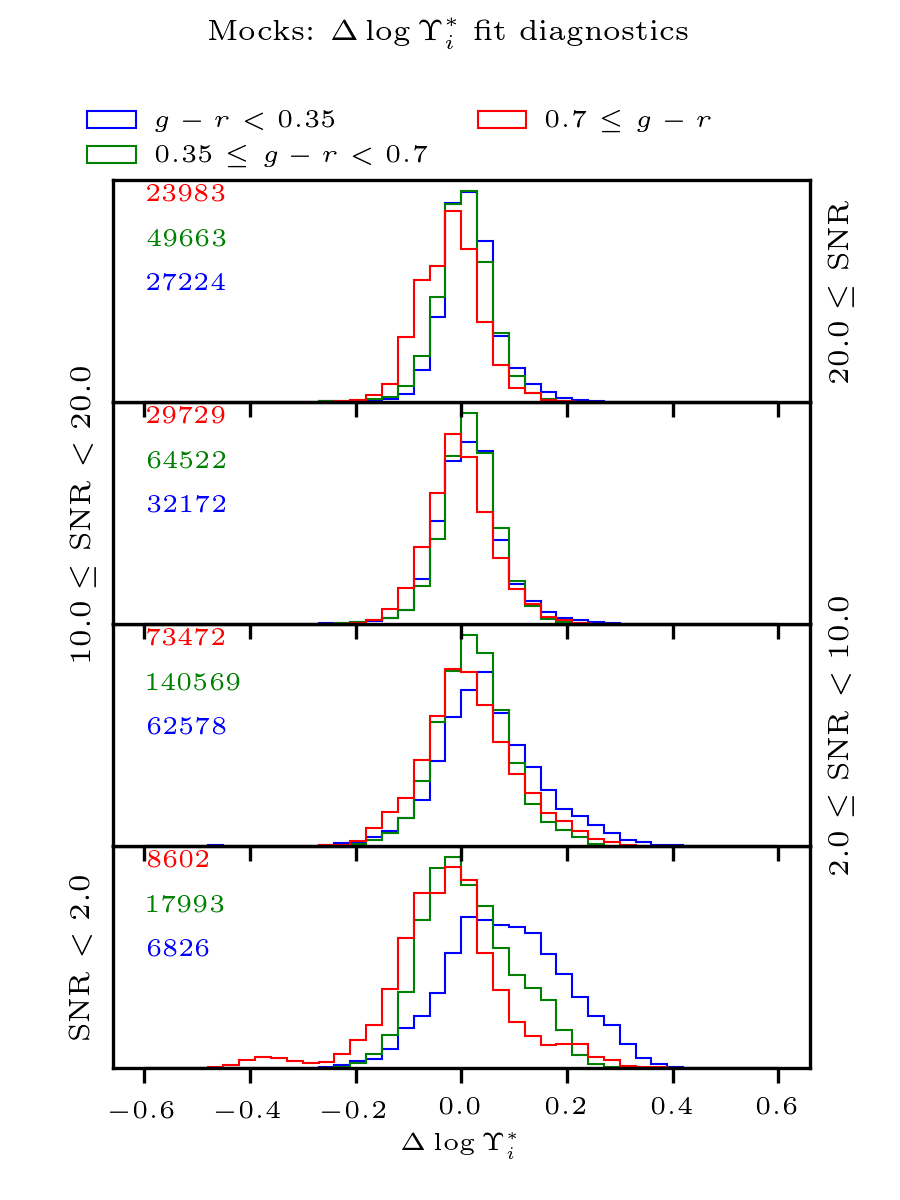}
    \caption{Distributions of deviations of PCA-inferred stellar mass-to-light ratio (\qtydev{\logml{i}}), binned into vertical subplots according to median signal-to-noise ratio, and then within each subplot according to $g-r$ color. Stellar mass-to-light ratio estimates become slightly more reliable with increasing signal-to-noise ratio, but do not improve significantly at SNR above 10, beyond $\qtydev{\logml{i}} \sim 0.1~{\rm dex}$.}
    \label{fig:mocks_snr_color_hist_devMLi}
\end{figure}

Using the mock observations for \ntestgalaxies galaxies (slightly less than 7\% of \mplv, and consisting of \ntestspaxels spaxels), we show in Figure \ref{fig:mocks_snr_color_hist_devMLi} the deviation \qtydev{\logml{i}} after binning separately by median signal-to-noise ratio (high SNR in top panel) and $g-r$ color (colored lines within one panel). At moderate to high signal-to-noise (SNR $>$ 10), the overall \qtydev{\logml{i}} profiles at fixed color do not change appreciably with increasing signal-to-noise. At lower signal-to-noise, the mode of red spectra moves to $\qtydev{\logml{i}} < 0$ (the mass-to-light ratio is underestimated), and the mode of blue spectra moves to $\qtydev{\logml{i}} > 0$ (the mass-to-light ratio is overestimated). The width of this distribution also decreases somewhat as signal-to-noise increases to moderate value.

\begin{table*}[p]
    \centering
    \begin{tabular}{||c|c|c|c|c||} \hline \hline
        Bin 1 (panel) range & Bin 2 (color) range & $P^{50}(\qtydev{\logml{i}})$ & $P^{50}(\qtydev{\logml{i}})$  - $P^{16}(\qtydev{\logml{i}})$ & $P^{84}(\qtydev{\logml{i}})$ - $P^{50}(\qtydev{\logml{i}})$ \\ \hline \hline
        [$-\infty$, 2.0] & [$-\infty$, 0.35] & $8.21 \times 10^{-2}$ & $1.04 \times 10^{-1}$ & $1.17 \times 10^{-1}$ \\ \hline
        [$-\infty$, 2.0] & [0.35, 0.7] & $5.51 \times 10^{-3}$ & $7.40 \times 10^{-2}$ & $1.05 \times 10^{-1}$ \\ \hline
        [$-\infty$, 2.0] & [0.7, $\infty$] & $-2.59 \times 10^{-2}$ & $9.04 \times 10^{-2}$ & $9.13 \times 10^{-2}$ \\ \hline
        [2.0, 10.0] & [$-\infty$, 0.35] & $4.51 \times 10^{-2}$ & $7.84 \times 10^{-2}$ & $9.92 \times 10^{-2}$ \\ \hline
        [2.0, 10.0] & [0.35, 0.7] & $2.15 \times 10^{-2}$ & $6.39 \times 10^{-2}$ & $6.71 \times 10^{-2}$ \\ \hline
        [2.0, 10.0] & [0.7, $\infty$] & $1.03 \times 10^{-2}$ & $7.44 \times 10^{-2}$ & $8.97 \times 10^{-2}$ \\ \hline
        [10.0, 20.0] & [$-\infty$, 0.35] & $1.47 \times 10^{-2}$ & $5.30 \times 10^{-2}$ & $5.51 \times 10^{-2}$ \\ \hline
        [10.0, 20.0] & [0.35, 0.7] & $1.62 \times 10^{-2}$ & $4.90 \times 10^{-2}$ & $5.03 \times 10^{-2}$ \\ \hline
        [10.0, 20.0] & [0.7, $\infty$] & $-1.89 \times 10^{-3}$ & $5.62 \times 10^{-2}$ & $5.94 \times 10^{-2}$ \\ \hline
        [20.0, $\infty$] & [$-\infty$, 0.35] & $1.19 \times 10^{-2}$ & $4.12 \times 10^{-2}$ & $4.87 \times 10^{-2}$ \\ \hline
        [20.0, $\infty$] & [0.35, 0.7] & $4.75 \times 10^{-3}$ & $4.52 \times 10^{-2}$ & $4.80 \times 10^{-2}$ \\ \hline
        [20.0, $\infty$] & [0.7, $\infty$] & $-1.73 \times 10^{-2}$ & $6.16 \times 10^{-2}$ & $5.11 \times 10^{-2}$ \\ \hline
    \end{tabular}
    \caption{Statistics of \qtydev{\logml{i}} for mock observations, separated by mean SNR and $g - r$ color: columns 3--5 respectively list the 50$^{\rm th}$ percentile value, the difference between the 84$^{\rm th}$ percentile value \& the 50$^{\rm th}$ percentile value, and the difference between the 50$^{\rm th}$ percentile value \& the 16$^{\rm th}$ percentile value.}
    \label{tab:mocks_snr_color_devMLi}
\end{table*}

\begin{figure*}
    \centering
    \begin{minipage}[t]{\columnwidth}
        \centering
        \includegraphics[width=\linewidth]{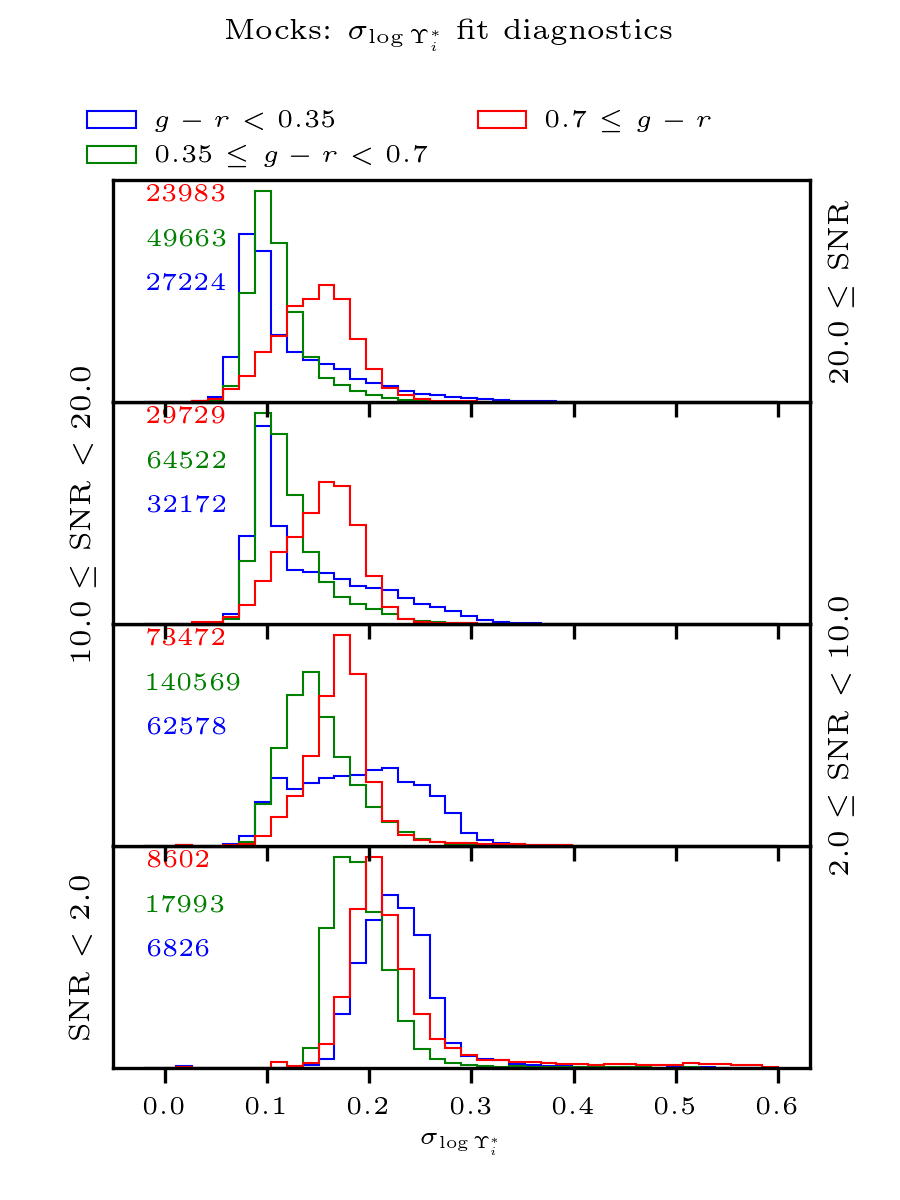}
        \caption{Distributions of uncertainty in PCA-inferred stellar mass-to-light ratio (\qtywid{\logml{i}}) for mock observations of synthetic spectra, binned into vertical subplots according to median signal-to-noise ratio, and then within each subplot according to $g-r$ color. The overall uncertainty does decrease with median signal-to-noise ratio: this effect is strongest for blue spectra and weakest for red (at low signal-to-noise, an acceptable fit to a blue spectrum allows for a significant amount of mass from old stars---this degeneracy weakens as signal-to-noise ratio rises).}
        \label{fig:mocks_snr_color_hist_widMLi}
    \end{minipage}
    \hfill
    \begin{minipage}[t]{\columnwidth}
        \centering
        \includegraphics[width=\linewidth]{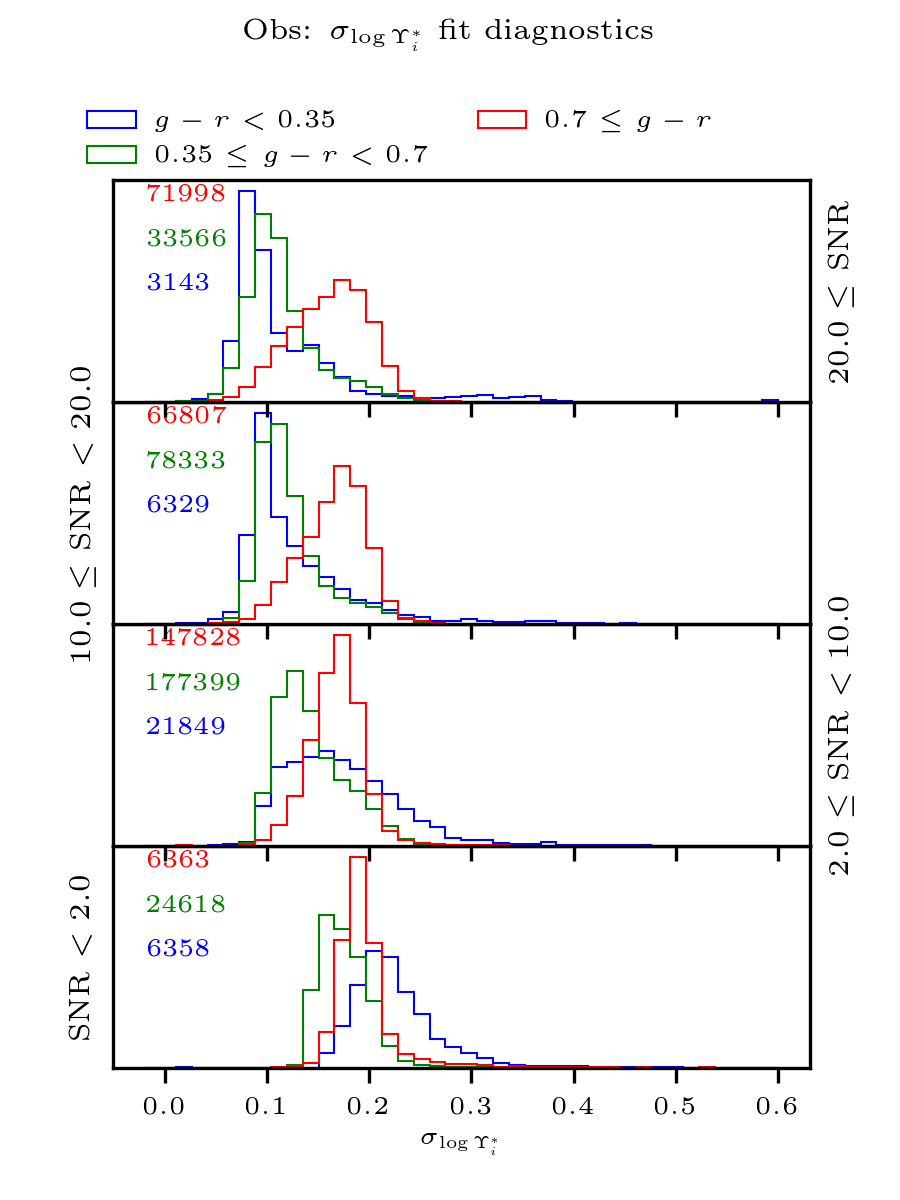}
        \caption{As Figure \ref{fig:mocks_snr_color_hist_widMLi}, except using analysis of real MaNGA galaxies, rather than mock observations of test data. The shapes and relative positions of individual color-SNR-binned distributions are qualitatively very similar to the distributions of mock observations in \ref{fig:mocks_snr_color_hist_widMLi}.}
        \label{fig:obs_snr_color_hist_widMLi}
    \end{minipage}
    
\end{figure*}

Next, we test the dependence of the quoted mass-to-light uncertainty (\qtywid{\logml{i}}) on optical ($g-r$) color, and the results are shown in Figure \ref{fig:mocks_snr_color_hist_widMLi}. Several effects manifest in this case, which we will address separately: first, at fixed signal-to-noise (within one panel), the moderate-color spectra have the lowest uncertainty, and the blue spectra have the highest, with red spectra falling somewhere in the middle. Naturally, the reddest spectra could be produced by either an intrinsically old stellar population or prevalent dust---in fact, \citet{bell_03} estimated the impact of dust for pure CMLRs as 0.1--0.2 dex, somewhat higher than the (approximately 0.05 dex) offset we observe between moderate-color and red spectra. The bluest spectra are the most uncertain because though the majority of the light originates from young, blue stars, most of the mass resides in small, dim stars. In other words, there is the potential for the mass-carrying population to have its signal washed out by the younger, brighter one. We believe the positive (0.05--0.1 dex) offset of \qtywid{\logml{i}} in blue spectra with respect to \qtywid{\logml{i}} for intermediate- and red-color spectra at signal-to-noise ratios less than 10 is a manifestation of this effect.

\begin{table*}[p]
    \centering
    \begin{tabular}{||c|c|c|c|c||} \hline \hline
        Bin 1 (panel) range & Bin 2 (color) range & $P^{50}(\qtywid{\logml{i}})$ & $P^{50}(\qtywid{\logml{i}})$  - $P^{16}(\qtywid{\logml{i}})$ & $P^{84}(\qtywid{\logml{i}})$ - $P^{50}(\qtywid{\logml{i}})$ \\ \hline \hline
        [$-\infty$, 2.0] & [$-\infty$, 0.35] & $2.25 \times 10^{-1}$ & $3.13 \times 10^{-2}$ & $3.25 \times 10^{-2}$ \\ \hline
        [$-\infty$, 2.0] & [0.35, 0.7] & $1.88 \times 10^{-1}$ & $2.30 \times 10^{-2}$ & $3.02 \times 10^{-2}$ \\ \hline
        [$-\infty$, 2.0] & [0.7, $\infty$] & $2.11 \times 10^{-1}$ & $2.47 \times 10^{-2}$ & $4.62 \times 10^{-2}$ \\ \hline
        [2.0, 10.0] & [$-\infty$, 0.35] & $1.90 \times 10^{-1}$ & $6.80 \times 10^{-2}$ & $6.28 \times 10^{-2}$ \\ \hline
        [2.0, 10.0] & [0.35, 0.7] & $1.46 \times 10^{-1}$ & $2.77 \times 10^{-2}$ & $3.99 \times 10^{-2}$ \\ \hline
        [2.0, 10.0] & [0.7, $\infty$] & $1.73 \times 10^{-1}$ & $2.98 \times 10^{-2}$ & $2.28 \times 10^{-2}$ \\ \hline
        [10.0, 20.0] & [$-\infty$, 0.35] & $1.21 \times 10^{-1}$ & $2.97 \times 10^{-2}$ & $9.19 \times 10^{-2}$ \\ \hline
        [10.0, 20.0] & [0.35, 0.7] & $1.13 \times 10^{-1}$ & $1.98 \times 10^{-2}$ & $3.70 \times 10^{-2}$ \\ \hline
        [10.0, 20.0] & [0.7, $\infty$] & $1.57 \times 10^{-1}$ & $4.10 \times 10^{-2}$ & $3.10 \times 10^{-2}$ \\ \hline
        [20.0, $\infty$] & [$-\infty$, 0.35] & $1.01 \times 10^{-1}$ & $2.07 \times 10^{-2}$ & $7.48 \times 10^{-2}$ \\ \hline
        [20.0, $\infty$] & [0.35, 0.7] & $1.05 \times 10^{-1}$ & $1.77 \times 10^{-2}$ & $3.01 \times 10^{-2}$ \\ \hline
        [20.0, $\infty$] & [0.7, $\infty$] & $1.49 \times 10^{-1}$ & $4.02 \times 10^{-2}$ & $3.41 \times 10^{-2}$ \\ \hline
    \end{tabular}
    \caption{Statistics of \qtywid{\logml{i}} for mock observations, separated by mean SNR and $g - r$ color: columns 3--5 respectively list the 50$^{\rm th}$ percentile value, the difference between the 84$^{\rm th}$ percentile value \& the 50$^{\rm th}$ percentile value, and the difference between the 50$^{\rm th}$ percentile value \& the 16$^{\rm th}$ percentile value.}
    \label{tab:mocks_snr_color_widMLi}
\end{table*}

\begin{table*}[p]
    \centering
    \begin{tabular}{||c|c|c|c|c||} \hline \hline
        Bin 1 (panel) range & Bin 2 (color) range & $P^{50}(\qtywid{\logml{i}})$ & $P^{50}(\qtywid{\logml{i}})$  - $P^{16}(\qtywid{\logml{i}})$ & $P^{84}(\qtywid{\logml{i}})$ - $P^{50}(\qtywid{\logml{i}})$ \\ \hline \hline
        [$-\infty$, 2.0] & [$-\infty$, 0.35] & $2.18 \times 10^{-1}$ & $2.81 \times 10^{-2}$ & $4.23 \times 10^{-2}$ \\ \hline
        [$-\infty$, 2.0] & [0.35, 0.7] & $1.72 \times 10^{-1}$ & $2.04 \times 10^{-2}$ & $2.70 \times 10^{-2}$ \\ \hline
        [$-\infty$, 2.0] & [0.7, $\infty$] & $1.91 \times 10^{-1}$ & $1.67 \times 10^{-2}$ & $1.99 \times 10^{-2}$ \\ \hline
        [2.0, 10.0] & [$-\infty$, 0.35] & $1.67 \times 10^{-1}$ & $4.74 \times 10^{-2}$ & $6.03 \times 10^{-2}$ \\ \hline
        [2.0, 10.0] & [0.35, 0.7] & $1.37 \times 10^{-1}$ & $2.43 \times 10^{-2}$ & $4.44 \times 10^{-2}$ \\ \hline
        [2.0, 10.0] & [0.7, $\infty$] & $1.69 \times 10^{-1}$ & $2.55 \times 10^{-2}$ & $2.16 \times 10^{-2}$ \\ \hline
        [10.0, 20.0] & [$-\infty$, 0.35] & $1.11 \times 10^{-1}$ & $2.15 \times 10^{-2}$ & $6.47 \times 10^{-2}$ \\ \hline
        [10.0, 20.0] & [0.35, 0.7] & $1.14 \times 10^{-1}$ & $1.86 \times 10^{-2}$ & $3.68 \times 10^{-2}$ \\ \hline
        [10.0, 20.0] & [0.7, $\infty$] & $1.69 \times 10^{-1}$ & $3.74 \times 10^{-2}$ & $2.62 \times 10^{-2}$ \\ \hline
        [20.0, $\infty$] & [$-\infty$, 0.35] & $9.54 \times 10^{-2}$ & $1.69 \times 10^{-2}$ & $6.45 \times 10^{-2}$ \\ \hline
        [20.0, $\infty$] & [0.35, 0.7] & $1.07 \times 10^{-1}$ & $2.18 \times 10^{-2}$ & $3.90 \times 10^{-2}$ \\ \hline
        [20.0, $\infty$] & [0.7, $\infty$] & $1.65 \times 10^{-1}$ & $4.37 \times 10^{-2}$ & $3.46 \times 10^{-2}$ \\ \hline
    \end{tabular}
    \caption{Statistics of \qtywid{\logml{i}} for real MaNGA observations, separated by mean SNR and $g - r$ color: columns 3--5 respectively list the 50$^{\rm th}$ percentile value, the difference between the 84$^{\rm th}$ percentile value \& the 50$^{\rm th}$ percentile value, and the difference between the 50$^{\rm th}$ percentile value \& the 16$^{\rm th}$ percentile value.}
    \label{tab:obs_snr_color_widMLi}
\end{table*}

In addition, at fixed color, an increase in signal-to-noise is not necessarily associated with a decrease in \qtywid{\logml{i}}. Rather, improvements seem to disappear (and possibly reverse at signal-to-noise greater than 20). In reality, there are several lower limits on \qtywid{\logml{i}}: the spectrophotometric uncertainty, which we model as independent of surface brightness, produces covariate noise at between the 1--3\% level, and has a spectral signature similar to a changing mass-to-light ratio. The rising uncertainty at $S/N > 20$ could also be understood in terms of how densely-populated the model grid is with respect to the uncertainty on the data: by increasing the signal-to-noise of the data, the $n$-dimensional volume subtended by a noise vector $N$ will decrease to the point where the parameter PDF is not well-sampled (this will be particularly problematic where a SFH's PC representation lies near an ``edge"). Interestingly, regardless of color or signal-to-noise, the RMS of the deviation (the width of the \qtydev{\logml{i}} distribution) is always of the same order as (and often a factor of up to two less than) the mean of the uncertainty (\qtywid{\logml{i}}). This means that on the whole, uncertainties in stellar mass-to-light ratio reflect the real statistical uncertainty. We examine this below by showing the distribution of \qtydevwid{\logml{i}}.

For the sake of comparison, we display \qtywid{\logml{i}} for the same sample of \emph{real} galaxies (Figure \ref{fig:obs_snr_color_hist_widMLi}). The distributions of these data are qualitatively similar to the case of the mock observations of synthetic spectra: regardless of color, stellar mass-to-light ratio uncertainty decreases as signal-to-noise ratio increases, but this effect is strongest for blue spectra and weakest for red. The most noticeable difference between fits to mock observations and real observations is at moderate signal-to-noise ratio: the mocks' distribution of uncertainty has a higher mode (0.25 dex, versus the observations' 0.15 dex). It is probable that this difference in behavior has to do with slight discrepancies in the distribution of training data with respect to real galaxies. The relative strengths of individual binned distributions are likewise determined by the details of the SFH library: for example, a larger proportion of observations at low signal-to-noise ratio are blue, since the lower surface-brightness outskirts of galaxies will have commensurately-lower signal-to-noise ratios. As a whole, though, the distributions of \qtywid{\logml{i}} for mocks and real observations are broadly similar when the same signal-to-noise and color ranges are compared. We believe this indicates both that the mocks are a faithful reconstruction of MaNGA observations; and that the actual distribution of SFHs in MaNGA spaxels is sufficiently similar to the training data to infer unobserved properties such as stellar mass-to-light ratio.

\begin{figure}
    \centering
    \includegraphics{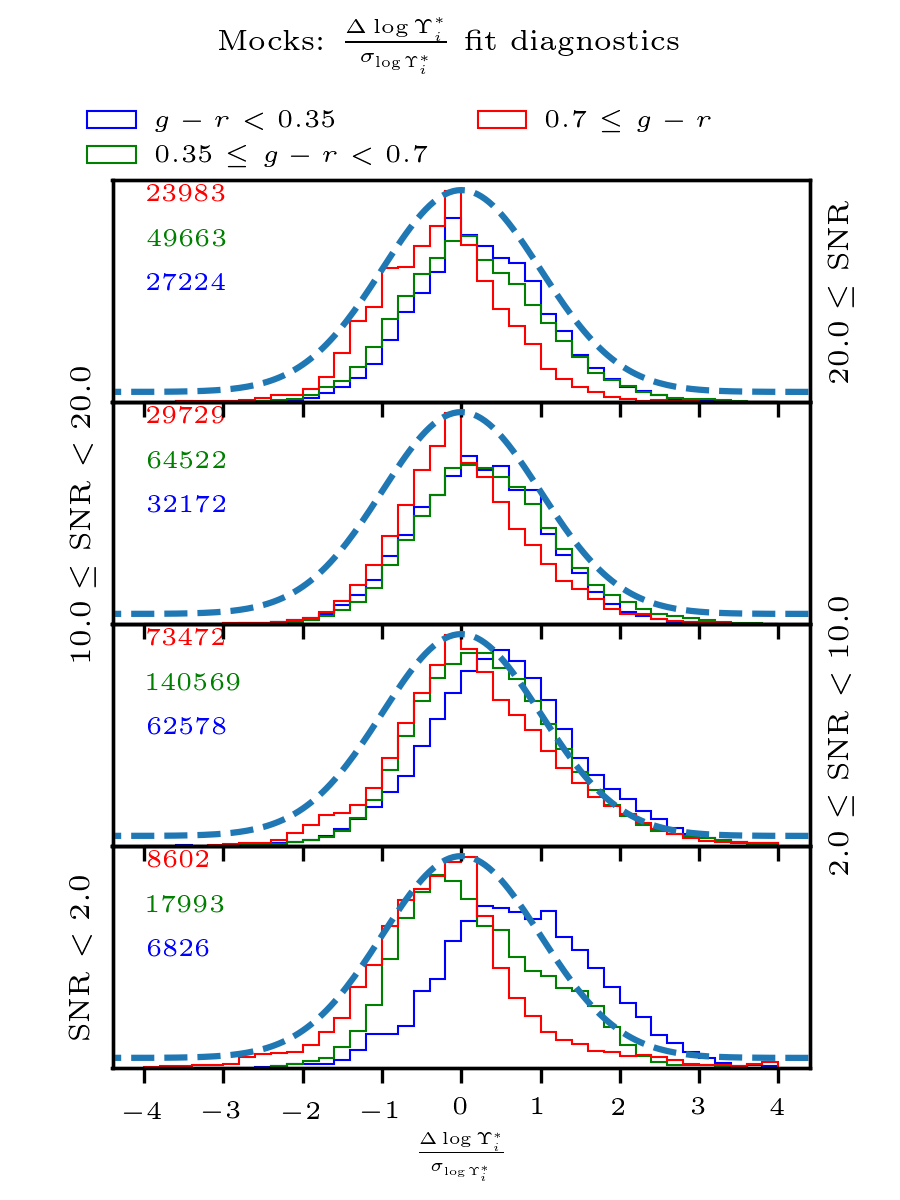}
    \caption{As Figure \ref{fig:mocks_snr_color_hist_widMLi}, except with distributions of \qtydevwid{\logml{i}}. Up to the small offset effects covariate with integrated color (see Figure \ref{fig:mocks_snr_color_hist_devMLi}), we find that the reported uncertainties in \logml{i} faithfully reflect the real deviation between the inferred stellar mass-to-light ratio and its true value. Overplotted in blue-gray is a normal distribution with dispersion of unity, which should correspond to the nominal case of uncertainties that match well with the actual accuracy of an estimate.}
    \label{fig:mocks_snr_color_hist_devwidMLi}
\end{figure}

We next consider the normalized deviations (\qtydevwid{\logml{i}}) with respect to the mocks, which are important for evaluating whether the provided mass-to-light ratio uncertainties are accurate. Figure \ref{fig:mocks_snr_color_hist_devwidMLi} illustrates the relatively steady accuracy of the \logml{i} estimates with respect to color and signal-to-noise (besides the effects on \qtydev{\logml{i}} already discussed). In all cases but low-signal-to-noise, blue spectra, the \qtydevwid{\logml{i}} distributions are relatively symmetrical, and do not exhibit significant power at high absolute values (which would indicate unreliable uncertainties in some region of parameter-space). Most distributions compare favorably to the ideal case of the uncertainty roughly matching the deviation (blue, dotted curve). In summary, from the above tests on both synthetic and real observations, we conclude that the PCA parameter estimation implemented here for \logml{i} achieves acceptable levels of accuracy and precision for use in estimating total stellar-mass.

\begin{table*}[p]
    \centering
    \begin{tabular}{||c|c|c|c|c||} \hline \hline
        Bin 1 (panel) range & Bin 2 (color) range & $P^{50}(\qtydevwid{\logml{i}})$ & $P^{50}(\qtydevwid{\logml{i}})$  - $P^{16}(\qtydevwid{\logml{i}})$ & $P^{84}(\qtydevwid{\logml{i}})$ - $P^{50}(\qtydevwid{\logml{i}})$ \\ \hline \hline
        [$-\infty$, 2.0] & [$-\infty$, 0.35] & $7.11 \times 10^{-1}$ & $9.11 \times 10^{-1}$ & $1.02 \times 10^{0}$ \\ \hline
        [$-\infty$, 2.0] & [0.35, 0.7] & $5.79 \times 10^{-2}$ & $7.61 \times 10^{-1}$ & $1.12 \times 10^{0}$ \\ \hline
        [$-\infty$, 2.0] & [0.7, $\infty$] & $-2.36 \times 10^{-1}$ & $8.42 \times 10^{-1}$ & $8.04 \times 10^{-1}$ \\ \hline
        [2.0, 10.0] & [$-\infty$, 0.35] & $5.16 \times 10^{-1}$ & $8.98 \times 10^{-1}$ & $9.70 \times 10^{-1}$ \\ \hline
        [2.0, 10.0] & [0.35, 0.7] & $3.00 \times 10^{-1}$ & $8.33 \times 10^{-1}$ & $9.56 \times 10^{-1}$ \\ \hline
        [2.0, 10.0] & [0.7, $\infty$] & $1.19 \times 10^{-1}$ & $8.41 \times 10^{-1}$ & $1.09 \times 10^{0}$ \\ \hline
        [10.0, 20.0] & [$-\infty$, 0.35] & $2.25 \times 10^{-1}$ & $8.29 \times 10^{-1}$ & $8.13 \times 10^{-1}$ \\ \hline
        [10.0, 20.0] & [0.35, 0.7] & $2.88 \times 10^{-1}$ & $8.12 \times 10^{-1}$ & $9.01 \times 10^{-1}$ \\ \hline
        [10.0, 20.0] & [0.7, $\infty$] & $-2.60 \times 10^{-2}$ & $7.03 \times 10^{-1}$ & $8.62 \times 10^{-1}$ \\ \hline
        [20.0, $\infty$] & [$-\infty$, 0.35] & $2.15 \times 10^{-1}$ & $7.67 \times 10^{-1}$ & $8.43 \times 10^{-1}$ \\ \hline
        [20.0, $\infty$] & [0.35, 0.7] & $9.37 \times 10^{-2}$ & $8.03 \times 10^{-1}$ & $9.31 \times 10^{-1}$ \\ \hline
        [20.0, $\infty$] & [0.7, $\infty$] & $-2.38 \times 10^{-1}$ & $7.90 \times 10^{-1}$ & $7.39 \times 10^{-1}$ \\ \hline
    \end{tabular}
    \caption{Statistics of \qtydevwid{\logml{i}} for mock observations, separated by mean SNR and $g - r$ color: columns 3--5 respectively list the 50$^{\rm th}$ percentile value, the difference between the 84$^{\rm th}$ percentile value \& the 50$^{\rm th}$ percentile value, and the difference between the 50$^{\rm th}$ percentile value \& the 16$^{\rm th}$ percentile value.}
    \label{tab:mocks_snr_color_devwidMLi}
\end{table*}

\begin{figure}
    \centering
    \includegraphics{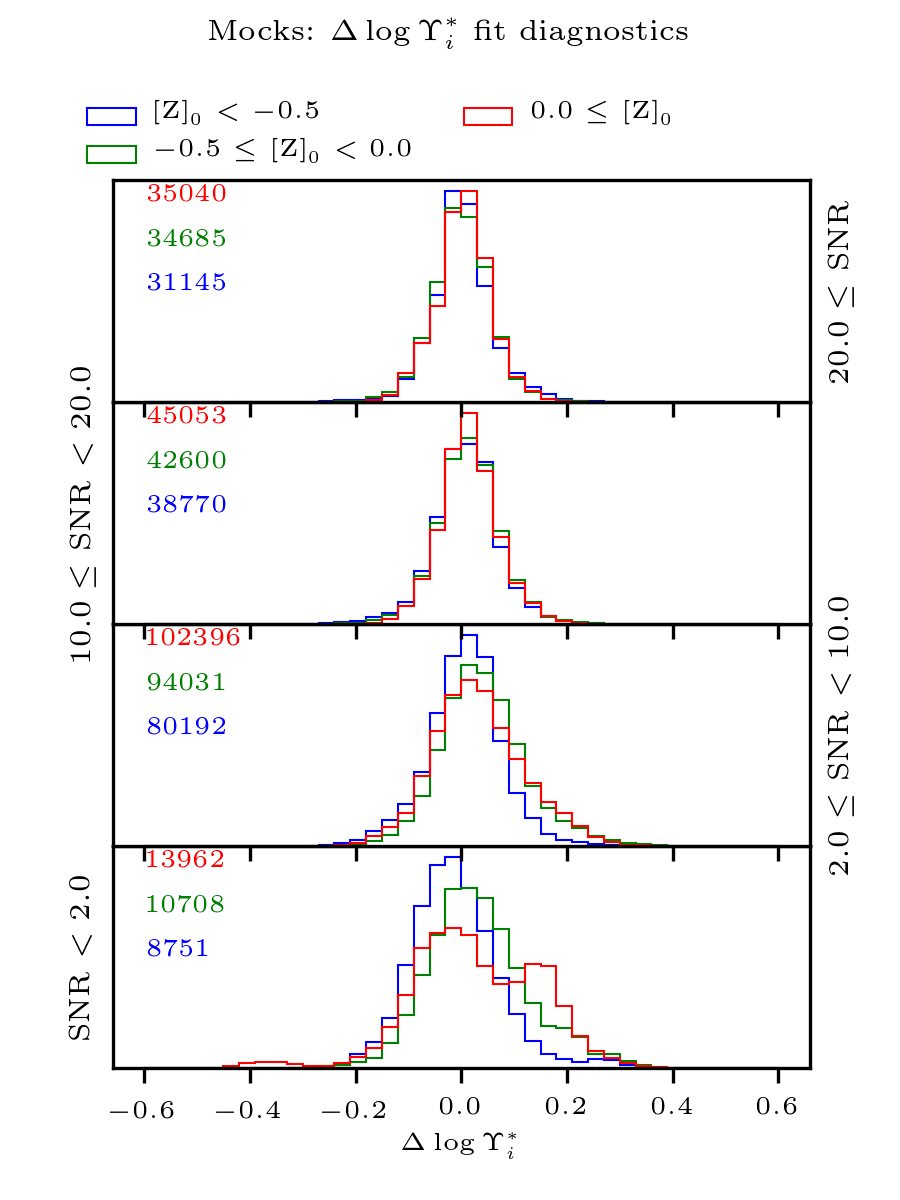}
    \caption{As Figure \ref{fig:mocks_snr_color_hist_devMLi}, except binning with respect to known ${\rm [Z]}$ rather than $g-r$ color. Other than at low signal-to-noise ratio and high metallicity (where deviations may reach 0.3 dex), there are minimal systematics in inferred $\log \Upsilon^*_i$ with respect to ${\rm [Z]}$.}
    \label{fig:mocks_snr_Z_hist_devMLi}
\end{figure}

\begin{figure}
    \centering
    \includegraphics{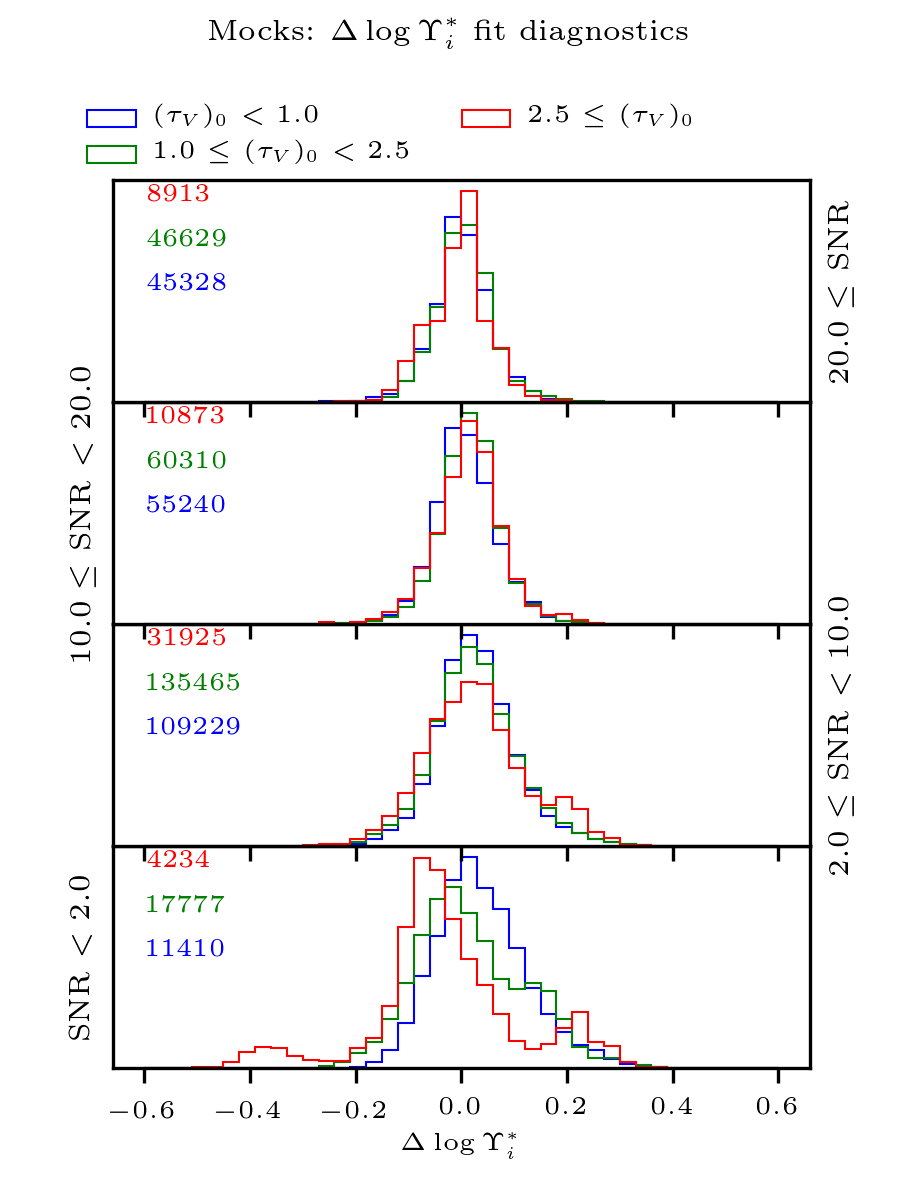}
    \caption{As Figure \ref{fig:mocks_snr_color_hist_devMLi}, except binning with respect to known $\tau_V$ rather than $g-r$ color. As before, at high signal-to-noise, performance of the \logml{i} estimate does not change strongly with attenuation; however, at lower signal-to-noise, high-attenuation spectra may have their stellar mass-to-light ratio overestimated by up to about 0.3 dex. Such cases are expected to be rare in the MaNGA data.}
    \label{fig:mocks_snr_tauV_hist_devMLi}
\end{figure}

In Section \ref{subsec:cmlrs}, we showed that our family of CSPs exhibit scatter about their best-fit CMLR which correlates in its magnitude with extreme stellar metallicity and attenuation. Here, we test the precision and accuracy of our \logml{i} estimates. when using stellar population absorption indices, heuristics like the ``3/2 rule" describe the covariance between mean stellar age and metallicity \citep{worthey_94}\footnote{The 3/2 rule is an observation stating that an increase (decrease) of a stellar population's age by a factor of three is almost indistinguishable from an increase (decrease) in metallicity by a factor of two.}. Similarly to what was observed with CMLRs, significant dust attenuation applied to an otherwise-young stellar population could conceivably effect an overestimate of its mass-to-light ratio. Figures \ref{fig:mocks_snr_Z_hist_devMLi} and \ref{fig:mocks_snr_tauV_hist_devMLi} respectively bin \qtydev{\logml{i}} by median signal-to-noise ratio and either $\tau_V$ or ${\rm [Z]}$ for mock observations.

\begin{table*}[p]
    \centering
    \begin{tabular}{||c|c|c|c|c||} \hline \hline
        Bin 1 (panel) range & Bin 2 (color) range & $P^{50}(\qtydev{\logml{i}})$ & $P^{50}(\qtydev{\logml{i}})$  - $P^{16}(\qtydev{\logml{i}})$ & $P^{84}(\qtydev{\logml{i}})$ - $P^{50}(\qtydev{\logml{i}})$ \\ \hline \hline
        [$-\infty$, 2.0] & [$-\infty$, -0.5] & $-1.69 \times 10^{-2}$ & $7.03 \times 10^{-2}$ & $8.05 \times 10^{-2}$ \\ \hline
        [$-\infty$, 2.0] & [-0.5, 0.0] & $2.47 \times 10^{-2}$ & $8.08 \times 10^{-2}$ & $9.85 \times 10^{-2}$ \\ \hline
        [$-\infty$, 2.0] & [0.0, $\infty$] & $1.74 \times 10^{-2}$ & $9.91 \times 10^{-2}$ & $1.38 \times 10^{-1}$ \\ \hline
        [2.0, 10.0] & [$-\infty$, -0.5] & $8.10 \times 10^{-3}$ & $6.63 \times 10^{-2}$ & $5.97 \times 10^{-2}$ \\ \hline
        [2.0, 10.0] & [-0.5, 0.0] & $3.56 \times 10^{-2}$ & $6.92 \times 10^{-2}$ & $7.96 \times 10^{-2}$ \\ \hline
        [2.0, 10.0] & [0.0, $\infty$] & $2.70 \times 10^{-2}$ & $7.58 \times 10^{-2}$ & $9.21 \times 10^{-2}$ \\ \hline
        [10.0, 20.0] & [$-\infty$, -0.5] & $8.30 \times 10^{-3}$ & $5.48 \times 10^{-2}$ & $5.30 \times 10^{-2}$ \\ \hline
        [10.0, 20.0] & [-0.5, 0.0] & $1.36 \times 10^{-2}$ & $5.46 \times 10^{-2}$ & $5.57 \times 10^{-2}$ \\ \hline
        [10.0, 20.0] & [0.0, $\infty$] & $1.35 \times 10^{-2}$ & $4.99 \times 10^{-2}$ & $5.28 \times 10^{-2}$ \\ \hline
        [20.0, $\infty$] & [$-\infty$, -0.5] & $6.53 \times 10^{-6}$ & $4.73 \times 10^{-2}$ & $5.23 \times 10^{-2}$ \\ \hline
        [20.0, $\infty$] & [-0.5, 0.0] & $2.05 \times 10^{-6}$ & $4.99 \times 10^{-2}$ & $5.04 \times 10^{-2}$ \\ \hline
        [20.0, $\infty$] & [0.0, $\infty$] & $5.05 \times 10^{-3}$ & $5.14 \times 10^{-2}$ & $4.64 \times 10^{-2}$ \\ \hline
    \end{tabular}
    \caption{Statistics of \qtydev{\logml{i}} for mock observations, separated by mean SNR and known stellar metallicity: columns 3--5 respectively list the 50$^{\rm th}$ percentile value, the difference between the 84$^{\rm th}$ percentile value \& the 50$^{\rm th}$ percentile value, and the difference between the 50$^{\rm th}$ percentile value \& the 16$^{\rm th}$ percentile value.}
    \label{tab:mocks_snr_Z_devMLi}
\end{table*}

In Figure \ref{fig:mocks_snr_Z_hist_devMLi}, we see that regardless of metallicity, reliability of \logml{i} estimates increase with signal-to-noise, and converge to $\qtydev{\logml{i}} \sim 0.1$ at $S/N \sim 20$. At high stellar metallicity and low signal-to-noise ratio, the distribution of \qtydev{\logml{i}} becomes significantly skewed (with the long tail at positive \qtydev{\logml{i}}, indicating an overestimate generally less than 0.15 dex). Though this deviation is not reflected in the associated uncertainties (i.e., \qtydevwid{\logml{i}} is high), it ceases at higher SNR. At lower metallicity, the typical deviation between known and estimated mass-to-light ratio remains unimodal across the entire SNR regime.

\begin{table*}[p]
    \centering
    \begin{tabular}{||c|c|c|c|c||} \hline \hline
        Bin 1 (panel) range & Bin 2 (color) range & $P^{50}(\qtydevwid{\logml{i}})$ & $P^{50}(\qtydevwid{\logml{i}})$  - $P^{16}(\qtydevwid{\logml{i}})$ & $P^{84}(\qtydevwid{\logml{i}})$ - $P^{50}(\qtydevwid{\logml{i}})$ \\ \hline \hline
        [$-\infty$, 2.0] & [$-\infty$, 1.0] & $2.95 \times 10^{-2}$ & $7.48 \times 10^{-2}$ & $9.21 \times 10^{-2}$ \\ \hline
        [$-\infty$, 2.0] & [1.0, 2.5] & $3.61 \times 10^{-3}$ & $8.48 \times 10^{-2}$ & $1.32 \times 10^{-1}$ \\ \hline
        [$-\infty$, 2.0] & [2.5, $\infty$] & $-4.19 \times 10^{-2}$ & $7.65 \times 10^{-2}$ & $1.33 \times 10^{-1}$ \\ \hline
        [2.0, 10.0] & [$-\infty$, 1.0] & $2.46 \times 10^{-2}$ & $6.45 \times 10^{-2}$ & $7.43 \times 10^{-2}$ \\ \hline
        [2.0, 10.0] & [1.0, 2.5] & $2.27 \times 10^{-2}$ & $7.12 \times 10^{-2}$ & $8.05 \times 10^{-2}$ \\ \hline
        [2.0, 10.0] & [2.5, $\infty$] & $2.23 \times 10^{-2}$ & $8.56 \times 10^{-2}$ & $1.03 \times 10^{-1}$ \\ \hline
        [10.0, 20.0] & [$-\infty$, 1.0] & $4.94 \times 10^{-3}$ & $5.15 \times 10^{-2}$ & $5.82 \times 10^{-2}$ \\ \hline
        [10.0, 20.0] & [1.0, 2.5] & $1.66 \times 10^{-2}$ & $5.12 \times 10^{-2}$ & $5.02 \times 10^{-2}$ \\ \hline
        [10.0, 20.0] & [2.5, $\infty$] & $1.86 \times 10^{-2}$ & $6.27 \times 10^{-2}$ & $5.30 \times 10^{-2}$ \\ \hline
        [20.0, $\infty$] & [$-\infty$, 1.0] & $-3.60 \times 10^{-7}$ & $4.81 \times 10^{-2}$ & $5.15 \times 10^{-2}$ \\ \hline
        [20.0, $\infty$] & [1.0, 2.5] & $4.13 \times 10^{-3}$ & $4.95 \times 10^{-2}$ & $4.76 \times 10^{-2}$ \\ \hline
        [20.0, $\infty$] & [2.5, $\infty$] & $9.44 \times 10^{-7}$ & $6.50 \times 10^{-2}$ & $4.50 \times 10^{-2}$ \\ \hline
    \end{tabular}
    \caption{Statistics of \qtydev{\logml{i}} for mock observations, separated by mean SNR and $\tau_V$: columns 3--5 respectively list the 50$^{\rm th}$ percentile value, the difference between the 84$^{\rm th}$ percentile value \& the 50$^{\rm th}$ percentile value, and the difference between the 50$^{\rm th}$ percentile value \& the 16$^{\rm th}$ percentile value.}
    \label{tab:mocks_snr_tauV_devMLi}
\end{table*}

In Figure \ref{fig:mocks_snr_tauV_hist_devMLi}, we see that high signal-to-noise spectra yield more or less equally-reliable estimates of \logml{i}, regardless of attenuation. As signal-to-noise decreases, spectra with intermediate attenuation tend to skew towards underestimating \logml{i}, while the distribution for high-attenuation spectra becomes much wider, in a way which is not reflected in the parameter uncertainty (\qtywid{\logml{i}}). Such spectra are expected to have little impact, though (by virtue of their high attenuation, such spectra have lower surface-brightness and contribute little to an estimate of a galaxy's total stellar mass).

In summary, while the underlying attenuation and stellar metallicity of a mock SFH does certainly impact the stellar mass-to-light ratio yielded by the PCA parameter estimation, the effects are relatively small for non-extreme cases. When \ntestgalaxies \emph{observed} galaxies are binned simultaneously by $\tau_V \mu$ and ${\rm [Z]}$, the vast majority have $\tau_V \mu < 1.0$ and $-0.5 < {\rm[Z]} < 0.1$, supporting the claim that the training data are more widely-distributed in parameter space than actual MaNGA galaxies are. That is, even with the extremely permissive priors on attenuation and stellar metallicity, the vast majority of fits to observations lie in the region of ${\rm [Z]}$--$\tau_V \mu$ space for which estimates of $\log \Upsilon^*_i$ behave the best. Similar figures illustrating \qtydev{\logml{i}} and \qtydevwid{\logml{i}} in bins of signal-to-noise ratio and either ${\rm [Z]}$ or $\tau_V \mu$ have been omitted for brevity's sake, and do not cause concern.

\section{Resolved stellar mass-to-light ratios: Discussion and Conclusion}
\label{sec:discussion}

In this work, we construct a set of \nSFHs synthetic SFHs (subsampled \nsubsample times in ${\rm [Z]}$, $\tau_V$, $\mu$, and $\sigma$), perform PCA on the resulting optical spectra, and use the resulting, lower-dimensional space to fit IFS observations from the SDSS-IV/MaNGA survey. Using those fits, we estimate resolved, $i$-band stellar mass-to-light ratios for galaxies in \mplv, with uncertainties which take into account model-dependent and age-metallicity degeneracies, as well as a spectrophotometric covariance estimated from multiply-observed galaxies. The parameter-estimation strategy chosen performs well when tested on synthetic, test data generated identically to the training data, at median signal-to-noise ratios between 2 and 20 (see Figures \ref{fig:mocks_snr_color_hist_devMLi}, \ref{fig:mocks_snr_color_hist_widMLi}, and \ref{fig:mocks_snr_color_hist_devwidMLi}). We note that deviations in this intermediate-signal-to-noise regime generally lie at the $\sim 0.1$ dex level or smaller, and are mostly uncorrelated with stellar metallicity and foreground attenuation. At higher and lower median signal-to-noise, extreme values of these two parameters correlate with mis-estimates of $\Upsilon^*_i$ at the $\sim 0.2$ dex level.

\begin{figure*}
    \centering
    \includegraphics[width=0.95\textwidth]{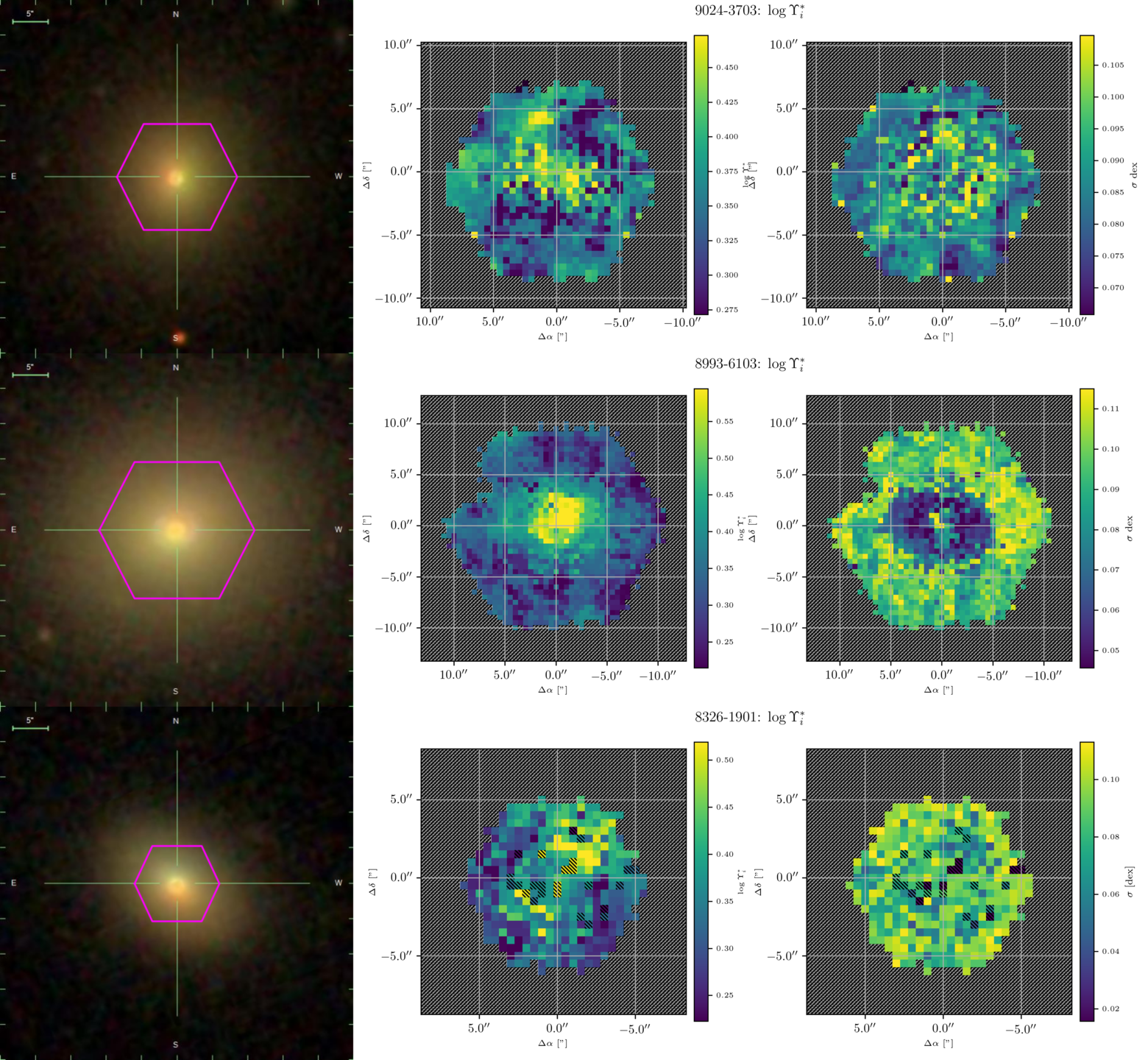}
    \caption{A selection of three early-type galaxies: In the left-hand column, the SDSS cutout with a purple hexagon denoting the approximate spatial grasp of the IFU. In the middle column, an image of the resolved estimate of $i$-band stellar mass-to-light ratio, taken as the 50$^{\rm th}$ percentile of the posterior PDF for a given spaxel. In the right-hand panel, a map of the adopted uncertainty in stellar mass-to-light ratio, taken as half the difference between the 16$^{\rm th}$ and 84$^{\rm th}$ percentiles of the posterior PDF. Spaxels with hatching top-left to bottom-right have poorly-sampled PDFs, and spaxels with hatching top-right to bottom-left have no data. Spaxels filled with dots had a numerical failure in the PC down-projection, which prevented an estimate from being made (very rare).}
    \label{fig:ETG_montage}
\end{figure*}

\begin{figure*}
    \centering
    \includegraphics[width=0.95\textwidth]{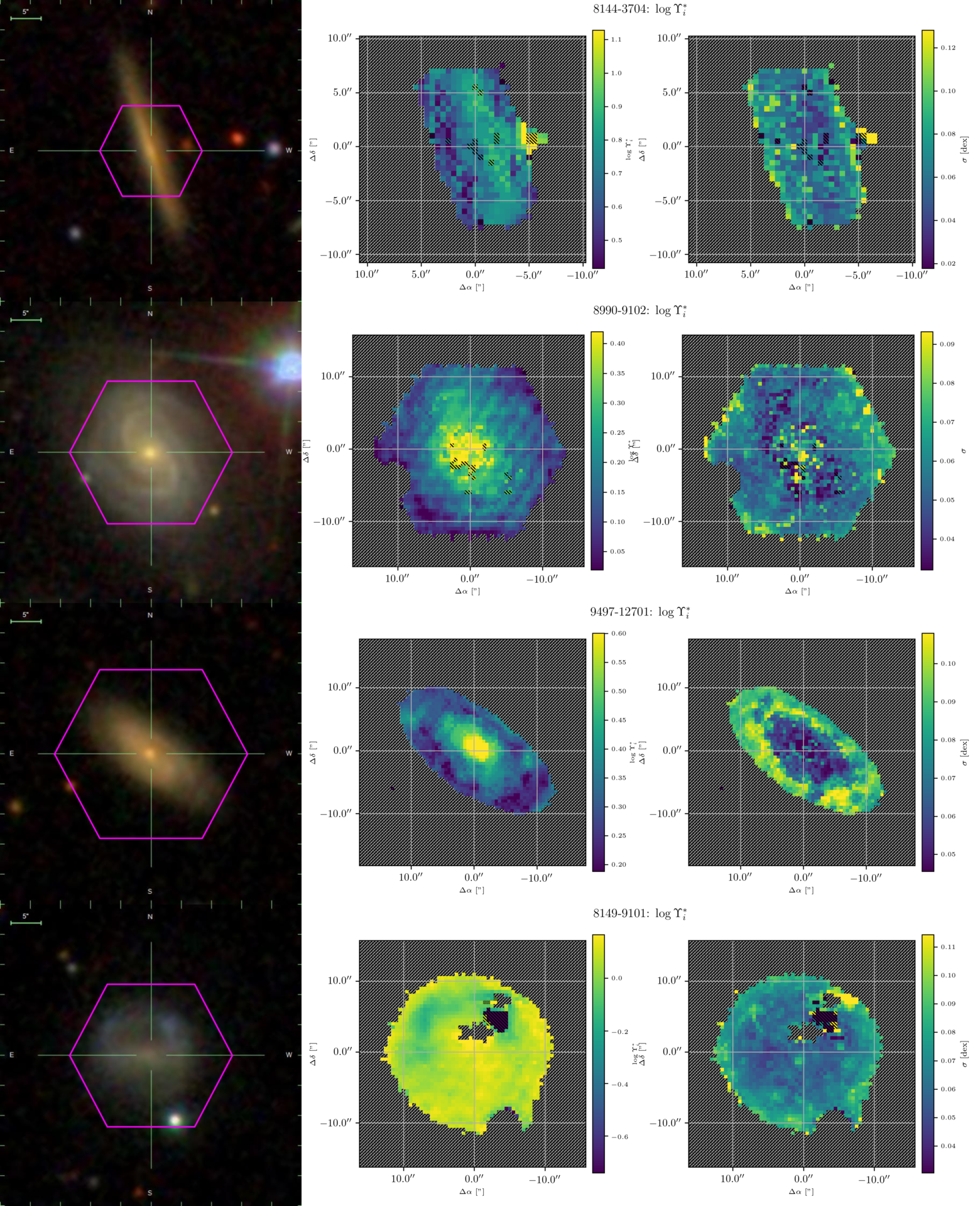}
    \caption{As Figure \ref{fig:ETG_montage}, but a selection of four late-type galaxies. 8114-3704 is seen edge-on \& exhibits a dust lane, 8990-9102 is star-forming and is seen nearly face-on, 9497-12701 is moderately-inclined and has a low SFR, and 8149-9101 is a low-mass dwarf galaxy with ongoing star-formation.}
    \label{fig:LTG_montage}
\end{figure*}

We include here sample maps of resolved stellar mass-to-light ratio for a sample of three early-type galaxies (Figure \ref{fig:ETG_montage}) and four late-type galaxies (Figure \ref{fig:LTG_montage}), including uncertainties \& data-quality masks, accompanied by a cutout image of the galaxy from legacy imaging.
The maps of resolved stellar mass-to-light ratio are spatially smooth, suggesting that the PC down-projection of two nearby spaxels indeed reflects the PSF of the data induced by the dithering and wavelength-rectification processes: The process of assembling individual science exposures (each at one of three positions on a galaxy's face, and subject to some distinct differential atmospheric refraction, attenuation by the atmosphere, etc.) detailed in \citet[][Section 9.2]{manga_drp} induces a spatial covariance between nearby spaxels. Therefore, one might expect that two spaxels that are nearby to one another might have similar estimates of mass-to-light ratio, beyond the degree to which the underlying stellar populations are similar. We do in fact qualitatively observe this smooth variation in the resolved stellar mass-to-light ratio.

\subsection{Remaining spectral-fitting systematics and degeneracies}

The use of a synthetic stellar library represents the most uncertain systematic in this work. Perhaps most importantly, the one-dimensional stellar atmospheres adopted for this work are fixed in abundance of $\alpha$ elements. In reality, ${\rm \frac{\alpha}{Fe}}$ is known to differ from the solar value in the central regions of early-type galaxies \citep{worthey_faber_92, matteucci_94, thomas_greggio_bender_99}. These are among the brightest spaxels in the survey. Indeed, such regions are occasionally observed to have poorly-determined estimates of \logml{i} (Section \ref{subsec:data_quality}). Fortunately, the MaStar project is now undertaking bright-time observations of stars using the BOSS spectrograph, the same instrument as is used for MaNGA galaxies. A resolution- and wavelength-range-matched sample of about 10,000 stars with a wide variety of stellar parameters will represent an important value-added deliverable of MaNGA, as well as a useful input to SPS codes. Secondarily, strong constraints on the prevalence and impact of non-standard stellar evolution scenarios (such as blue stragglers and blue horizontal branch) will inform future choices of SPS inputs.

\subsection{Public Data and Future Work}
The resolved estimates of stellar mass-to-light ratio treated in detail in this work will be included in the next public data release of SDSS-IV as a value-added catalog (VAC). In \citetalias{pace_19b_pca} (next in this series), we:

\begin{itemize}
    \item Further evaluate the resolved stellar mass-to-light estimates of MaNGA galaxies by transforming them to maps of stellar mass surface density and comparing to radial averages of dynamical mass surface density from the DiskMass Survey \citep{diskmass_i}.
    \item Devise a method of aperture-correcting estimates of resolved stellar mass in order to obtain estimates of total galaxy stellar mass, which will also be released to the community.
    \item Compare the PCA-derived estimates of total galaxy stellar mass to those from integrated photometry.
    \item Evaluate the factors contributing to a mass deficit in IFU-summed spectra, relative to summing stellar masses in individual spaxels.
\end{itemize}

Also provided will be light-weight \texttt{python} scripting utilities to assist in accessing the resolved mass-to-light maps\footref{footnote:software_link}. Resolved maps of additional parameters (such as dust, SFH burst diagnostics, and stellar metallicity) may be released as part of future scientific analyses (they may also be obtained from the authors, upon request).

\section*{Acknowledgements}
ZJP acknowledges the support of US National Science Foundation East Asia Pacific Summer Institute (EAPSI) Grant OISE-1613857, operated in cooperation with the Ministry of Science \& Technology (MOST) and the China Science \& Technology Exchange Center (CSTEC). ZJP expresses gratitude for the kind hospitality of Nanjing University, where a large fraction of this work took place. ZJP also acknowledges Astro Hack Week 2016, which produced valuable discussions relevant to this work. ZJP and CT acknowledge NSF CAREER Award AST-1554877. Y. C. acknowledges support from the National Key R\&D Program of China (No. 2017YFA0402700), the National Natural Science Foundation of China (NSFC grants 11573013, 11733002). MAB acknowledges NSA Award AST-1517006. The authors thank Charlie Conroy \& Ben Johnson for fast-tracking the release of the C3K templates and improvements to \texttt{FSPS}'s \texttt{python} bindings. This research made use of \texttt{Astropy}, a community-developed core \texttt{python} package for astronomy \citep{astropy}; \texttt{matplotlib} \citep{matplotlib}, an open-source \texttt{python} plotting library; and \texttt{statsmodels} \citep{seabold2010statsmodels}, a \texttt{python} library for econometrics and statistical modeling.

The authors are grateful to the anonymous referee, whose constructive input was very helpful in improving the overall quality and readability of this manuscript.

Funding for the Sloan Digital Sky Survey IV has been provided by the Alfred P. Sloan Foundation, the U.S. Department of Energy Office of Science, and the Participating Institutions. SDSS acknowledges support and resources from the Center for High-Performance Computing at the University of Utah. The SDSS web site is www.sdss.org.

SDSS is managed by the Astrophysical Research Consortium for the Participating Institutions of the SDSS Collaboration including the Brazilian Participation Group, the Carnegie Institution for Science, Carnegie Mellon University, the Chilean Participation Group, the French Participation Group, Harvard-Smithsonian Center for Astrophysics, Instituto de Astrof\'{i}sica de Canarias, The Johns Hopkins University, Kavli Institute for the Physics and Mathematics of the Universe (IPMU) / University of Tokyo, the Korean Participation Group, Lawrence Berkeley National Laboratory, Leibniz Institut f\"{u}r Astrophysik Potsdam (AIP), Max-Planck-Institut f\"{u}r Astronomie (MPIA Heidelberg), Max-Planck-Institut f\"{u}r Astrophysik (MPA Garching), Max-Planck-Institut f\"{u}r Extraterrestrische Physik (MPE), National Astronomical Observatories of China, New Mexico State University, New York University, University of Notre Dame, Observat\'{o}rio Nacional / MCTI, The Ohio State University, Pennsylvania State University, Shanghai Astronomical Observatory, United Kingdom Participation Group, Universidad Nacional Aut\'{o}noma de M\'{e}xico, University of Arizona, University of Colorado Boulder, University of Oxford, University of Portsmouth, University of Utah, University of Virginia, University of Washington, University of Wisconsin, Vanderbilt University, and Yale University.

\software{Astropy \citep{astropy}, FSPS \& FSPS-C3K \citep{fsps_1,fsps_2,fsps_3, fsps_c3k}, matplotlib \citep{matplotlib}, python-FSPS \citep{pyfsps_DFM}, statsmodels \citep{seabold2010statsmodels}}

\bibliography{main}

\appendix

\section{The MaNGA Instrumental Line-Spread Function}
\label{apdx:lsf}

The line-spread function of a spectrograph will introduce an additional ``blur" in the spectral dimension, beyond that produced by astrophysical velocity dispersion. Additionally, the width of the (presumed-Gaussian) resolution element varies across the instrument by nearly a factor of two. This has the potential to severely hamper the quality and reliability of the SFH fits. For example, assuming a constant kernel width across the full MaNGA wavelength range will:

\begin{itemize}
    \item Artificially increase measured velocity dispersions. This effect will be a pervasive and systematic bias, and will not simply increase the width of the velocity dispersion PDF.
    \item Potentially introduce other systematics related to how certain absorption lines reflect different stellar populations.
\end{itemize}

A more subtle effect results from the process of de-redshifting a spectrum into the rest frame. The width of the (observed-frame) LSF, in pixels, will be modified by a factor of $\frac{1}{1 + z}$, such that higher-redshift galaxies will seem to experience less instrumental blurring in the rest-frame. If all observed galaxies were assumed to have a redshift of zero, then this would be a $sim 10\%$ effect; however, by assuming a fiducial redshift of .04, this effect becomes less important, on average. A slight redshift bias may persist, which could be solved by producing many (redshift-dependent) PCA solutions. In practice, repeated (expensive) LSF convolutions and SVD operations defeats the speed gains of the PCA parameter fitting, and so a single fiducial redshift is deemed sufficient.

\section{Constructing synthetic observations using held-out test data}
\label{apdx:fakedata}

Here we address PCA's ability to recover properties of synthetic spectra, derived from known SFHs and stellar properties. We use held-out test data generated identically to the CSP model library to construct synthetic datacubes, with similar statistical properties to observed MaNGA galaxies, as described in Appendix \ref{apdx:fakedata}. The process is as follows:

\begin{enumerate}
    \itemsep0em
    \item Obtain a full-resolution model spectrum, along with the properties used to generate it (``truth")
    \item Read in MaNGA DRP and DAP products, which will be used to generate the remaining galaxy properties such as cosmological redshift and velocity field, without having to model them explicitly.
    \item Convolve model spectrum with fiducial instrument dispersion (interpolated to the correct wavelength grid), after adjusting for the cosmological redshift of the source \citep{cappellari_17}
    \item Redshift the model according to the velocity field from the DAP products (making a high-resolution cube in the observed frame)
    \item Scale each spaxel according to the total $r$-band flux map from the DRP products
    \item Down-sample the observed-frame model spectra onto the MaNGA instrument's wavelength grid
    \item Add noise according to the inverse-variance arrays from the DRP products
    \item Mask additional spaxels where velocity field is not well-defined
    \item Write out synthetic DRP \& DAP datacubes and ``ground-truth" values for the parameters PCA will estimate.    
\end{enumerate}

The PCA parameter estimation method is then used in the same manner as on the science data. One example is shown below in Figure \ref{fig:full_diag_fake}

\begin{figure}
    \centering
    \includegraphics[width=\textwidth,height=\textheight,keepaspectratio]{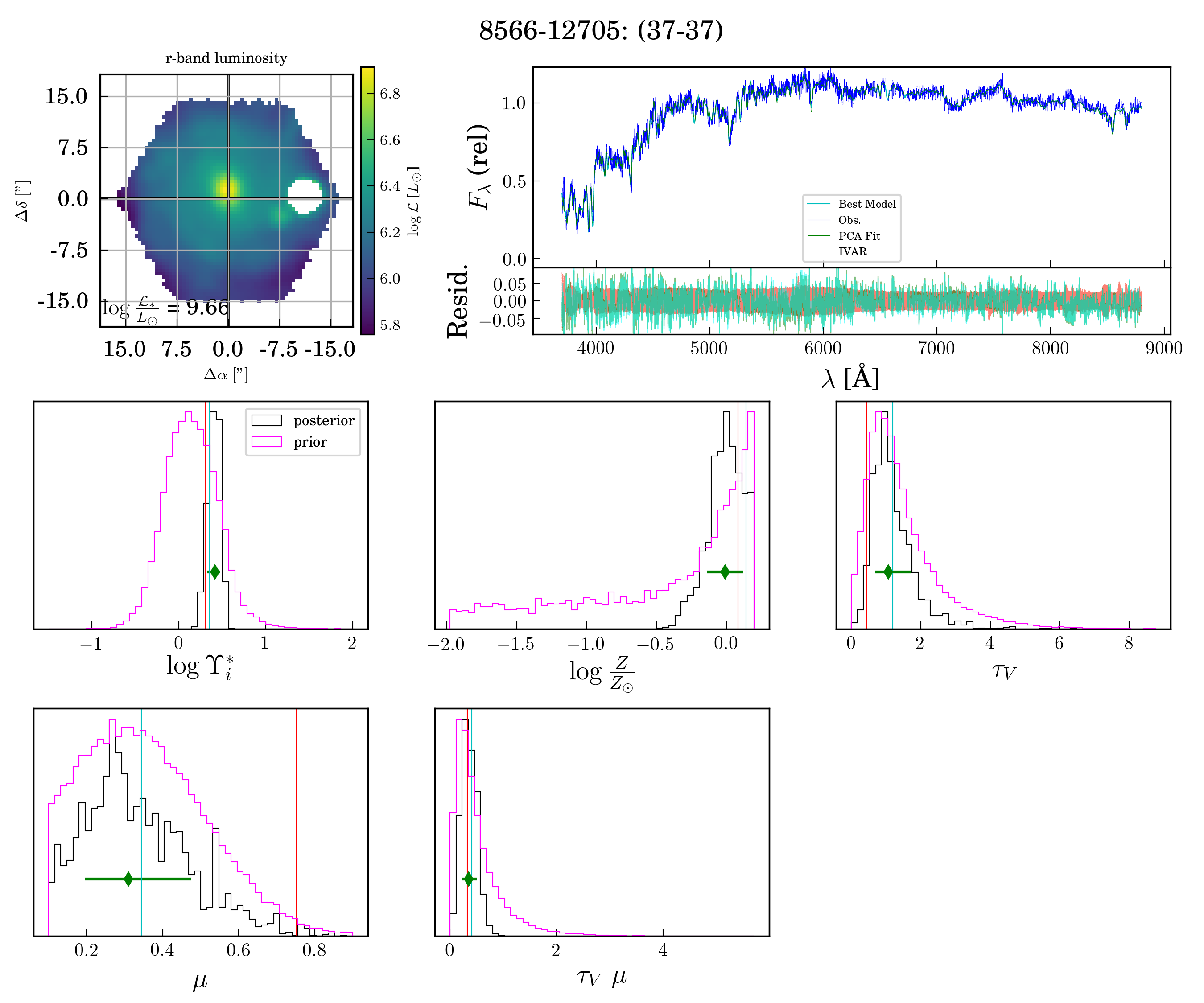}
    \caption{Full diagnostic figure for synthetic data based on the test galaxy 8566-12705. Same format as Figure \ref{fig:sample_fit}, with the addition of a vertical, red line on the parameter histograms denoting the actual value of the parameter.}
    \label{fig:full_diag_fake}
\end{figure}


\listofchanges

\end{document}